\documentclass[11pt,letterpaper]{article}

\usepackage{mathtools} \mathtoolsset{showonlyrefs=true}
\usepackage{accents}
\usepackage{amsfonts,amsmath,amssymb,amsthm,bm}
\usepackage{array}
\usepackage[english]{babel}
\usepackage{bbm}
\usepackage[font=sl,labelfont=bf]{caption}
\usepackage[usenames,dvipsnames]{color} 
\usepackage{colortbl}
\usepackage{enumerate}
\usepackage{float} \restylefloat{table}
\usepackage[left=1in,top=1in,bottom=1in,right=1in]{geometry}
\usepackage{graphicx}
\usepackage[colorlinks=true,pdfstartview=FitV,linkcolor=blue,citecolor=blue,urlcolor=blue]{hyperref}
\usepackage{mathrsfs}
\usepackage[framemethod=default]{mdframed}
\usepackage{subcaption}
\usepackage{url}
\usepackage{verbatim}
\usepackage{multirow}

\newcommand{\toDist}{\overset{\mathcal{D}}{\longrightarrow}}

\newcommand{\eqDist}{\stackrel{\mathcal{D}}{=}}

\newtheorem{theorem}{Theorem}
\newtheorem{proposition}[theorem]{Proposition}
\newtheorem{lemma}[theorem]{Lemma}

\newtheorem{assumption}[theorem]{Assumption}
\theoremstyle{definition}

\theoremstyle{remark}
\newtheorem{remark}[theorem]{Remark}

\numberwithin{equation}{section}
\numberwithin{theorem}{section}

\def\cF{\mathcal{F}}

\def\cN{\mathcal{N}}

\def\bE{\mathbb{E}}

\def\bP{\mathbb{P}}

\def\bR{\mathbb{R}}

\def\sF{\mathscr{F}}

\let\originalleft\left
\let\originalright\right
\renewcommand{\left}{\mathopen{}\mathclose\bgroup\originalleft}
\renewcommand{\right}{\aftergroup\egroup\originalright}

\newcommand{\sgn}{{\rm sgn}}
\newcommand{\wh}{\widehat}
\newcommand{\wt}{\widetilde}

\newcommand{\Var}{\textnormal{Var}} %
\newcommand{\Cov}{\textnormal{Cov}}

\definecolor{Red}{rgb}{1,0,0} 
\definecolor{DR}{rgb}{0.7,0.3,0} 

\definecolor{Green}{rgb}{0.2,0.5,0.2} 
\definecolor{Blue}{rgb}{0,0,1} 
\definecolor{PG}{rgb}{0.6,0.6,0.6} 

\definecolor{Purple}{rgb}{0.5,0.00,1}

\title{Efficient Volatility Estimation for L\'evy Processes with Jumps of Unbounded Variation}

\author{B. Cooper Boniece\thanks{{Department of Mathematics, University of Utah, Salt Lake City, UT 84112, USA ({\tt bcboniece@math.utah.edu}).}} \and Jos\'e E. Figueroa-L\'opez\thanks{{Department of Mathematics and Statistics, Washington University in St. Louis, St. Louis, MO 63130, USA ({\tt figueroa-lopez@wustl.edu}). Research supported in part by the NSF Grants: DMS-2015323, DMS-1613016.}} \and Yuchen Han\thanks{{Department of Mathematics and Statistics, Washington University in St. Louis, St. Louis, MO 63130, USA ({\tt y.han@wustl.edu}).}}}
\date{July 15, 2021}

\date{\today}

\begin{document}

\maketitle

\begin{abstract}
Statistical inference for stochastic processes based on high-frequency observations has been an active research area for more than a decade. One of the most well-known and widely studied problems is that of estimation of the quadratic variation of the continuous component of an It\^o semimartingale with jumps. Several rate- and variance-efficient estimators have been proposed in the literature when the jump component is of bounded variation. However, to date, very few methods can deal with jumps of unbounded variation. By developing new high-order expansions of the truncated moments of a L\'evy process, we construct a new rate- and variance-efficient estimator for a class of L\'evy processes of unbounded variation, whose small jumps behave like those of a stable L\'evy process with  Blumenthal-Getoor index less than $8/5$. The proposed method is based on a two-step debiasing procedure for the truncated realized quadratic variation of the process. Our Monte Carlo experiments indicate that the method outperforms other efficient alternatives in the literature in the setting covered by our theoretical framework. 
\end{abstract}

\section{Introduction}
Statistical inference for stochastic processes based on high-frequency observations has attracted considerable attention in the literature for more than a decade.  Among the many problems studied so far, none has received more attention than that of the estimation of the continuous or predictable quadratic variation of an It\^o semimartingale $X=\{X_{t}\}_{t\geq{}0}$. Specifically, if 
\[
	X_t:=X^{c}_t+J_t:=\int_0^t b_t dt+\int_0^t \sigma_s dW_s+J_t,\qquad t\in[0,T],
\]
where $W=\{W_t\}_{t\geq{}0}$ is a Wiener process and $J=\{J_{t}\}_{t\geq{}0}$ is a pure-jump It\^o semimartingale, the estimation target is $IV_T=\int_0^T\sigma^2_s ds$. This quantity, also known as the \emph{integrated volatility} or \emph{integrated variance} of $X$, has many applications, especially in finance. When $X$ is observed at times $0=t_0<t_1<\ldots<t_n=T$, in the absence of jumps, an efficient estimator of $IV_T$ is given by the realized quadratic variation $\widehat{IV}_T=\sum_{i=1}^{n}(X_{t_i}-X_{t_{i-1}})^2$ in the so-called high-frequency asymptotic regime; i.e., when $\max_{i}(t_i-t_{i-1})\to{}0$ and $T\equiv t_n$ is fixed.  In the presence of jumps, $\widehat{IV}_T$ is no longer even consistent for $IV_T$, instead converging to $IV_T+\sum_{s\leq T}(\Delta X_s)^2$, where $\Delta X_s:= X_s-X_{s^-}$ denotes the jump at time $s$.  To accommodate jump behavior, several estimators have been proposed, among which the most well-known are the truncated realized quadratic variation  and the multipower variations. We focus on the first class, which, unlike the second, is both rate- and variance-efficient, {in the Cramer-Rao lower bound sense,} when jumps are of bounded variation under certain additional conditions.%

The truncated realized quadratic variation (TRQV), first introduced by Mancini in \cite{mancini2001} and \cite{mancini2004}, is defined as 
\begin{align}\label{eq:TRV0}
\wh{C}_{n}(\varepsilon)=\sum_{i=1}^{n}(\Delta_i^n X )^2{\bf 1}_{\{|\Delta_i^n X|\leq\varepsilon\}},
\end{align}
where $\varepsilon=\varepsilon_n>0$ is a tuning parameter converging to $0$ at a suitable rate. Above, $\Delta_{i}^{n}X:=X_{t_{i}}-X_{t_{i-1}}$ is the $i$-th increment of $(X_{t})_{t\geq 0}$ based on evenly spaced observations $X_{t_{0}},\ldots,X_{t_{n}}$ over a fixed time interval $[0,T]$ (i.e., $t_{i}=ih_n$ with $h_{n}=T/n$). 
It is shown in \cite{mancini2009non} that TRQV is consistent when either the jumps have finite activity or stem from an infinite-activity L\'evy process. 
 For a semimartingale model with L\'evy jumps of  bounded variation, \cite{cont2011nonparametric} showed that the TRQV admits a feasible central limit theorem (CLT), provided that $\varepsilon_n = ch_n^\beta$ with some\footnote{N.b.: in \cite{cont2011nonparametric}, a different parameterization of the threshold parameter $\beta$ is used.} $\beta \in [\frac{1}{4-Y}, \frac{1}{2})$, where $Y\in [0,1)$ denotes the corresponding Blumenthal-Getoor index.  
In \cite{jacod2008asymptotic}, consistency was established for a general It\^o semimartingale $X$, and a corresponding CLT is given when the jumps of $X$ are of  bounded variation. In that case, the TRQV attains the optimal rate and asymptotic variance of $\sqrt{ h_n}$ and $2\int_{0}^{T}\sigma_s^4ds$, respectively.

In the presence of jumps of unbounded variation, the situation is notably different, and the problem is comparatively much less studied. In \cite{cont2011nonparametric}, it is shown that when jumps stem from a L\'evy process with stable-like small-jumps of infinite variation, the TRQV estimator $\wh{C}_n(\varepsilon)$ converges to $IV_T$ at a rate slower than $\sqrt{h_n}$. Further, in \cite{mancini2011speed} it is shown that when the jump component $J$ is a $Y$-stable L\'evy process and $\varepsilon_n=h_n^\beta$ with $\beta\in(0,1/2)$, the decomposition $\widehat{C}_{n}( \varepsilon_n)-IV_T=\sqrt{h_n}Z_n+\mathcal{R}_n$ holds, where $Z_{n}$ converges stably in law to $\mathcal{N}(0,2\int_0^T\sigma_s^4 ds)$, while $\mathcal{R}_n$ is precisely of order $\varepsilon_n^{2-Y}$ in the sense that $\mathcal{R}_n=O_P(\varepsilon_n^{2-Y})$ and $\varepsilon_n^{2-Y}=O_P(\mathcal{R}_n)$ (which decays too slowly to allow for efficiency when $Y>1$). In \cite{AmorinoGloter20}, the smoothed version of the TRQV estimator $\widehat{C}^{Sm}_{n}(\varepsilon)=\sum_{i=1}^{n}(X_{t_i}-X_{t_{i-1}})^2\varphi((X_{t_i}-X_{t_{i-1}})/\varepsilon)$ is considered\footnote{The authors in \cite{AmorinoGloter20}, in fact, consider the more general estimation of $\int_0^T f(X_s) \sigma_s^2 ds$ for functions $f$ of polynomial growth, for which $IV_T$ is a special case.}, where $\varphi\in C^{\infty}$ vanishes in $\mathbb{R}\backslash{}(-2,2)$ and $\varphi(x)=1$ for $x\in(-1,1)$. In that case, using the truncation level $\varepsilon_n:=h_n^\beta$, it is shown that $\widehat{C}^{Sm}_{n}( \varepsilon_n)-IV_T=\sqrt{h_n}Z_n+\mathcal{R}_n$ with $\mathcal{R}_n$ such that $\varepsilon_n^{-(2-Y)}\mathcal{R}_n\to c_{Y}\int \varphi(u)|u|^{1-Y}du$, for a constant $c_Y$. By taking $\varphi$ such that $\int \varphi(u)|u|^{1-Y}du=0$, a ``bias-corrected" estimator was considered under the additional condition that $Y<4/3$. Specifically, the resulting estimator is such that, for any $\tilde{\varepsilon}>0$, $\widehat{C}^{Sm}_{n}( \varepsilon_n)-IV_T=o_{P}(h_n^{1/2-\tilde\varepsilon})$, ``nearly" attaining the optimal statistical error $O_{P}(h_n^{1/2})$. Unfortunately, the construction of such an estimator requires knowledge or accurate estimation of the jump intensity index $Y$, and no feasible CLT was proved when jumps are of unbounded variation even assuming $Y$ is known. 

Until very recently, in the case of jumps of unbounded variation, the only rate- and variance-efficient estimator of the integrated volatility known in the literature  was that proposed by Jacod and Todorov \cite{JacodTodorov:2014}, under the additional condition that the jump intensity index $Y<3/2$ or the process $X$ has a symmetric jump component\footnote{The authors in \cite{JacodTodorov:2014} also constructed an estimator that is rate-efficient even in the presence of  asymmetric jumps, but its asymptotic variance is twice as big as the optimal value $2\int_{0}^{T}\sigma_s^4ds$.}. Their estimator is based on locally estimating the volatility from the empirical characteristic function of the process' increments over disjoint time intervals shrinking to $0$, but still containing an increasing number of observations. It requires two debiasing steps, which are simpler to explain for a L\'evy process $X$ with symmetric $Y$-stable jump component $J$. The first debiasing step is meant to reduce the bias introduced when attempting to estimate $\log\mathbb{E}(\cos(\omega X_{h_n}/\sqrt{h_n}))$ with $\log\big\{\frac{1}{n}\sum_{i=1}^{n}\cos\big(\omega \Delta_i^n X/\sqrt{h_n}\big)\big\}$. The second debiasing step is aimed at eliminating the second term 
in the expansion $-2\log\mathbb{E}(\cos(\omega X_{h_n}/\sqrt{h_n}))
/\omega^2=\sigma^2+2|\gamma|^{Y} \omega^{Y-2}h_n^{1-Y/2}+O(h_n)$, which otherwise diverges when multiplied by the optimal scaling $h_{n}^{-1/2}$. Using an extension of this approach, Jacod and Todorov were able to apply these techniques to a more general class of It\^o semimartingales in \cite{jacod2016efficient}, even allowing any $Y<2$, though only rate-efficient, but not variance-efficient, estimators were ultimately constructed. It is stated therein %, but not formally proved, 
that both rate- and variance-efficiency can be achieved for symmetric jump components for this more general class of semimartingales.

Recently, Mies \cite{Mies:2020}   proposed an efficient estimation method for L\'{e}vy processes based on the generalized method of moments. Specifically, for some suitable functions $f_{1},f_{2},\dots,f_{m}$ and a scaling factor $u_{n}\rightarrow\infty$, \cite{Mies:2020}  {proposed to search} for the parameter values $\wh{\boldsymbol{\theta}}=(\wh{\theta}_{1},\ldots,\wh{\theta}_{m})$ such that
\begin{align}\label{eq:MF00}
\frac{1}{n}\sum_{i=1}^{n}f_{j}\big(u_{n}\Delta_{i}^{n}X\big)-\bE_{\wh{\boldsymbol{\theta}}}\Big(f_{j}\big(u_{n}\Delta^{n}_{i}\wt{ X}\big)\Big)=0,\quad j=1,\ldots,m,
\end{align}
where $\wt{X}$ is the superposition of a Brownian motion and independent stable L\'{e}vy processes closely approximating $X$ in a certain sense. The distribution measure $\bP_{\boldsymbol{\theta}}$ of  $\wt{X}$ depends on some parameters $\boldsymbol{\theta}=(\theta_1,\ldots,\theta_m)$, one of which is the volatility $\sigma$ of $X$, and $\bE_{\boldsymbol{\theta}}(\cdot)$ denotes the expectation with respect to $\bP_{\boldsymbol{\theta}}$. %
Though in principle this method is efficient,  it suffers from several important drawbacks. First, its finite-sample performance critically depends on the chosen moment functions $f_{1},\ldots,f_{m}$. Secondly, its implementation is computationally expensive and may lead to numerical issues since it involves solving a system of nonlinear equations. Moreover,  in {addition} to the required use of a numerical solver to determine the {values} of $\widehat{\pmb{\theta}}$ in \eqref{eq:MF00}, the expectations contained therein need to be numerically evaluated since the moments $\bE_{\boldsymbol{\theta}}(f_{j}(u_{n}\Delta^{n}_{i}\wt{ X}))$ are typically not explicit. This fact introduces numerical errors that complicates its performance.

In this paper, we consider a new method to estimate the volatility of a L\'evy process under a semiparametric specification of the jump component. To the best of our knowledge, our method, together with of those in \cite{JacodTodorov:2014} and \cite{Mies:2020}, are the only efficient methods to deal with L\'evy processes with jumps of unbounded variation. The idea is natural. We simply apply debiasing steps similar to those of \cite{JacodTodorov:2014} to the TRQV of  \cite{mancini2009non}. To give the heuristics as to why this strategy works, let us consider a small-time expansion of the truncated second moment $\mathbb{E}(X_{h_n}^2{\bf 1}_{\{|X_{h_n}|\leq{}\varepsilon_n\}})$ under the asymptotic regime $\varepsilon_n/\sqrt{h_n}\to\infty$. Using a variety of techniques, including a change of probability measure, Fourier-based methods, and small-large jump decompositions, we show the following expansion:
\begin{align*}
	\mathbb{E}\left[X^2_{h_n}{\bf 1}_{\{|X_{h_n}|\leq{}\varepsilon_n\}}\right]&=\sigma^2h_n+c_1 h_n\varepsilon_n^{2-Y} +c_2 h_n^2 \varepsilon^{-Y}_n+ O\left(h_n^3 \varepsilon_n^{-Y-2}\right) + O\left(h_n^{2}\varepsilon_n^{2-2Y} \right),
\end{align*}
for certain constants $c_1,c_2\neq{}0$. Based on this expansion, it is easy to see that the rescaled bias ${\mathbb{E}[h_n^{-1/2}(\wh{C}_{n}(\varepsilon)-\sigma^2T)]}$ satisfies
\begin{align}
	\mathbb{E}\left[h_n^{-\frac{1}{2}}\left(\wh{C}_{n}(\varepsilon)-\sigma^2T\right)\right]&=
	T c_1 h_n^{-\frac{1}{2}}\varepsilon_n^{2-Y} +Tc_2 h_n^{\frac{1}{2}} \varepsilon^{-Y}_n+ O\left(h_n^{\frac{3}{2}} \varepsilon_n^{-Y-2}\right) + O\left(h_n ^{\frac{1}{2}}\varepsilon_n^{2-2Y} \right), \label{e:bias_expansion}
\end{align}
which suggests the necessity of the condition $h_n^{-1/2}\varepsilon_n^{2-Y} =o(1)$ for a feasible CLT for $\widehat C_n(\varepsilon)$ at the rate $\sqrt{h_n}$. However, together with the (necessary) condition $\varepsilon_n/\sqrt{h_n}\to\infty$, this can happen only if $Y<1$, and removal of the  first terms in \eqref{e:bias_expansion} is necessary for efficient estimation, when jumps are of unbounded variation.  To that end,  note that for any $\zeta>1$, 
\begin{align*}
	\mathbb{E}\left(\wh{C}_{n}(\zeta\varepsilon)-\wh{C}_{n}(\varepsilon)\right)&= c_1\zeta^{2-Y}-1)\varepsilon_n^{2-Y}+{\rm h.o.t.},\\
	\mathbb{E}\left(\wh{C}_{n}(\zeta^2\varepsilon)-2\wh{C}_{n}(\zeta\varepsilon)+\wh{C}_{n}(\varepsilon)\right)&=
	 c_1(\zeta^{2-Y}-1)^2\varepsilon_n^{2-Y}+{\rm h.o.t.},
\end{align*}
where ``h.o.t." means high order terms. The above formulas motivate the ``bias-corrected" estimator
\begin{align}\label{Bias1Corr}
	\wh{C}'_{n}(\varepsilon;\zeta):=\wh{C}_{n}(\varepsilon)-\frac{\left(\wh{C}_{n}(\zeta\varepsilon)-\wh{C}_{n}(\varepsilon)\right)^2}{\wh{C}_{n}(\zeta^2\varepsilon)-2\wh{C}_{n}(\zeta\varepsilon)+\wh{C}_{n}(\varepsilon)},
\end{align}
which is the essence of the debiasing procedure of \cite{JacodTodorov:2014}. 
As we shall see, the story is more complicated than what the simple heuristics above suggest. 
Our main result shows that, for a class of L\'evy processes with stable-like small jumps and some additional mild conditions, the estimator \eqref{Bias1Corr} is indeed rate- and variance-efficient provided that $Y<4/3$. Furthermore, if $Y<8/5$, a second bias correcting step will achieve both rate- and variance-efficiency (for the case $ 8/5\leq Y<2$, see remark at the end of Section \ref{Sec:DebiasMthd}). As mentioned above, our estimator provides a simple alternative method to those of \cite{JacodTodorov:2014} and \cite{Mies:2020}. Furthermore, our Monte Carlo experiments indicate improved performance for the important class of CGMY L\'evy processes (cf. \cite{CarrGemanMadanYor:2002}). 

The rest of this paper is organized as follows. Section \ref{Sec:Model} introduces the framework and assumptions as well as some known preliminary results from the literature.  Section \ref{Sec:DebiasMthd} introduces the debiasing method and main results of the paper. Section \ref{constCGMY} illustrates the performance of our method via Monte Carlo simulations and compares it to the method in \cite{JacodTodorov:2014}.  The proofs are deferred to three appendix sections.

\section{Setting and background}\label{Sec:Model}
%\BCBcomment{The title change for this section is a suggestion just to avoid the implication that we have ``results'' in this section (previously, it said "the model and preliminary results." To slightly shorten the intro and make it expositionally further from Section 2 of from your earlier work \cite{gong2021}, I dropped mention of ``the usual conditions." I also alphabetized the references, so the citation numbering is slightly different.}

In this section, we introduce the model and main assumptions, some notation, and collect several facts needed from the literature. 
Our setting is similar to that of \cite[Section 2]{FigueroaLopezGongHoudre:2016} and is described here in detail for completeness.  We consider a 1-dimensional L\'evy process $X=(X_{t})_{t\in\bR_{+}}$ defined on a complete filtered probability space $(\Omega,\sF,(\sF_{t})_{t\in\bR_{+}},\bP)$ that can be decomposed as% of the form
\begin{align}\label{eq:MnMdlX}
X_{t}=\sigma W_{t}+J_{t},\quad t\in\bR_{+},
\end{align}
%\BCBcomment{To shorten the intro and make it different from your earlier work, I dropped mention of ``the usual conditions." }%The material herein follows closely Section 2 in \cite{FigueroaLopezGongHoudre:2016} and is summarized here for completeness.
%Throughout, we set $\bR_{+}:=[0,\infty)$ and $\bR_{0}:=\bR\backslash\{0\}$.   All processes \BCBnewnew{we consider} are \BCBnewnew{defined} on a complete filtered probability space $(\Omega,\sF,\bF,\bP)$, where $\bF:=(\sF_{t})_{t\in\bR_{+}}$ satisfies the usual conditions (see \cite{Protter}). We study estimation of a L\'evy process $X=(X_{t})_{t\in\bR_{+}}$ of the form
%\begin{align}\label{eq:MnMdlX}
%X_{t}=\sigma W_{t}+J_{t},\quad t\in\bR_{+},
%\end{align}
where $W:=(W_{t})_{t\in\bR_{+}}$ is a Wiener process and $J:=(J_{t})_{t\in\bR_{+}}$ is an independent pure-jump L\'{e}vy  process with L\'{e}vy triplet $(b,0,\nu)$.  The L\'{e}vy measure $\nu$ is assumed to admit a density $s:\bR_{0}\rightarrow\bR_{+}$ of the form
\begin{align}\label{eq:levyden}
s(x):=\frac{d\nu}{dx}:=\big(C_{+}{\bf 1}_{(0,\infty)}(x)+C_{-}{\bf 1}_{(-\infty,0)}(x)\big)q(x)\,|x|^{-1-Y},\quad x\in\bR_{0}.
\end{align}
Here, $C_{\pm}>0$, $Y\in(1,2)$, and $q:\bR_{0}\rightarrow\bR_{+}$ is a bounded Borel-measurable function satisfying the following assumptions:
\begin{assumption}\label{assump:Funtq}
\hfill

\begin{itemize}
\item [(i)] $q(x)\rightarrow 1$, as $x\rightarrow 0$;
\item [(ii)] there exist $\alpha_{\pm}\neq 0$ such that
    \begin{align*}
    \int_{(0,1]}\big|q(x)-1-\alpha_{+}x\big|x^{-Y-1}dx+\int_{[-1,0)}\big|q(x)-1-\alpha_{-}x\big|x^{-Y-1}dx<\infty;
    \end{align*}
\item [(iii)] $\displaystyle{\limsup_{|x|\rightarrow\infty}\frac{|\ln q(x)|}{|x|}<\infty}$;
\item [(iv)] for any $\varepsilon>0$, $\displaystyle{\inf_{|x|<\varepsilon}q(x)>0}$;
\item [(v)] $\displaystyle{\int_{|x|>1}q(x)^{2}|x|^{-1-Y}dx<\infty}$.
\end{itemize}
\end{assumption}
\noindent These processes are sometimes called ``stable-like L\'evy processes" and were studied in \cite{FigueroaLopezGongHoudre:2016,FigueroaLopezOlafsson:2019}. In simple terms, condition (i) above says that the small jumps of the L\'evy process $X$ behave like those of a $Y$-stable L\'evy process with L\'evy measure
\begin{align}\label{Dfntildenu0}
\wt{\nu}(dx):=\big(C_{+}{\bf 1}_{(0,\infty)}(x)+C_{-}{\bf 1}_{(-\infty,0)}(x)\big)|x|^{-Y-1}dx.
\end{align}
The condition $Y\in(1,2)$ implies that $J$ has unbounded variation in that sense that 
\[
	\sum_{i=1}^{n}|J_{t_i}-J_{t_{i-1}}|\to{}\infty \quad\textnormal{a.s.,}
\]
as the partition $0=t_0<t_1<\dots<t_n=T$ is such that $\max\{t_i-t_{i-1}\}\to{}0$.
 
 As in \cite{FigueroaLopezGongHoudre:2016}, it will be important for our analysis to apply a density transformation technique \cite[Section 6.33]{Sato:1999} to ``transform" the process $J$ into a stable L\'evy process. Concretely, we can change the probability measure from $\bP$ to another locally absolutely continuous measure $\wt{\bP}$, under which $W$ is still a standard Brownian motion independent of $J$, but, under $\wt{\bP}$, $J$ has L\'evy triplet $(\tilde{b},0,\tilde{\nu})$, where $\wt{\nu}(dx)$ is given as in \eqref{Dfntildenu0} and $\wt{b}:=b+\int_{0<|x|\leq 1} x(\tilde{\nu}-\nu)(dx)$. For future reference, we recall that (see \cite{FigueroaLopezGongHoudre:2016}), for any $t\in\bR_{+}$, 
\begin{align}\label{eq:DenTranTildePP}
U_t:=\ln
\frac{d\wt{\bP}\big|_{\sF_{t}}}{d\bP\big|_{\sF_{t}}}=\lim_{\varepsilon\rightarrow 0}\left(\sum_{s\in(0,t]:|\Delta J_{s}|>\varepsilon}\varphi(\Delta J_{s})+t\int_{|x|>\varepsilon}\big(e^{-\varphi(x)}-1\big)\tilde{\nu}(dx)\right),
\end{align}
where $\varphi(x)=-\ln q(x)$. Under $\wt{\bP}$, the centered process $S:=(S_{t})_{t\in\bR_{+}}$, given by 
\begin{equation}\label{e:def_St}
S_{t}:=J_{t}-t\wt{\gamma},\quad\wt{\gamma}:=\wt{\bE}(J_{1})=\wt{b}+\int_{|x|>1}x\,\wt{\nu}(dx),
\end{equation}
is a strictly $Y$-stable process with its scale, skewness, and location parameters given by $[(C_{+}+C_{-})\Gamma(-Y)|\cos(\pi Y/2)|]^{1/Y}$, $(C_{+}-C_{-})/(C_{+}+C_{-})$, and $0$, respectively.
We recall the following bounds for the tail probabilities and density of $S_1$ under $\wt\bP$, which hereafter we denote $p_S$:
\begin{align}
%&p_{S}(z)\sim C_{\pm}|z|^{-Y-1},\quad\text{as }\,z\rightarrow\pm\infty,\,\,\,\text{respectively},\\
\label{eq:1stOrderEstDenZ}
p_{S}(\pm z)&\leq\wt{K}\,z^{-Y-1},\\
\label{eq:2ndOrderEstDenZ} \Big|p_{S}(\pm z)-C_{\pm} z^{-Y-1}\Big|&\leq\wt{K}\big(z^{-Y-1}\wedge z^{-2Y-1}\big),\\
\label{eq:1stOrderEstTailZ} \wt{\bP}\big(\!\pm\!S_{1}>z\big)&\leq\wt{K}z^{-Y},\\
\label{eq:2ndOrderEstTailZ} \bigg|\wt{\bP}\big(\!\pm\!S_{1}>z\big)-\frac{C_{\pm}}{Y}z^{-Y}\bigg|&\leq\wt{K}z^{-2Y},
\end{align}
for a constant $\widetilde{K}<\infty$. The inequalities above can be derived from the asymptotics in Section 14 of \cite{Sato:1999} (see \cite{FigueroaLopezGongHoudre:2016} for more details).

It will be useful later on to express the processes $S=(S_{t})_{t\in\bR_{+}}$ and $U:=(U_{t})_{t\in\bR_{+}}$ in terms of the jump measure $N(dt,dx)$ of the process $J$ and its compensator  $\wt{\nu}(dx)dt$ under $\wt{\bP}$. Specifically, with $\wt{N}(dt,dx):=N(dt,dx)-\wt{\nu}(dx)dt$, we have:
\begin{align}\label{eq:DecompJZpm}
J_{t}&=S_{t}+t\wt{\gamma}:=S^{+}_{t}+S^{-}_{t}+t\wt{\gamma},\\
\label{eq:DecompUpm}
U_{t}&=\wt{U}_{t}+\eta t:=\int_{0}^{t}\int_{\bR_{0}}\varphi(x)\wt{N}(ds,dx)+t\eta,
\end{align}
where
\begin{align*}
S^{+}_{t}:=\int_{0}^{t}\!\int_{(0,\infty)}x\wt{N}(dt,dx),\quad S^{-}_{t}:=\int_{0}^{t}\!\int_{(-\infty,0)}x\wt{N}(dt,dx),\quad\eta:=\int_{\bR_{0}}\big(e^{-\varphi(x)}-1+\varphi(x)\big)\wt{\nu}(dx).
\end{align*}
Under $\widetilde \bP$, $S^{+}:=(S^{+}_{t})_{t\in\bR_{+}}$ and $-S^{-}:=(-S^{-}_{t})_{t\in\bR_{+}}$ are independent one-sided $Y$-stable processes with scale, skewness, and location parameters given by $[C_{\pm}|\Gamma(-Y)\cos(\pi Y/2)|]^{{1/Y}}$, $1$, and $0$, respectively, so that 
(cf. \cite[Theorem 2.6.1]{Zolotarev:1986})
\begin{align}\label{eq:ExpMomentZpm}
\wt{\bE}\Big(e^{\mp S_{t}^{\pm}}\Big)=\exp\bigg(C_{\pm}\Gamma(-Y)\cos\bigg(\frac{\pi Y}{2}\bigg)\text{sgn}(1-Y)t\bigg)<\infty.
\end{align}

\section{ Main results}\label{Sec:DebiasMthd}
In this section, we construct an efficient estimator for the volatility parameter $\sigma^2$ based on the well-studied estimator TRQV \eqref{eq:TRV0}. Throughout, we assume the process $X=\{X_{t}\}_{t\geq{}0}$ is sampled at $n$ evenly spaced observations, $X_{t_1}, X_{t_2}, \ldots, X_{t_n}$, during a fixed time interval $[0,T]$, where for $i=0,\dots,n$,  $t_{i}=t_{i,n}= ih_{n}$ with $h_n=T/n$. As usual, we define the increments of the process as $\Delta_{i}^{n}X:=X_{t_{i}}-X_{t_{i-1}}$, $i=1,\dots,n$ and, for simplicity, assume that $T=1$.
In what follows, we sometimes approximate {$X$} by a L\'{e}vy process {$\wt{X}:=(\wt{X}_{t})_{t\in\bR_{+}}$} with characteristic triplet  $(0,\sigma^{2},\wt{\nu})$, where $\wt{\nu}$ is defined in \eqref{Dfntildenu0}.  Note that, under $\wt\bP$,
\begin{align*}
\wt{X}\eqDist\sigma W+ S,%
\end{align*}
where $\eqDist$ denotes equality in law, and, for notational simplicity,  we assume $\wt{X}$ has the same distribution under both $\wt \bP$ and $\bP$. 
As mentioned in the introduction, the TRQV estimator $\wh{C}_n(\varepsilon)$ is not efficient since it possesses a bias that vanishes at a rate slower than $n^{-1/2}$, the rate at which the ``centered'' TRQV, 
\[
	\overline{C}_{n}(\varepsilon)=\sum_{i=1}^{n}\left\{\big(\Delta_{i}^{n}X\big)^{2}{\bf 1}_{\{|\Delta_{i}^{n}X|\leq\varepsilon\}}- \bE\big(X^2_{h_n}{\bf 1}_{\{|X_{h_n}|\leq\varepsilon\}}\big)\right\},
\]
converges to Gaussianity.  To overcome this, our idea is to apply the debiasing procedure of \cite{JacodTodorov:2014} to the TRQV.  Unlike the referred work, 
our procedure is simpler, as it does not require an extra debiasing step to correct the nonlinear nature of the logarithmic transformation  employed therein nor does it require a symmetrization step to deal with asymmetric L\'evy measures.

Before constructing our estimator, we first establish an infeasible CLT for an approximately centered estimator, where the truncated moment $\bE\big(X^2_{h_n}{\bf 1}_{\{|X_{h_n}|\leq\varepsilon\}}\big)$ is replaced by the quantity $\wt\bE\big(\widetilde{X}^2_{h_n}{\bf 1}_{\{|\widetilde{X}_{h_n}|\leq\varepsilon\}}\big)$.  Below and throughout the rest of the paper, we use the usual notation $a_{n}\ll b_n$, whenever $a_n/b_n\to0$ as $n\to\infty$. %
\begin{proposition}\label{prop1}
Suppose that $1<Y<8/5$ and $h_n^{\frac{4}{8+Y}}\ll \varepsilon_n\ll  h_n^{\frac{1}{4-Y}}$. Then, as $n\to\infty$,
\begin{align}\label{limitCLT}
Z_{n}(\varepsilon_n):=\sqrt{n}\left(\sum_{i=1}^{n}\left[\big(\Delta_{i}^{n}X\big)^{2}{\bf 1}_{\{|\Delta_{i}^{n}X|\leq \varepsilon_n\}}- \bE\left(\widetilde{X}^2_{h_n}{\bf 1}_{\{|\widetilde{X}_{h_n}|\leq  \varepsilon_n\}}\right)\right]\right)\toDist \mathcal{N}(0, 2\sigma^4).
\end{align}

\end{proposition}
The role of the next result is twofold.  First, it establishes the asymptotic normality of $\widehat C_n(\varepsilon)$ using a fully specified (unobservable) centering quantity $A(\varepsilon,h)$ rather than the inexplicit centering term $\bE\big(\widetilde{X}^2_{h_n}{\bf 1}_{\{|\widetilde{X}_{h_n}|\leq  \varepsilon_n\}}\big)$ of Proposition \ref{prop1} (the quantity $A(\varepsilon,h)$  will subsequently be estimated and removed through our proposed debiasing method).  {Secondly}, subject to this centering, it provides the joint asymptotic behavior of $\widehat C_n(\varepsilon)$ and the difference $\widehat C_n(\zeta\varepsilon) -\widehat C_n(\varepsilon)$, for some $\zeta>1$; this is the main technical result from which we deduce the efficiency of our debiased estimator.
\begin{theorem}\label{thm:singleclt}
Suppose that $1<Y<8/5$ and $h_n^{\frac{4}{8+Y}}\ll \varepsilon_n\ll  h_n^{\frac{1}{4-Y}}$. Let 
\begin{align}
 \wt{Z}_n(\varepsilon) :=  \sqrt{n}\left(\widehat C_n(\varepsilon) - \sigma^2 - A(\varepsilon,h)  \right),\label{e:def_Zn}%
\end{align}
where \begin{align}\label{e:def_a(eps,h)}
    A(\varepsilon,h) 
    =\, & \frac{C_+ + C_-}{2-Y}\varepsilon^{2-Y} - (C_{+} + C_{-})\frac{(Y+1)(Y+2)}{2Y} \sigma^2 h \varepsilon^{-Y}.
\end{align}
 Then, for arbitrary $\zeta>1$,
\begin{equation}\label{e:joint_CLT_expression}
\begin{pmatrix}
\wt{Z}_n( \varepsilon_n)\\
u_n^{-1}\left(\wt Z_n( \zeta\varepsilon_n) - \wt Z_n( \varepsilon_n)\right)
\end{pmatrix}
\toDist  \mathcal{N}\left(
\begin{pmatrix}
0\\0
\end{pmatrix}, 
\begin{pmatrix}
2\sigma^4 & 0\\
0 & \frac{C_+ + C_-}{4-Y}(\zeta^{4-Y}-1)
\end{pmatrix}\right),
\end{equation}
 as $n\to\infty$, where $u_n := h_{n}^{-\frac{1}{2}}\varepsilon_{n}^{\frac{4-Y}{2}}\to 0$.
\end{theorem}
\begin{remark}
Note that \eqref{e:joint_CLT_expression} implies that
\begin{align}\label{SmplImpFML}
	\varepsilon_n^{-Y/2}\left(\frac{\widehat C_n(\zeta\varepsilon)-\widehat C_n(\varepsilon)}{\varepsilon^{2-Y}}-\frac{C_++C_-}{2-Y}(\zeta^{2-Y}-1)\right)\toDist  \mathcal{N}\left(0,\frac{C_+ + C_-}{4-Y}(\zeta^{4-Y}-1)\right).
\end{align}
In particular,
\begin{align}\label{SmplImpFMLb}
	 \frac{\widehat C_n(\zeta\varepsilon)-\widehat C_n(\varepsilon)}{\varepsilon^{2-Y}}\stackrel{\mathbb{P}}{\longrightarrow}{}\frac{C_++C_-}{2-Y}(\zeta^{2-Y}-1).
\end{align}
Expression \eqref{SmplImpFMLb} plays a role in our numerical implementation in Section \ref{constCGMY}. %
\end{remark}

We are now in a position to introduce our proposed estimator.  Let $\zeta_1,\zeta_2>1$, and set
\begin{align}
    \wt C_n '(\varepsilon, \zeta_1) &= \widehat C_n(\varepsilon)-\frac{\left(\widehat C_n(\zeta
    _1\varepsilon)-\widehat C_n(\varepsilon)\right)^2}{\widehat C_n(\zeta_1^2\varepsilon)-2\widehat C_n(\zeta_1 \varepsilon) + \widehat C_n(\varepsilon)},\label{e:debiased1}\\
    \wt C_n ''(\varepsilon, \zeta_2,\zeta_1) &= \wt C_n '(\varepsilon, \zeta_1)-\frac{\left(\wt C_n '(\zeta_2\varepsilon, \zeta_1)-\wt C_n '(\varepsilon, \zeta_1)\right)^2}{\wt C_n '(\zeta_2^2\varepsilon, \zeta_1)-2 \wt C_n '(\zeta_2\varepsilon, \zeta_1) + \wt C_n '(\varepsilon, \zeta_1)},\label{e:debiased2}
\end{align}
The next theorem is the main result of the paper.  It establishes the rate- and variance-efficiency of the two-step debiased estimator $ \wt C_n ''(\varepsilon, \zeta_2,\zeta_1)$ provided $Y<8/5$.
\begin{theorem} \label{thm:debiasclt}
Suppose that $1<Y<8/5$ and  $h_n^{\frac{4}{8+Y}}\ll \varepsilon_n\ll  h_n^{\frac{1}{4-Y}\vee \frac{1}{2+Y/2}}$. Then, as $n\to\infty$,
\begin{align*}
     \sqrt{n}\left(\wt C_n ''(\varepsilon, \zeta_2,\zeta_1) - \sigma^2 \right) \toDist  \mathcal{N}(0, 2\sigma^4).
\end{align*}
\end{theorem}
 It is customary to use power thresholds of the form $\varepsilon_n = c_0 h_n^\beta$, where $c_0>0$ and $\beta>0$ are some constants\footnote{ Though, it is shown in  \cite{gong2021} that under the setting of Section \ref{Sec:Model}, the optimal threshold $\varepsilon_n^{*}$ is such that $\varepsilon_n^{*}\sim\sqrt{(2-Y)\sigma^{2}h_n\ln(1/h_n)}$.}. In that case, the assumption   $h_n^{\frac{4}{8+Y}}\ll \varepsilon_n\ll  h_n^{\frac{1}{4-Y}\vee \frac{1}{2+Y/2}}$ in Theorem \ref{thm:debiasclt} becomes
 \begin{align}\label{IDWBHT}
   \left( \frac{1}{4-Y} \vee \frac{1}{2+Y/2}\right) < \beta < \frac{4}{8+Y}.
\end{align}
Note that the value of $\beta=5/12$ satisfies the above constraint for any value {$Y$} of the possible range considered by Theorem \ref{thm:debiasclt}.

\begin{remark}\label{DRemH}
	As a consequence of the proof of Theorem \ref{thm:debiasclt}, it follows that if $1<Y<4/3$, then only one debiasing step is needed to achieve efficiency. That is, for $1<Y<4/3$, we already have 
	\begin{align*}
     \sqrt{n}\left(\wt C_n'(\varepsilon, \zeta_1) - \sigma^2 \right) \toDist  \mathcal{N}(0, 2\sigma^4),
\end{align*}
whenever  $h_n^{\frac{1}{2Y}}\ll \varepsilon_n\ll  h_n^{\frac{1}{2+Y/2}}$.
	If $4/3\leq Y<8/5$, the proof of Theorem \ref{thm:debiasclt} shows a second debiasing step is required. These two facts suggest that further debiasing steps similar to \eqref{e:debiased1}-\eqref{e:debiased2} could {lead to} an extension of this method to handle values of $Y$ larger than $8/5$. %
	This conjecture requires significant further analysis beyond the scope of the present paper and, hence, we leave it for future research.
\end{remark}

\section{Monte Carlo performance for CGMY L\'evy processes} \label{constCGMY}
{In this section, we study the performance of the two-step debiasing procedure introduced in the previous section to the case of a L\'evy process with a CGMY jump component $J$ (cf. \cite{CarrGemanMadanYor:2002}). Specifically, we work with simulated data from the model \eqref{eq:MnMdlX}
where $\{J_t\}_{t\geq{}0}$ is a CGMY process, independent of the Brownian motion $\{W_{t}\}_{t\geq{}0}$, with L\'evy measure having a $q$-function, in the notation of \eqref{eq:levyden}, of the form:
\begin{align*}
q(x)=e^{-Mx}{\bf 1}_{(0,\infty)}(x)+e^{Gx}{\bf 1}_{(-\infty,0)}(x),
\end{align*}
and $C_{+}=C_{-}=C$. Thus, the conditions of Assumption \ref{assump:Funtq} are satisfied with $\alpha_{+}=-M$ and $\alpha_{-}=G$.
We adopt the parameter setting
\begin{align}\label{USPH0}
C=0.028,\quad G=2.318,\quad M=4.025,
\end{align}
which are similar to those used in \cite{FigueroaLopezOlafsson:2019}\footnote{\cite{FigueroaLopezOlafsson:2019} considers the asymmetric case  $\nu(dx)=C_{\sgn (x)}\bar{q}(x)|x|^{-1-Y}\,dx$ with $C_+=0.015$ and $C_-=0.041$. Here, we take $C=(C_++C_-))/2$ in order to simplify the simulation of the model.  The parameter values of $C_+$, $C_-$, $G$, and $M$ used in \cite{FigueroaLopezOlafsson:2019} were taken from \cite{Kawai}, who calibrated the tempered stable model using market option prices.}, and take $\sigma=0.2, 0.4$ and $Y=1.25, 1.35, 1.5, 1.7$, respectively. We take $T=1$ year and $n=252(6.5)(12)$, which corresponds to a frequency of $5$ minutes (assuming $252$ trading days and a $6.5$-hour trading period each day). 

In a fashion similar to \cite{JacodTodorov:2014}, for the threshold $\varepsilon = c_0h^\beta$,  we take {$c_0 = \sigma_{BV}$, where
\[
\sigma_{BV}^2= \frac{\pi}{2} \sum_{i=2}^n |\Delta_{i-1}^nX||\Delta_i^nX|,
\] 
which} is the standard Bipower variation estimator of $\sigma^2$ introduced by \cite{Barndorff}. {For the value of $\beta$ we take $\beta = \frac{5}{12}$, which, as mentioned above, satisfies the condition \eqref{IDWBHT} for any $Y\in(1,8/5)$.}
We compare the performance of the following estimators:
\begin{enumerate} 
    \item TRQV: $\wh{C}_{n}(\varepsilon)=\sum_{i=1}^{n}\big(\Delta_{i}^{n}X\big)^{2}{\bf 1}_{\{|\Delta_{i}^{n}X|\leq\varepsilon\}}$;
    
    \item 1-step debiasing estimator removing positive bias:
    \begin{align}
    &\widetilde C_{n,pb}'(\varepsilon, \zeta_1, p_1) = \widehat C_n(\varepsilon)- \eta_1 \left(\widehat C_n(\zeta_1\varepsilon)-\widehat C_n(\varepsilon)\right),%
    \\
    &\eta_1 = \frac{\widehat C_n(p_1\, \zeta
    _1\varepsilon)-\widehat C_n(p_1\, \varepsilon)}{\widehat C_n(p_1\, \zeta_1^2\varepsilon)-2\widehat C_n(p_1\, \zeta_1 \varepsilon) + \widehat C_n(p_1\, \varepsilon)} \vee 0,%
    \end{align}
with $\zeta_1=1.45$ and $p_1=0.6$, which were selected to achieve favorable estimation performance. If $\wt C_{n,pb} '(\varepsilon, \zeta_1,p_1)$ is negative, we recompute $\eta_1$ with $\varepsilon = 2 \varepsilon/3$.  This method is inspired by \cite{JacodTodorov:2014} and is motivated by the following decomposition of the bias correction term of \eqref{e:debiased1} into a product of two factors: 
    \[\frac{\left(\widehat C_n(\zeta
    _1\varepsilon)-\widehat C_n(\varepsilon)\right)}{\widehat C_n(\zeta_1^2\varepsilon)-2\widehat C_n(\zeta_1 \varepsilon) + \widehat C_n(\varepsilon)} \times \left(\widehat C_n(\zeta
    _1\varepsilon)-\widehat C_n(\varepsilon)\right),\]
    where, due to \eqref{SmplImpFMLb}, the first term estimates $(\zeta_1^{2-Y} - 1)^{-1}$, which is positive. So, we should expect $\eta_1>0$.
    
    \item 2-step debiasing estimator removing negative bias:
    \begin{align*}
    &\widetilde C_{n,nb} ''(\varepsilon, \zeta_2, \zeta_1,p_2,p_1) = \widetilde C_{n,pb} '(\varepsilon, \zeta_1, p_1)-  \eta_2 \left(\left(\widetilde C_{n,pb} '(\zeta_2\varepsilon, \zeta_1, p_1)-\widetilde C_{n,pb} '(\varepsilon, \zeta_1, p_1)\right) \vee 0 \right),\\
    &\eta_2 = \frac{\widetilde C_{n,pb} '(p_2\, \zeta_2\varepsilon,\, \zeta_1,\, p_1)-\widetilde C_{n,pb} '(p_2\, \varepsilon,\, \zeta_1,\, p_1)}{\widetilde C_{n,pb} '(p_2\, \zeta_2^2\varepsilon,\, \zeta_1,\, p_1)-2 \widetilde C_{n,pb} '(p_2\, \zeta_2\varepsilon,\, \zeta_1,\, p_1) + \widetilde C_{n,pb} '(p_2\, \varepsilon, \,\zeta_1,\, p_1)} \wedge 0,
\end{align*}
with $\zeta_1=1.45$, $\zeta_2=1.2$, $p_1=0.6$, and  $p_2=0.75$.
    If $\wt C_{n,nb} ''(\varepsilon, \zeta_2, \zeta_1,p_2,p_1)$ is negative, we recompute $\eta_2$ with $\varepsilon= 2 \varepsilon/3$.  The reason for this adjustment is the fact that $\eta_2$ is expected to be negative since it serves as estimate of $(\zeta_2^{-Y} - 1)^{-1}$. The values of the tuning parameters $\zeta_1, \zeta_2, p_1, p_2$ were selected for favorable estimation performance.

\end{enumerate}

We further compare the simulated performance of the above estimators to the estimator implemented in the Monte Carlo study in \cite{JacodTodorov:2014}. Specifically, we use the equation (5.3) in the paper \cite{JacodTodorov:2014}:
\begin{align*}
    \wh C_{\text{JT}, 53}(u_n, \zeta) &= \widehat C_{\text{JT}}(u_n)- \eta \left( \left(\widehat C_{\text{JT}}(\zeta u_n)-\widehat C_{\text{JT}}(u_n)\right)\wedge 0\right)\\
    \eta &= \frac{\widehat C_{\text{JT}}(p_0\, \zeta u_n)-\widehat C_{\text{JT}}(p_0\, u_n)}{\widehat C_{\text{JT}}(p_0\, \zeta^2u_n)-2\widehat C_{\text{JT}}(p_0\, \zeta u_n) + \widehat C_{\text{JT}}(p_0\, u_n)} \wedge 0,
\end{align*}
where $\widehat C_{\text{JT}}$ denotes their ``nonsymmetrized" two-step debiased estimator. For the parameter settings, we take
$$
	\zeta=1.5, \quad u_{n}=(\ln(1/h_{n}))^{-1/30}/\sigma_{BV}, \quad p_0=0.5,
$$
where the values of $\zeta$ and $u_n$ were those  suggested by \cite{JacodTodorov:2014} and the value of $p_0$ was chosen for favorable estimation performance.
Note that, since a L\'evy model has constant volatility, it is not necessary to localize the estimator and, hence, we treat the 1-year data as one block, which corresponds to $k_n=252(6.5)(12)$ in the notation of \cite{JacodTodorov:2014}. 

The simulation results are summarized in Tables \ref{table:s2y125}-\ref{table:s4y17}. %
 We report the sample means, standard deviations {(SDs)}, the average and {SD} of relative errors,  the MSEs, and median of absolute deviations (MADs) of each of the four estimators described above, based on 1000 simulations.

 When $\sigma=0.2$ and $Y=1.25$ or $1.35$, or when $\sigma=0.4$ and $Y=1.25$, $1.35$, or $1.5$, %
$\widetilde C_{n,nb} ''$ significantly outperforms $\widetilde C_{n,pb}'$. Also, $\widetilde C_{n,nb} ''$   has superior performance compared to $\widehat C_{\text{JT}, 53}$ as measured by MSE and MAD. Table \ref{table:s2y15} also shows that, though when $\sigma=0.2$ and $Y=1.5$, $\widetilde C_{n,nb} ''$ has larger MSE and MAD than $\widehat{C}_n$ and $\widetilde C_{n,pb}'$, it still performs better than $\widehat C_{\text{JT}, 53}$: the MSE and MAD of $\widetilde C_{n,nb} ''$  in this case are approximately {53\% and 73\%} of those of $\widehat C_{\text{JT}, 53}$, respectively.
 Tables \ref{table:s2y17} and \ref{table:s4y17} show that, when $Y=1.7$, which is not covered by our theoretical framework (c.f. Remark \ref{DRemH}), $\widetilde C_{n,nb} ''$ has larger MSE and MAD than $\widehat C_{\text{JT}, 53}$. %
Overall, we conclude that our debiasing procedure outperforms $\widehat C_{\text{JT}, 53}$ when $Y \in (1, 1.5]$}.  %

\begin{table}[ht]
\centering
\begin{tabular}{ccccccc}
 \hline
& \begin{tabular}{c} Sample Mean\end{tabular} & \begin{tabular}{c} Sample SD \end{tabular} & \begin{tabular}{c} Mean of \\ Relative Error \end{tabular} & \begin{tabular}{c} SD of \\ Relative Error \end{tabular} & \begin{tabular}{c} MSE \end{tabular} & \begin{tabular}{c} MAD \end{tabular}\\
\hline
  $\wh{C}_{n}$ & 0.034962 & 0.000455 & -0.1260 & 0.0114 & 2.56E-05 & 5.05E-03 \\ 
  $\wt C_{n,pb}'$ & 0.034962 & 0.000455 & -0.1260 & 0.0114 & 2.56E-05 & 5.05E-03 \\ 
  $\wt C_{n,nb}''$ & 0.040360 & 0.000487 & 0.0090 & 0.0122 & 3.67E-07 & 4.31E-04 \\ 
  $\wh C_{\text{JT}, 53}$ & 0.041260 & 0.000748 & 0.0315 & 0.0187 & 2.15E-06 & 1.32E-03 \\ 
  \hline
\end{tabular}
\caption{\small Estimation based on simulated $5$-minute observations of $1000$ paths over a $1$ year time horizon. The parameters are $Y=1.25$ and $\sigma=0.2$.}
\label{table:s2y125}
\end{table}

\begin{table}[ht]
\centering
\begin{tabular}{ccccccc}
 \hline
& \begin{tabular}{c} Sample Mean\end{tabular} & \begin{tabular}{c} Sample SD \end{tabular} & \begin{tabular}{c} Mean of \\ Relative Error \end{tabular} & \begin{tabular}{c} SD of \\ Relative Error \end{tabular} & \begin{tabular}{c} MSE \end{tabular} & \begin{tabular}{c} MAD \end{tabular}\\
\hline
  $\wh{C}_{n}$ & 0.036021 & 0.000475 & -0.0995 & 0.0119 & 1.61E-05 & 3.96E-03 \\ 
  $\wt C_{n,pb}'$ & 0.036021 & 0.000475 & -0.0995 & 0.0119 & 1.61E-05 & 3.96E-03 \\ 
  $\wt C_{n,nb}''$ & 0.041365 & 0.000519 & 0.0341 & 0.0130 & 2.13E-06 & 1.36E-03 \\ 
  $\wh C_{\text{JT}, 53}$ & 0.042505 & 0.001467 & 0.0626 & 0.0367 & 8.43E-06 & 2.61E-03 \\ 
  \hline
\end{tabular}
\caption{\small Estimation based on simulated $5$-minute observations of $1000$ paths over a $1$ year time horizon. The parameters are $Y=1.35$ and $\sigma=0.2$.}
\label{table:s2y135}
\end{table}

\begin{table}[ht]
\centering
\begin{tabular}{ccccccc}
 \hline
& \begin{tabular}{c} Sample Mean\end{tabular} & \begin{tabular}{c} Sample SD \end{tabular} & \begin{tabular}{c} Mean of \\ Relative Error \end{tabular} & \begin{tabular}{c} SD of \\ Relative Error \end{tabular} & \begin{tabular}{c} MSE \end{tabular} & \begin{tabular}{c} MAD \end{tabular}\\
  \hline
  $\wh{C}_{n}$  & 0.040159 & 0.000517 & 0.0040 & 0.0129 & 2.93E-07 & 3.52E-04 \\ 
  $\wt C_{n,pb}'$ & 0.040159 & 0.000517 & 0.0040 & 0.0129 & 2.93E-07 & 3.52E-04 \\ 
  $\wt C_{n,nb}''$ & 0.045790 & 0.000556 & 0.1448 & 0.0139 & 3.38E-05 & 5.78E-03 \\ 
  $\wh C_{\text{JT}, 53}$ & 0.047771 & 0.001907 & 0.1943 & 0.0477 & 6.40E-05 & 7.89E-03 \\ 
  \hline
\end{tabular}
\caption{\small Estimation based on simulated $5$-minute observations of $1000$ paths over a $1$ year time horizon. The parameters are $Y=1.5$ and $\sigma=0.2$.}
\label{table:s2y15}
\end{table}

\begin{table}[ht]
\centering
\begin{tabular}{ccccccc}
 \hline
& \begin{tabular}{c} Sample Mean\end{tabular} & \begin{tabular}{c} Sample SD \end{tabular} & \begin{tabular}{c} Mean of \\ Relative Error \end{tabular} & \begin{tabular}{c} SD of \\ Relative Error \end{tabular} & \begin{tabular}{c} MSE \end{tabular} & \begin{tabular}{c} MAD \end{tabular}\\
  \hline
  $\wh{C}_{n}$ & 0.065650 & 0.000859 & 0.6413 & 0.0215 & 6.59E-04 & 2.56E-02 \\ 
  $\wt C_{n,pb}'$ & 0.065650 & 0.000859 & 0.6413 & 0.0215 & 6.59E-04 & 2.56E-02 \\ 
  $\wt C_{n,nb}''$ & 0.075250 & 0.000979 & 0.8813 & 0.0245 & 1.24E-03 & 3.53E-02 \\ 
  $\wh C_{\text{JT}, 53}$ & 0.071881 & 0.001498 & 0.7970 & 0.0374 & 1.02E-03 & 3.20E-02 \\ 
  \hline
\end{tabular}
\caption{\small Estimation based on simulated $5$-minute observations of $1000$ paths over a $1$ year time horizon. The parameters are $Y=1.7$ and $\sigma=0.2$.}
\label{table:s2y17}
\end{table}

\begin{table}[ht]
\centering
\begin{tabular}{ccccccc}
 \hline
& \begin{tabular}{c} Sample Mean\end{tabular} & \begin{tabular}{c} Sample SD \end{tabular} & \begin{tabular}{c} Mean of \\ Relative Error \end{tabular} & \begin{tabular}{c} SD of \\ Relative Error \end{tabular} & \begin{tabular}{c} MSE \end{tabular} & \begin{tabular}{c} MAD \end{tabular}\\
  \hline
  $\wh{C}_{n}$ & 0.136732 & 0.001754 & -0.1454 & 0.0110 & 5.44E-04 & 2.33E-02 \\ 
  $\wt C_{n,pb}'$ & 0.136732 & 0.001754 & -0.1454 & 0.0110 & 5.44E-04 & 2.33E-02 \\ 
  $\wt C_{n,nb}''$ & 0.158973 & 0.001940 & -0.0064 & 0.0121 & 4.82E-06 & 1.48E-03 \\ 
  $\wh C_{\text{JT}, 53}$ & 0.161206 & 0.002464 & 0.0075 & 0.0154 & 7.52E-06 & 1.86E-03 \\ 
   \hline
\end{tabular}
\caption{\small Estimation based on simulated $5$-minute observations of $1000$ paths over a $1$ year time horizon. The parameters are $Y=1.25$ and $\sigma=0.4$.}
\label{table:s4y125}
\end{table}

\begin{table}[ht]
\centering
\begin{tabular}{ccccccc}
 \hline
& \begin{tabular}{c} Sample Mean\end{tabular} & \begin{tabular}{c} Sample SD \end{tabular} & \begin{tabular}{c} Mean of \\ Relative Error \end{tabular} & \begin{tabular}{c} SD of \\ Relative Error \end{tabular} & \begin{tabular}{c} MSE \end{tabular} & \begin{tabular}{c} MAD \end{tabular}\\
 \hline
  $\wh{C}_{n}$ & 0.138156 & 0.001821 & -0.1365 & 0.0114 & 4.80E-04 & 2.18E-02 \\ 
  $\wt C_{n,pb}'$ & 0.138156 & 0.001821 & -0.1365 & 0.0114 & 4.80E-04 & 2.18E-02 \\ 
  $\wt C_{n,nb}''$ & 0.160316 & 0.001983 & 0.0020 & 0.0124 & 4.03E-06 & 1.37E-03 \\ 
  $\wh C_{\text{JT}, 53}$ & 0.162448 & 0.002571 & 0.0153 & 0.0161 & 1.26E-05 & 2.90E-03 \\ 
   \hline
\end{tabular}
\caption{\small Estimation based on simulated $5$-minute observations of $1000$ paths over a $1$ year time horizon. The parameters are $Y=1.35$ and $\sigma=0.4$.}
\label{table:s4y135}
\end{table}

\begin{table}[ht]
\centering
\begin{tabular}{ccccccc}
 \hline
& \begin{tabular}{c} Sample Mean\end{tabular} & \begin{tabular}{c} Sample SD \end{tabular} & \begin{tabular}{c} Mean of \\ Relative Error \end{tabular} & \begin{tabular}{c} SD of \\ Relative Error \end{tabular} & \begin{tabular}{c} MSE \end{tabular} & \begin{tabular}{c} MAD \end{tabular}\\
  \hline
  $\wh{C}_{n}$ & 0.143553 & 0.001849 & -0.1028 & 0.0116 & 2.74E-04 & 1.65E-02 \\ 
  $\wt C_{n,pb}'$ & 0.143553 & 0.001849 & -0.1028 & 0.0116 & 2.74E-04 & 1.65E-02 \\ 
  $\wt C_{n,nb}''$ & 0.165903 & 0.002111 & 0.0369 & 0.0132 & 3.93E-05 & 5.90E-03 \\ 
  $\wh C_{\text{JT}, 53}$ & 0.167605 & 0.002659 & 0.0475 & 0.0166 & 6.49E-05 & 7.69E-03 \\ 
   \hline
\end{tabular}
\caption{\small Estimation based on simulated $5$-minute observations of $1000$ paths over a $1$ year time horizon. The parameters are $Y=1.5$ and $\sigma=0.4$.}
\label{table:s4y15}
\end{table}

\begin{table}[ht]
\centering
\begin{tabular}{ccccccc}
 \hline
& \begin{tabular}{c} Sample Mean\end{tabular} & \begin{tabular}{c} Sample SD \end{tabular} & \begin{tabular}{c} Mean of \\ Relative Error \end{tabular} & \begin{tabular}{c} SD of \\ Relative Error \end{tabular} & \begin{tabular}{c} MSE \end{tabular} & \begin{tabular}{c} MAD \end{tabular}\\
 \hline  
  $\wh{C}_{n}$ & 0.171569 & 0.002136 & 0.0723 & 0.0134 & 1.38E-04 & 1.16E-02 \\ 
  $\wt C_{n,pb}'$ & 0.171569 & 0.002136 & 0.0723 & 0.0134 & 1.38E-04 & 1.16E-02 \\ 
  $\wt C_{n,nb}''$ & 0.197880 & 0.002409 & 0.2367 & 0.0151 & 1.44E-03 & 3.78E-02 \\ 
  $\wh C_{\text{JT}, 53}$ & 0.179758 & 0.017441 & 0.1235 & 0.1090 & 6.95E-04 & 2.43E-02 \\ 
   \hline
\end{tabular}
\caption{\small Estimation based on simulated $5$-minute observations of $1000$ paths over a $1$ year time horizon. The parameters are $Y=1.7$ and $\sigma=0.4$.}
\label{table:s4y17}
\end{table}
\newpage

\appendix
\section{Proofs for main theorems}

Throughout the proofs, we routinely make use of the following fact: under the assumption  $h_n^{\frac{4}{8+Y}}\ll \varepsilon_n$ made throughout Section \ref{Sec:DebiasMthd}, it holds  $e^{-\frac{\varepsilon^2_n}{2\sigma^2h_n}}\ll h_n^s$ for any $s>0$ . For notational simplicity, we also often omit the subscript $n$ in $h_n$ and $\varepsilon_n$.

\begin{proof}[Proof of Proposition \ref{prop1}]
Note that the variable $Z_n(\varepsilon)$ of \eqref{limitCLT}
 can be decomposed as follows:
\begin{align}
    Z_{n}(\varepsilon)
    &= \sqrt{n}\,\sum_{i=1}^{n}\left(\big(\Delta_{i}^{n}X\big)^{2}\,{\bf 1}_{\{|\Delta_{i}^{n}X|\leq\varepsilon\}}- \bE\big(\big(\Delta_{i}^{n}X\big)^{2}\,{\bf 1}_{\{|\Delta_{i}^{n}X|\leq\varepsilon\}}\big) \right) \\
    &\quad\qquad\quad + n^{3/2} \left(\bE\big(\big(\Delta_{i}^{n}X\big)^{2}\,{\bf 1}_{\{|\Delta_{i}^{n}X|\leq\varepsilon\}}\big) - \bE\big(\widetilde{X}^2_{h}\,{\bf 1}_{\{|\widetilde{X}_{h}|\leq\varepsilon\}}\big)\right)\\
    &=: T_1 + T_2.
\end{align}
Define $\xi_i^n := \sqrt{n}\left(\big(\Delta_{i}^{n}X\big)^{2}{\bf 1}_{\{|\Delta_{i}^{n}X|\leq\varepsilon\}}-\bE\big(\big(\Delta_{i}^{n}X\big)^{2}\,{\bf 1}_{\{|\Delta_{i}^{n}X|\leq\varepsilon\}}\big)\right)$ and observe that $\xi_i^n$, $1\leq i\leq n$ are independent and $\bE\left(\xi_i^n\right)=0$. Note that Lemma \ref{prop:EX2} implies that
\begin{align}\label{asymptotic_variance_00}
\Var \left(\left(\Delta_i^n X\right)^{2}\,{\bf 1}_{\{|\Delta_i^n X|\leq\varepsilon\}}\right) = 2\sigma^4 h^2+ \frac{C_+ + C_-}{4-Y}h\varepsilon^{4-Y} + o\left(h\varepsilon^{4-Y}\right),
\end{align}
 and, therefore, 
\begin{align}
    {\rm Var}(Z_{n}(\varepsilon))&=\sum_{i=1}^n \bE\left((\xi_i^n)^2\right) = n^2\, \Var \left(\left(\Delta_i^n X\right)^{2}\,{\bf 1}_{\{|\Delta_i^n X|\leq\varepsilon\}}\right) \\
    &= 2\sigma^4 + \frac{C_+ + C_-}{4-Y}\,h^{-1}\varepsilon^{4-Y} + o\left(h^{-1}\varepsilon^{4-Y}\right)\longrightarrow{} 2\sigma^4. \label{e:asymptotic_variance_T1}
\end{align}
 Also, for any $\delta >0$,
note that 
\begin{align*}
    \bE\left(|\xi_i^n|^2 \, {\bf 1}_{\{|\xi_i^n|>\delta\}}\right) \leq  \delta^{-2} \, \bE\left(\big|\xi_i^n\big|^4\right)&= \delta^{-2} n^2\, \bE\left(\left(\big(\Delta_{i}^{n}X\big)^{2}{\bf 1}_{\{|\Delta_{i}^{n}X|\leq\varepsilon\}}-\bE\big(\big(\Delta_{i}^{n}X\big)^{2}\,{\bf 1}_{\{|\Delta_{i}^{n}X|\leq\varepsilon\}}\big)\right)^4\right)\\
    &= \delta^{-2} h^{-2} \left(O\left(h^4\right) + O\left(h\varepsilon^{8-Y}\right) \right),
\end{align*}
where in the last two lines we again apply Lemma \ref{prop:EX2}.
Thus, $\sum_{i=1}^n \bE\left(|\xi_i^n|^2 \, {\bf 1}_{\{|\xi_i^n|>\delta\}}\right)=o(1)$ and, 
  by the Lindeberg-Feller CLT, $T_1\toDist \cN(0,2\sigma^4)$.
Next, we deal with $T_2$.  By Proposition \ref{lemma:2E}, for all small $\delta>0$,
\begin{align}
    T_2 &
    =O\left(h^{\frac{3}{2}}\varepsilon^{-Y-2}\right)+O\big(\varepsilon^{1-\frac{Y}{2}}\big) + O\left(h^{\frac{1}{2}}\varepsilon^{2-2Y}\right) + O(h^{-\frac{1}{2}-\delta} \varepsilon^{3-Y+\delta Y}).
\end{align}
The  second term of $T_2$ clearly converges to $0$ as $h\to 0$.   For the remaining terms, based on the assumption $h_n^{\frac{4}{8+Y}}\ll \varepsilon_n\ll  h_n^{\frac{1}{4-Y}}$, we have:
\begin{align*}
	&h^{\frac{3}{2}}\varepsilon^{-Y-2}\ll h^{\frac{3}{2}}\Big(h^{\frac{4}{8+Y}}\Big)^{-Y-2}=h^{\frac{5(8/5-Y)}{2(8+Y)}}\to{}0,\\
	&h^{\frac{1}{2}}\varepsilon^{2-2Y}\ll h^{\frac{1}{2}}\Big(h^{\frac{4}{8+Y}}\Big)^{2-2Y}=h^{\frac{15(8/5-Y)}{2(8+Y)}}\to{}0,\\
	&h^{-\frac{1}{2}+\delta}\varepsilon^{3-Y+\delta Y}\ll h^{-\frac{1}{2}+\delta}\Big(h^{\frac{1}{4-Y}}\Big)^{3-Y+\delta Y}
	=h^{\frac{2-Y+8\delta}{2(4-Y)}}\to{}0,
\end{align*}
if $\delta$ is chosen small enough. We then conclude the result.
\end{proof}
\bigskip
\bigskip
\begin{proof}[Proof of Theorem \ref{thm:singleclt}]
Throughout the proof, let
\[
	\Phi_n:=\, u_n^{-1}\left(\wt{Z}_n(\zeta \varepsilon)-\wt{Z}_n(\varepsilon)\right).\]
First note that, by Propositions \ref{lemma:Estable} and \ref{lemma:2E}, for all small $\delta>0$,
\begin{align*}
    \bE(\wt Z_n(\varepsilon))
    &= n^{3/2}\left(\bE\left(X_h^2{\bf 1}_{\{|X_h|\leq\varepsilon\}}\right) -\bE{\big(\widetilde{X}^2_{h}\,{\bf 1}_{\{|\widetilde{X}_{h}|\leq\varepsilon\}}\big)}\right) + n^{3/2}\left(\bE{\big(\widetilde{X}^2_{h}\,{\bf 1}_{\{|\widetilde{X}_{h}|\leq\varepsilon\}}\big)} - \sigma^2 h - h A(\varepsilon,h) \right)\\
    &=O\left(h^{\frac{3}{2}}\varepsilon^{-Y-2}\right)+O\big(\varepsilon^{1-\frac{Y}{2}}\big) + O\left(h^{\frac{1}{2}}\varepsilon^{2-2Y}\right) + O(h^{-\frac{1}{2}-\delta} \varepsilon^{3-Y+\delta Y}).
\end{align*}
Then, recalling that $u_n= h^{-\frac{1}{2}}\varepsilon^{\frac{4-Y}{2}} \to0$,
\begin{align*}
    \bE\left(\Phi_n\right) &= h^{\frac{1}{2}}\varepsilon^{-\frac{4-Y}{2}}\, \bE\left(\wt{Z}_n(\zeta \varepsilon)-\wt{Z}_n(\varepsilon)\right)\\
    &=O\left(h^{2}\varepsilon^{-\frac{Y+8}{2}}\right)+O\left(h^{\frac{1}{2}}\varepsilon^{-1}\right) + O\left(h\varepsilon^{-\frac{3Y}{2}}\right) + O\left(h^{-\delta} \varepsilon^{\frac{2-Y+2\delta Y}{2}}\right).
\end{align*}
Our Assumption, $h_n^{\frac{4}{8+Y}}\ll \varepsilon_n\ll  h_n^{\frac{1}{4-Y}}$, implies that the terms above converge to $0$ since
\begin{align*}
	&h^{2}\varepsilon^{-\frac{Y+8}{2}}\ll h^{2}\Big(h^{\frac{4}{8+Y}}\Big)^{-\frac{Y+8}{2}}=1,\quad h^{\frac{1}{2}}\varepsilon^{-1}\ll h^{\frac{1}{2}}\Big(h^{\frac{4}{8+Y}}\Big)^{-1}=h^{\frac{Y}{2(8+Y)}}\to{}0,\\
	&h\varepsilon^{-\frac{3Y}{2}}\ll h\Big(h^{\frac{4}{8+Y}}\Big)^{-\frac{3Y}{2}}=h^{\frac{10(8/5-Y)}{2(8+Y)}}\to{}0,\\
	&h^{-\delta} \varepsilon^{\frac{2-Y+2\delta Y}{2}}
	\ll h^{-\delta}\Big(h^{\frac{1}{4-Y}}\Big)^{\frac{2-Y+2\delta Y}{2}}
	=h^{\frac{(2-Y)(1-4\delta)}{2(4-Y)}}\to{}0,
\end{align*}
if $\delta$ is small enough. By \eqref{e:asymptotic_variance_T1}, we know $\Var(\wt Z_n(\varepsilon)) =\Var(Z_n(\varepsilon)) \rightarrow 2\sigma^4$. Next, by Lemma \ref{prop:EX2}, for $\zeta>1$, 
\begin{align}
    & \Cov\left(\left(\Delta_i^n X\right)^{2}\,{\bf 1}_{\{|\sigma W_{h}+J_{h}|\leq\varepsilon\}},\, \left(\Delta_i^n X\right)^{2}\,{\bf 1}_{\{|\sigma W_{h}+J_{h}|\leq\zeta\varepsilon\}}\right)\\
    &\quad = \bE \left(\Delta_i^n X\right)^{4}\,{\bf 1}_{\{|\sigma W_{h}+J_{h}|\leq\varepsilon\}} - \bE \big(\left(\Delta_i^n X\right)^{2}\,{\bf 1}_{\{|\sigma W_{h}+J_{h}|\leq\varepsilon\}}\big) \bE\big( \left(\Delta_i^n X\right)^{2}\,{\bf 1}_{\{|\sigma W_{h}+J_{h}|\leq  \zeta \varepsilon\}}\big)\\
    &\quad= 2 \sigma^{4}h^{2} + \frac{C_+ + C_-}{4-Y}\,h\varepsilon^{4-Y} + O\left(h\varepsilon^{4-\frac{Y}{2}}\right) + O\left(h^{2}\varepsilon^{2-Y}\right)\\ %
   &\quad =2 \sigma^{4}h^{2} + \frac{C_+ + C_-}{4-Y}\,h\varepsilon^{4-Y} + o(h\varepsilon^{4-Y}), \label{e:Cov(eps,zeta*eps)}
\end{align}
where the last line follows since $h^{\frac{4}{8+Y}}\ll \varepsilon$ implies $h^{2}\varepsilon^{2-Y} \ll h \varepsilon^{4-Y}$.   
Using expressions \eqref{asymptotic_variance_00} and \eqref{e:Cov(eps,zeta*eps)}, we get:
\begin{align*}
    &\Var\left(\left(\Delta_i^n X\right)^{2}\,{\bf 1}_{\{|\Delta_i^n X|\leq\varepsilon\}} - \left(\Delta_i^n X\right)^{2}\,{\bf 1}_{\{|\Delta_i^n X|\leq\zeta\varepsilon\}}\right)\\
    &\quad= \Var\left(\left(\Delta_i^n X\right)^{2}\,{\bf 1}_{\{|\Delta_i^n X|\leq\varepsilon\}}\right) + \Var\left(\left(\Delta_i^n X\right)^{2}\,{\bf 1}_{\{|\Delta_i^n X|\leq\zeta\varepsilon\}}\right)\\
    & \quad\qquad - 2\,\Cov\left(\left(\Delta_i^n X\right)^{2}\,{\bf 1}_{\{|\Delta_i^n X|\leq\varepsilon\}},\, \left(\Delta_i^n X\right)^{2}\,{\bf 1}_{\{|\Delta_i^n X|\leq\zeta\varepsilon\}}\right)\\
    &\quad= \frac{C_+ + C_-}{4-Y}\,h\varepsilon^{4-Y}(\zeta^{4-Y}-1) + o(h\varepsilon^{4-Y}). 
\end{align*}
Hence, 
\begin{align*}
    \Var(\Phi_n) &= u_n^{-2}\, h^{-1} \Var\left(\wh{C}_n(\zeta\varepsilon) - \wh{C}_n(\varepsilon)\right) \\
    &= u_n^{-2}\, h^{-1} \sum_{i=1}^n \Var\left(\left(\Delta_i^n X\right)^{2}\,{\bf 1}_{\{|\Delta_i^n X|\leq\varepsilon\}} - \left(\Delta_i^n X\right)^{2}\,{\bf 1}_{\{|\Delta_i^n X|\leq\zeta\varepsilon\}}\right) \\
    &= \frac{C_+ + C_-}{4-Y}\, (\zeta^{4-Y}-1) + o(1).
\end{align*}
Now, clearly $\Cov\left(\wt Z_n(\varepsilon), \wt Z_n(\zeta \varepsilon) \right)= h^{-1}\Cov\left(\widehat C_n(\varepsilon),\widehat C_n(\zeta\varepsilon) \right)$.  Also, using expressions \eqref{asymptotic_variance_00}  and \eqref{e:Cov(eps,zeta*eps)}, we see 
$$
\Cov\left(\widehat C_n(\varepsilon),\widehat C_n(\zeta\varepsilon) \right)- \Var\left(\widehat C_n(\varepsilon) \right) = o(\varepsilon^{4-Y}).
$$
Therefore,
\begin{align*}
    \Cov\left(\wt Z_n(\varepsilon), \Phi_n \right) &= u_n^{-1}\, \Cov\left(\wt Z_n(\varepsilon), \wt Z_n(\zeta \varepsilon) \right) - u_n^{-2}\, \Var(\wt Z_n(\varepsilon))\\
    &= u_n^{-1}h^{-1} \Big(\Cov\left(\widehat C_n(\varepsilon),\widehat C_n(\zeta\varepsilon) \right)- \Var\left(\widehat C_n(\varepsilon) \right)\Big)\\
    &= u_n^{-1} \,h^{-1} \times o(\varepsilon^{4-Y})
    = o(1),
\end{align*}
where the last equality follows from our assumption $\varepsilon_n\ll  h_n^{\frac{1}{4-Y}}$.
Thus,
\[\Var\left(\begin{pmatrix}
\wt Z_n(\varepsilon)\\
\Phi_n %
\end{pmatrix}\right)
\rightarrow 
\begin{pmatrix}
2\sigma^4 & 0\\
0 & \frac{C_+ + C_-}{4-Y}(\zeta^{4-Y}-1)
\end{pmatrix}.
\]
We now turn to apply the Lindeberg-Feller CLT. To that end, define
\[
{\bf B}_{i,n} :=  n \begin{pmatrix}
z_{i,n}\\
\phi_{i,n}
\end{pmatrix}:=n \begin{pmatrix}
\left(\left(\Delta_i^n X\right)^{2}\,{\bf 1}_{\{|\Delta_i^n X|\leq\varepsilon\}} - h \sigma^2 - h A(\varepsilon, h)\right)\\
u_n^{-1} \left(\left(\Delta_i^n X\right)^{2}\,{\bf 1}_{\{|\Delta_i^n X|\leq \zeta \varepsilon\}} - h A(\zeta \varepsilon, h) -\left(\Delta_i^n X\right)^{2}\,{\bf 1}_{\{|\Delta_i^n X|\leq \varepsilon\}} + h A(\varepsilon, h)\right)
\end{pmatrix},\]
so \[\begin{pmatrix}
\wt Z_n(\varepsilon)\\
\Phi_n %
\end{pmatrix} = 
\sqrt n  \
\sum_{i=1}^n \begin{pmatrix} z_{i,n}\\
\phi_{i,n}
\end{pmatrix} =
\sqrt{n}\left(\frac{1}{n}\sum_{i=1}^n {\bf B}_{i,n}\right).\]
We claim that, for any $\delta>0$,%
\begin{equation}\label{e:lindeberg_conditions_Bin}
\lim_{n\rightarrow\infty} \sum_{i=1}^n \bE\left(\left\Vert{\bf B}_{i,n}/\sqrt{n}\right\Vert^2 \, {\bf 1}_{\{\left\Vert{\bf B}_{i,n}\right\Vert\geq \delta\sqrt{n}\}}\right)=0.
\end{equation}
Indeed, observe
\begin{align}
 \bE\left(\left\Vert{\bf B}_{i,n}/\sqrt{n}\right\Vert^2 \, {\bf 1}_{\{\left\Vert{\bf B}_{i,n}\right\Vert\geq \delta\sqrt{n}\}}\right)&\leq \delta^{-2}n^{-2} \bE\left(\left\Vert{\bf B}_{i,n}\right\Vert^4\right)\\
 & = \delta^{-2}n^2\big(\bE z_{1,n}^4 + \bE \phi_{1,n}^4 + 2\bE (z_{1,n}^2\phi_{1,n}^2)\big).\label{e:B_i,n_bound}
\end{align}
From Lemma \ref{prop:EX2} and our assumption $h_n^{\frac{4}{8+Y}}\ll \varepsilon_n$, it holds that for each $k\geq 1$,
\begin{equation}\label{e:truncated_square_moment_asymptotics}
\bE\left(\left(\Delta_i^n X\right)^{2k}\,{\bf 1}_{\{|\sigma W_{h}+J_{h}|\leq\varepsilon\}}\right) = (2k-1)!!\,\sigma^{2k} h^k + \frac{C_{+}+C_{-}}{2k-Y}\,h\varepsilon^{2k-Y}+ o(h\varepsilon^{2k-Y}).
\end{equation}
Thus, for some constant $K$,
\begin{equation}\label{e:z_1,n_estimate}
\bE z_{1,n}^4 \leq K \Big( \bE \left(\Delta_i^n X\right)^{8}\,{\bf 1}_{\{|\Delta_i^n X|\leq\varepsilon\}} + h^4(\sigma^2 + A(\varepsilon,h))^4\Big) = O(h^{4}).
\end{equation}
Likewise,
\begin{equation}\label{e:phi_1,n_estimate}
\bE \phi_{1,n}^4 \leq K u_n^{-4}\bigg(\bE \Big[\left(\Delta_i^n X\right)^{2}\,{\bf 1}_{\{|\Delta_i^n X|\leq \zeta \varepsilon\}} - \left(\Delta_i^n X\right)^{2}\,{\bf 1}_{\{|\Delta_i^n X|\leq \varepsilon\}} \Big]^4 + h^4 \big( A(\varepsilon, h) - A(\zeta \varepsilon, h)\big)^4\bigg).
\end{equation}
From the definition of $A(\varepsilon,h)$ it follows that $A(\varepsilon, h) - A(\zeta \varepsilon, h) = O(\varepsilon^{2-Y})$, and repeated use of the elementary identity ${\bf 1}_{\{|\Delta_i^n X|\leq \varepsilon\}} {\bf 1}_{\{|\Delta_i^n X|\leq \varepsilon\zeta\}} ={\bf 1}_{\{|\Delta_i^n X|\leq \varepsilon\}}$ shows that
\begin{align*}
	&\bE \Big[\left(\Delta_i^n X\right)^{2}\,{\bf 1}_{\{|\Delta_i^n X|\leq \zeta \varepsilon\}} - \left(\Delta_i^n X\right)^{2}\,{\bf 1}_{\{|\Delta_i^n X|\leq \varepsilon\}} \Big]^4\\
	&\quad =\bE\left(\left(\Delta_i^n X\right)^{8}\,{\bf 1}_{\{|\Delta_i^n X|\leq \zeta \varepsilon\}} - \left(\Delta_i^n X\right)^{8}\,{\bf 1}_{\{|\Delta_i^n X|\leq \varepsilon\}}\right) = O(h\varepsilon^{8-Y}),
\end{align*}
where the last equality is a consequence of \eqref{e:truncated_square_moment_asymptotics}.  
Hence, based on expression \eqref{e:phi_1,n_estimate}, 
we obtain
\begin{align*}
\bE \phi_{1,n}^4 &\leq K u_n^{-4}\Big( h\varepsilon^{8-Y}  +h^4\varepsilon^{8-4Y}  \Big) \leq K  h^2\varepsilon^{2Y-8} (h\varepsilon^{8-Y} + o(h\varepsilon^{8-Y}))=O(h^3\varepsilon^{Y}).
\end{align*}
Also, by Cauchy-Schwarz, $\bE (z_{1,n}^2\phi_{1,n}^2) \leq  \sqrt{ O(h^4)\times  O(h^3  \varepsilon^{Y})} = o(h^3{ \varepsilon^{Y}})$. Thus, turning back to \eqref{e:B_i,n_bound}, we see
\begin{align*}
\bE\left(\left\Vert{\bf B}_{i,n}/\sqrt{n}\right\Vert^2 \, {\bf 1}_{\{\left\Vert{\bf B}_{i,n}\right\Vert\geq \delta\sqrt{n}\}}\right) &\leq \delta^{-2}h^{-2}\left( O(h^{4}) + O(h^3 \varepsilon^{Y}+ o(h^3\varepsilon^{Y}))\right) \leq O(h\varepsilon^{Y})= o(n^{-1})
\end{align*}
which establishes \eqref{e:lindeberg_conditions_Bin}.  By the Lindeberg-Feller CLT, the statement \eqref{e:joint_CLT_expression} holds.
\end{proof}
\bigskip
\begin{proof}[Proof of Theorem \ref{thm:debiasclt}] 
First, we set up some notation.  %
Define
\begin{align*}
a_1(\varepsilon) &:=  \frac{C_+ + C_-}{2-Y}\varepsilon^{2-Y}=: {\kappa_1}\varepsilon^{2-Y},\\
a_2(\varepsilon)&:= a_2(\varepsilon,h):= - (C_{+} + C_{-})\frac{(Y+1)(Y+2)}{2Y} \sigma^2 h \varepsilon^{-Y}=:{\kappa_2}h \varepsilon^{-Y}\\
    \Phi_n&:= u_n^{-1}\left(\wt{Z}_n(\zeta_1\varepsilon)-\wt{Z}_n(\varepsilon) \right) = O_p(1),\\
        \Psi_n&:={u_n^{-1}\left(\wt{Z}_n(\zeta^2_1\varepsilon)-2\wt{Z}_n(\zeta_1\varepsilon)+\wt{Z}_n(\varepsilon)\right)} = O_p(1),
\end{align*}
where the stochastic boundedness of $\Phi_n$, $\Psi_n$ is a consequence of Theorem \ref{thm:singleclt}.
The proof is obtained in two steps.
\bigskip
\noindent
\textbf{Step 1.} We first analyze the behavior of $\wt C_n '(\varepsilon, \zeta_1)={\wh C_n (\varepsilon)}-\widehat a_1(\varepsilon)$, where
$$
\widehat a_1(\varepsilon) := \frac{\left(\widehat C_n(\zeta
    _1 \varepsilon)-\widehat C_n( \varepsilon)\right)^2}{\widehat C_n(\zeta_1^2 \varepsilon)-2\widehat C_n(\zeta_1  \varepsilon) + \widehat C_n( \varepsilon)}.
$$
  If we let {$\eta_1(\zeta)=\zeta^{2-Y}-1$} and {$\eta_2(\zeta)=\zeta^{-Y}-1$}, then, for $i=1,2$,  we can rewrite
   \begin{align}
{a_i(\zeta_1 \varepsilon)-a_i(\varepsilon)=\eta_i(\zeta_1) a_i(\varepsilon),\quad a_i(\zeta_1^2\varepsilon)-2a_i(\zeta_1\varepsilon))+a_i(\varepsilon)=
  \eta_i^2(\zeta_1) a_i(\varepsilon).}
  \end{align}
  {For simplicity, we often omit the dependence of $\eta_i(\zeta)$ on $\zeta$.}  Also,  by definition, $\widehat C_n(\varepsilon) =  \sqrt{h} \wt Z_n(\varepsilon) + \sigma^2 + A(h,\varepsilon)$ and $A(\varepsilon,h)=a_1(\varepsilon)+a_2(\varepsilon,h)$. Therefore, if we write
   $$\widetilde a_1(\varepsilon) :={a_1(\varepsilon) + \wt\eta_2 a_2(\varepsilon):= a_1(\varepsilon)+ \frac{2\eta_1\eta_2-\eta_2^2}{\eta_1^2}a_2(\varepsilon)},$$%
    we may express
\begin{align} 
	{\widehat a_1(\varepsilon)}&=\frac{(\eta_1 a_1(\varepsilon) + \eta_2 a_2(\varepsilon) +  \sqrt h u_n \Phi_n)^2}{\eta_1^2 a_1(\varepsilon)+{\eta_2^2 a_2(\varepsilon)}+ \sqrt h u_n \Psi_n }\\
&=\widetilde a_1(\varepsilon) + \frac{ {\eta_2^2(1-\wt \eta_2)}a^2_2(\varepsilon) + \sqrt h \left(2 u_n \Phi_n(\eta_1 a_1(\varepsilon) + \eta_2 a_2(\varepsilon)) - \widetilde a_1(\varepsilon) u_n \Psi_n \right) + h u_n^2\Phi_n^2}{\eta_1^2 a_1(\varepsilon)+ {\eta_2^2 a_2(\varepsilon)}+ \sqrt h u_n \Psi_n }\\
&=\widetilde a_1(\varepsilon)+ \sqrt h \times \frac{O( h^{3/2}\varepsilon^{-2Y}) +  O_p(u_n \varepsilon^{2-Y})+ O_p(\sqrt h u_n^{2})   }{\eta_1^2 a_1(\varepsilon)+ {o(a_1(\varepsilon)) + O_p(\sqrt h u_n)} }\\
&=\widetilde a_1(\varepsilon)+ \sqrt h \times \frac{O(h^{3/2}\varepsilon^{-2-Y}) +  O_p(u_n )+ O_p(h^{-1/2}\varepsilon^2)   }{\eta_1^2b_1 + {o(1)} +  O_p(\varepsilon^{Y/2}) }\\
&=\widetilde a_1(\varepsilon)+ {\sqrt h \times O_p(u_n),}
\end{align}
 where in the last equality we use our assumption $h_n^{\frac{4}{8+Y}}\ll \varepsilon_n$ to conclude that {$h^{3/2}\varepsilon^{-2-Y}\ll u_n$}.
Then, we see that
\begin{align}
\wt  C'_n(\varepsilon,\zeta_1) &=  \widehat C_n(\varepsilon) - \widehat a_1(\varepsilon)\\
& =  \sqrt{h} \wt Z_n(\varepsilon) + \sigma^2 + A(h,\varepsilon)-\wt a_1(\varepsilon) + O_p(h^{1/2}u_n)\\
& =  \sqrt{h} \wt Z_n(\varepsilon) + \sigma^2 + a'_2(\varepsilon) + O_p(h^{1/2}u_n),\label{e:C'_breakdown}
\end{align}
where {$a_2'(\varepsilon)=(1-\wt\eta_2)a_2(\varepsilon)$}.
So, 
\begin{align}
\wt{Z}_n'( \varepsilon) &:= \sqrt{n}\left(\wt C_n '(\varepsilon, \zeta_1) - \sigma^2 - a'_2(\varepsilon)   \right)={ \wt{Z}_n( \varepsilon)}  +O_p(u_n) ,\label{e:step1_essential_asymptotics}
\end{align}
where the $O_p(u_n)$ term is a consequence of expression \eqref{e:C'_breakdown}. Then, by Theorem \ref{thm:singleclt},
\begin{equation}\label{e:Z'_clt}
\wt{Z}'_n( \varepsilon) = \wt{Z}_n(\varepsilon) + O_p(u_n) \toDist \mathcal{N}(0, 2\sigma^4).
\end{equation}
since $u_n\to 0$ by our Assumption $\varepsilon_n\ll  h_n^{\frac{1}{4-Y}}$. {Note that if $\varepsilon_n\gg h_n^{\frac{1}{2Y}}$, then $\sqrt{n}a_{2}'\ll 1$ and we conclude that $\sqrt{n}\left(\wt C_n '(\varepsilon, \zeta_1) - \sigma^2\right)\toDist \mathcal{N}(0, 2\sigma^4)$}.

\noindent
\textbf{Step 2.} Now we analyze the behavior of  ${\wt C_n ''(\varepsilon, \zeta_2,\zeta_1)}= \wt C_n '(\varepsilon, \zeta_1) - \widehat a'_2(\varepsilon,\zeta_1,\zeta_2)$, where
$$
\widehat a'_2(\varepsilon,\zeta_1,\zeta_2):= \frac{\left(\wt C_n '(\zeta_2\varepsilon, \zeta_1)-\wt C_n '(\varepsilon, \zeta_1)\right)^2}{\wt C_n '(\zeta_2^2\varepsilon, \zeta_1)-2 \wt C_n '(\zeta_2\varepsilon, \zeta_1) + \wt C_n '(\varepsilon, \zeta_1)},
$$
For simplicity, we omit the dependence on $\zeta_1$ and $\zeta_2$ in $\wt C_n '(\varepsilon, \zeta_1), C_n ''(\varepsilon, \zeta_1,\zeta_2)$, etc.
First, analogous to  $\Phi_n,\Psi_n$ defined in Step 1, we define
\begin{align*}
    \Phi_n' :=& u_n^{-1}\left(\wt{Z}_n'(\zeta_2\varepsilon)-\wt{Z}'(\varepsilon)\right)= O_p(1),\\
        \Psi'_n :=& u_n^{-1}\left(\wt{Z}'_n(\zeta^2_2\varepsilon)-2\wt{Z}'_n(\zeta_2\varepsilon)+\wt{Z}'_n(\varepsilon)\right) = O_p(1),
\end{align*}
where the stochastic boundedness of $\Phi'_n,\Psi'_n$ follows from \eqref{e:step1_essential_asymptotics}.
Now,  by definition, $ \widetilde C'_n(\varepsilon) =  \sqrt{h} \wt Z'_n(\varepsilon) + \sigma^2 + a'_2(\varepsilon)$.  Also, with {$\eta_2'(\zeta)=\zeta^{-Y}-1$}, the term ${a_2'(\varepsilon)=(1-\wt\eta_2)a_2(\varepsilon)}=:b_2' h \varepsilon^{-Y}$ satisfies
   \begin{align}
{\eta_2'(\zeta_2) a_2'(\varepsilon)= a_2'(\zeta_2 \varepsilon)-a'_2(\varepsilon),\quad 
  (\eta_2')^2(\zeta_2) a'_2(h,\varepsilon)=a'_2(\zeta_2^2\varepsilon)-2a'_2(\zeta_2\varepsilon))+a'_2(\varepsilon)}.
  \end{align}
  Therefore, we may express
\begin{align*} \widehat a_2'(\varepsilon)&=\frac{(\eta_2' a_2'(\varepsilon) +  \sqrt h u_n \Phi'_n)^2}{(\eta_2')^2 a_2'(\varepsilon)+ \sqrt h u_n \Psi_n' }\\
&= a_2'(\varepsilon) + \frac{ \sqrt h a_2'(\varepsilon) u_n(2\Phi'_n   -  \Psi'_n )  + h u_n^2(\Phi_n')^2}{(\eta_2')^2 a_2'(\varepsilon)+ \sqrt h u_n \Psi_n' }\\
&= a_2'(\varepsilon)+ \sqrt h \times \frac{  O_p(u_n )+ O_p(h^{-3/2}\varepsilon^4)   }{(\eta_2')^2b_2' + o_p(1) }\\
&= a_2'(\varepsilon)+ \sqrt h \times O_p(u_n),%
\end{align*}
{where, in the last equality we used that $\varepsilon_n\ll h_n^{1/(2+Y/2)}$ to conclude that $h^{-3/2}\varepsilon\ll u_n$.} 
Finally, 
\begin{align}
	\sqrt{n}(\wt C_n ''({\varepsilon, \zeta_2,\zeta_1})-\sigma^2)&= h^{-1/2}\left(\wt C_n '(\varepsilon, \zeta_1)-\sigma^2 - \widehat a'_2(\varepsilon,\zeta_1,\zeta_2)\right)\\
	&= h^{-1/2}\left(\wt C_n '(\varepsilon, \zeta_1)-\sigma^2 - a'_2(\varepsilon)\right)+O(u_n)\\
	&=\wt Z'_n(\varepsilon) + O_p(u_n)\\
	&\toDist \cN(0, 2\sigma^4),
\end{align}
where the last limit follows from  \eqref{e:Z'_clt} and the fact that $u_n\to 0$.  
\end{proof}

\section{Key auxiliary results}\label{s:proofs_of_lemmas}

 Throughout this section, we use the notation
\begin{equation}\label{e:def_phi,Phi-bar}
\phi(x):=\frac{1}{\sqrt{2\pi}}e^{-x^{2}/2},\quad\overline{\Phi}(x):=\int_{x}^{\infty}\phi(x)\,dx,\quad x\in\bR.
\end{equation}
We also recall the Fourier transform and its inverse transform:
\begin{align}\label{FourierTrsf}
(\cF g)(z):=\frac{1}{\sqrt{2\pi}}\int_{\bR}g(x)e^{-izx}\,dx,\quad\big(\cF^{-1}g\big)(x):=\frac{1}{\sqrt{2\pi}}\int_{\bR}g(z)e^{izx}\,dz.
\end{align}
In addition, in all asymptotic expressions, unless otherwise stated, we assume that as $h\to 0$,
\begin{equation}\label{e:basic_asymptotics}
\varepsilon\to0 \quad \text{such that}\quad \varepsilon \gg \sqrt h.
\end{equation}

The following lemma is a refinement of Theorem 3.1 in \cite{gong2021}.
\begin{proposition}\label{lemma:Estable}
As $h=1/n\rightarrow 0$, 
\begin{align}
\bE\left( \wt X_h^2 \,{\bf 1}_{\{|\wt X_h|\leq \varepsilon\}}\right)&= \sigma^2 h + A(\varepsilon,h)h +  O\left(h^3 \varepsilon^{-Y-2}\right) + O\left(h^{2}\varepsilon^{2-2Y} \right)\\
& \quad + O\left(h^{\frac{1}{2}}\varepsilon e^{-\frac{\varepsilon^2}{2\sigma^2 h}}\right)+ 
{O\left(h^{\frac{1}{2}} \varepsilon^{3-Y}  e^{-\frac{\varepsilon^2}{8\sigma^2 h}} \right)},
\end{align} 
where $A(\varepsilon,h)$ is as in \eqref{e:def_a(eps,h)}. 
\end{proposition}
\begin{proof}
Throughout the proof, we set $\bar{C}:=  C_++C_-$. Recall that under $\wt \bP$, $\wt X\stackrel{\mathcal{D}}{=}  \sigma W + S$, and that $\widetilde X$ is identically distributed under $\wt \bP$ and $\bP$. Thus, we may write
\begin{align}
\bE\left( \wt X_h^2 \,{\bf 1}_{\{|\wt X_h|\leq \varepsilon\}}\right) 
\label{eq:Decompb1epsstable} &=\sigma^{2}\,\wt \bE\Big(W_{h}^{2}\,{\bf 1}_{\{|\sigma W_{h}+S_{h}|\leq\varepsilon\}}\Big)\!+2\sigma \wt \bE\Big(W_h S_{h}\,{\bf 1}_{\{|\sigma W_{h}+S_{h}|\leq\varepsilon\}}\Big)\!+\!\wt\bE\Big(S_{h}^{2}\,{\bf 1}_{\{|\sigma W_{h}+S_{h}|\leq\varepsilon\}}\Big). %
\end{align}
The sum of the first two terms of \eqref{eq:Decompb1epsstable} is denoted $\mathcal{T}_{1,n}$. 
By the symmetry of $W_{1}$, we have:
\begin{align} 
\nonumber
\mathcal{T}_{1,n}&=
\sigma^2 h- \sigma^2 h{\wt\bE}\Big(W_{1}^{2}[{\bf 1}_{\{\sigma\sqrt{h}W_{1}+S_{h}>\varepsilon\}}+{\bf 1}_{\{\sigma\sqrt{h}W_{1}-S_{h}>\varepsilon\}}]\Big) +2\sigma\sqrt{h}\wt\bE\Big(W_1 S_{h}\,{\bf 1}_{\{|\sigma\sqrt{h} W_{1}+S_{h}|\leq\varepsilon\}}\Big)\\
\label{eq:12decomp}
&=\sigma^{2}h-\sigma\sqrt{h}\varepsilon\wt\bE\left(\phi\bigg(\frac{\varepsilon+S_{h}}{\sigma\sqrt{h}}\bigg)+\phi\bigg(\frac{\varepsilon-S_{h}}{\sigma\sqrt{h}}\bigg)\right)\\
&\qquad+\sigma\sqrt{h}\wt\bE\left(S_{h}\bigg(\phi\bigg(\frac{\varepsilon+S_{h}}{\sigma\sqrt{h}}\bigg)-\phi\bigg(\frac{\varepsilon-S_{h}}{\sigma\sqrt{h}}\bigg)\bigg)\right)-\sigma^{2}h\wt\bE\left( \overline{\Phi}\bigg(\frac{\varepsilon+S_{h}}{\sigma\sqrt{h}}\bigg)+ \overline{\Phi}\bigg(\frac{\varepsilon-S_{h}}{\sigma\sqrt{h}}\bigg)\right),
\end{align}
where in the last equality we condition on $S_{h}$ and apply the identities ${\wt\bE}\left(W_{1}^{2}{\bf 1}_{\{W_{1}>x\}}\right)=x\phi(x)+\overline{\Phi}(x)$ and ${\wt\bE}\left(W_{1}{\bf 1}_{\{x_1<W_{1}<x_2\}}\right)=\phi(x_1)-\phi(x_2)$.
From \eqref{eq:EJk}, we have $\wt{\bE}\phi\left(\frac{\varepsilon\pm {S_h}}{\sigma\sqrt{h}}\right)=\frac{\sigma\sqrt{h}}{2\pi} \mathcal{I}(0)$ and
\begin{align}
    &\wt{\bE}\left(\left(\mp  S_h \right)\,\phi\bigg(\frac{\varepsilon\pm S_h}{\sigma\sqrt{h}}\bigg)\right)\\
    &\quad=  \varepsilon \frac{\sigma\sqrt{h}}{2\pi}\left(\mathcal{I}(0)+(-1)^{1}\,2 \left(i\varepsilon\right)^{-1} \left(\frac{\sigma^2 h}{2}\right) \mathcal{I}(1)\right)\\
    &\quad= \varepsilon\wt\bE\phi\left( \frac{\varepsilon \pm S_{h} }{\sigma\sqrt{h}}\right)
    +O \left(h^{5/2} \varepsilon^{-Y-2}\right)+ O \left(\varepsilon e^{-\frac{\varepsilon^2}{2\sigma^2 h}} \right)+ O\left(h^{\frac{4-Y}{2Y}}e^{-\delta h^{\frac{Y-2}{Y}}}\right), 
    \label{eq:ESphi}
\end{align}
where in the last equality we used \eqref{e:moments_of_FT_expression}. Therefore, plugging \eqref{eq:ESphi} into \eqref{eq:12decomp}, we obtain that
\begin{align}\nonumber
\mathcal{T}_{1,n}
&= \sigma^2 h  - 2 \sigma\sqrt{h}\, \varepsilon\, {\wt \bE}\left(\phi\bigg(\frac{\varepsilon-S_{h}}{\sigma\sqrt{h}}\bigg) +\phi\bigg(\frac{\varepsilon+S_{h}}{\sigma\sqrt{h}}\bigg)\right) - \sigma^2 h \, {\wt \bE}\left(\overline{\Phi}\bigg(\frac{\varepsilon -S_{h}}{\sigma\sqrt{h}}\bigg) + \overline{\Phi}\bigg(\frac{\varepsilon +S_{h}}{\sigma\sqrt{h}}\bigg)\right)\\
&\qquad +  O\left(h^3 \varepsilon^{-Y-2}\right)+O\bigg(\varepsilon\sqrt{h}e^{-\frac{\varepsilon^{2}}{2\sigma^{2}h}}\bigg)
+ O\left(h^{\frac{2}{Y}}e^{-\delta h^{\frac{Y-2}{Y}}}\right)
 \label{eq:w2}
\end{align}
By Lemma \ref{lemmaC1},
\begin{align}%
{\wt\bE}\left(\phi\bigg(\frac{\varepsilon\pm S_{h}}{\sigma\sqrt{h}}\bigg)\right)&={{C}_{\pm}\sigma h ^{\frac{3}{2}} \varepsilon^{-Y-1} + O\left(h^{\frac{5}{2}}\varepsilon^{-Y-3}\right)+O\bigg(e^{-\frac{\varepsilon^{2}}{2\sigma^{2}h}}\bigg).}  
\label{eq:Ephi}
\end{align}
Also, by Lemma \ref{lemma:lambda}, we have
\begin{align} \label{eq:Ephibar}
    {\wt \bE}\left(\overline{\Phi}\left( \frac{\varepsilon \pm S_{h} }{\sigma\sqrt{h}}\right)\right) &= \frac{C_\mp}{Y}h\varepsilon^{-Y}  + O\left(h^2\varepsilon^{-2-Y}\right)+{O\left(h\varepsilon^{2-2Y}\right)}\\
    &\quad+{O\left(h^{1/2}\varepsilon^{-1}e^{-\frac{\varepsilon^2}{2\sigma^2 h}}\right)+O\left(h^{-1/2} \varepsilon^{3-Y}  e^{-\frac{\varepsilon^2}{8\sigma^2 h}} \right)}.\nonumber
\end{align}
Finally, {we combine \eqref{eq:w2}, \eqref{eq:Ephi}, \eqref{eq:Ephibar}, Lemma \ref{lemma:S2k_new}, and \eqref{eq:Decompb1epsstable}} to accomplish the proof.
\end{proof}

\begin{proposition}\label{lemma:2E}
Suppose $\varepsilon \gg \sqrt{h}$. Then, 
as $h=1/n\rightarrow 0$, for all small $\delta>0$,
\begin{align}\nonumber
    \bE\left( X_h^2 \,{\bf 1}_{\{|X_h|\leq \varepsilon\}} - \wt X_h^2 \,{\bf 1}_{\{|\wt X_h|\leq \varepsilon\}}\right)&=O\left(h^{3}\varepsilon^{-Y-2}\right)+O\big(h^{\frac{3}{2}}\varepsilon^{1-\frac{Y}{2}}\big) + O\left(h^2\varepsilon^{2-2Y}\right) + O(h^{1-\delta} \varepsilon^{3-Y+\delta Y})  \\
    &\quad+ O\left(\sqrt{h}\varepsilon e^{-\frac{\varepsilon^2}{2\sigma^2h}}\right) + {O\left(\sqrt{h}\varepsilon^{3-Y}  e^{-\frac{\varepsilon^2}{8\sigma^2 h}} \right)}.
    \nonumber
\end{align}
\end{proposition}
\begin{proof}
We set $\bar{C}:=C_++C_-$ and again consider the simple decomposition%
\begin{align}
\bE\left(X_h^2\, {\bf 1}_{\{|X_h^2|\leq\varepsilon\}}\right)
 &=\bE\Big(A_n(\sigma W_{h},J_h)\Big)+
 2\bE\Big(B_n(\sigma W_{h},J_h)\Big)+\bE\Big(C_n(\sigma W_{h},J_h)\Big),
 \label{eq:Decompb1epsdiff}
\end{align}
where $A_n(w,s)=w^2{\bf 1}_{\{|w+s|\leq{}\varepsilon_n\}}$, $B_n(w,s)=w s{\bf 1}_{\{|w+s|\leq{}\varepsilon_n\}}$, and
$C_n(w,s)=s^2{\bf 1}_{\{|w+s|\leq{}\varepsilon_n\}}$.
We prove the result in two steps.

\bigskip
\noindent
\textbf{Step 1.} We first analyze the difference between each of the first two terms in \eqref{eq:Decompb1epsdiff} and their counterparts under $\wt \bP$. By \eqref{eq:DenTranTildePP} and \eqref{eq:DecompUpm},
\begin{align}
\nonumber
\mathcal{R}_{n,1}
&:=\bE(A_n(\sigma W_{h},J_h))-\wt\bE(A_n(\sigma W_{h},S_h))+
 2\{\bE(B_n(\sigma W_{h},J_h))-\wt\bE(B_n(\sigma W_{h},S_h))\}\\
&=e^{-\eta h} \left(-\sigma^2h \,\wt{\bE}\Big(W_{1}^{2}\,{\bf 1}_{\{|\sigma\sqrt{h}W_{1}+J_{h}|>\varepsilon\}}\Big) + 2\sigma \sqrt{h}\,\wt{\bE}\Big(W_{1}J_{h}{\bf 1}_{\{|\sigma\sqrt{h}W_{1}+J_{h}|\leq\varepsilon\}}\Big)\right)\\
&\quad -\left(- \sigma^2 h{\wt\bE}\Big(W_{1}^{2}\,{\bf 1}_{\{|\sigma\sqrt{h}W_{1}+S_{h}|>\varepsilon\}}\Big) + 2\sigma\sqrt{h}\wt\bE\Big(W_1 S_{h}\,{\bf 1}_{\{|\sigma\sqrt{h} W_{1}+S_{h}|\leq\varepsilon\}}\Big)\right)\\
&\quad -\sigma^2he^{-\eta h}\,\wt{\bE}\Big(\Big(e^{-\wt{U}_{h}}-1\Big)W_{1}^{2}\,{\bf 1}_{\{|\sigma\sqrt{h}W_{1}+J_{h}|>\varepsilon\}}\Big) \\
&\quad + 2\sigma\sqrt{h}\,e^{-\eta h}\,\wt{\bE}\Big(\Big(e^{-\wt{U}_{h}}-1\Big)W_{1}J_{h}{\bf 1}_{\{|\sigma\sqrt{h}W_{1}+J_{h}|\leq\varepsilon\}}\Big)\\
&=:e^{-\eta h} \, I_{1}(h) - I_{2}(h) -\sigma^2he^{-\eta h}I_{3}(h) + 2\sigma\sqrt{h} e^{-\eta h}\,I_{4}(h).
\label{eq:Decompb1eps1J} 
\end{align}
{Note that, in terms of the term $\mathcal{T}_{1,n}$ of the proof of Proposition \ref{lemma:Estable}, $I_2=-\sigma^{2}h+\mathcal{T}_{1,n}$. Therefore, combining \eqref{eq:w2}-\eqref{eq:Ephibar}, we get}
\begin{align*}%
I_2(h)  &= -\bar{C}\frac{2Y+1}{Y}\sigma^2 h ^{2} \varepsilon^{-Y} +  O\left(h^3 \varepsilon^{-Y-2}\right)+{O\left(h^2\varepsilon^{2-2Y}\right)}\\
&\quad+{O\bigg(\sqrt{h}\varepsilon e^{-\frac{\varepsilon^2}{2\sigma^2 h}}\bigg) 
+O\left(\sqrt{h}\varepsilon^{3-Y}  e^{-\frac{\varepsilon^2}{8\sigma^2 h}} \right)}.
\end{align*}
For $I_1(h)$, first note that, as in (\ref{eq:12decomp}), we have:
\begin{align} \nonumber
I_{1}(h) &= - \sigma\sqrt{h}\, \varepsilon\, \wt\bE\left(\phi\bigg(\frac{\varepsilon-J_{h}}{\sigma\sqrt{h}}\bigg) + \phi\bigg(\frac{\varepsilon+J_{h}}{\sigma\sqrt{h}}\bigg)\right) - \sigma\sqrt{h}\, \wt\bE\left(J_h\phi\bigg(\frac{\varepsilon-J_{h}}{\sigma\sqrt{h}}\bigg) - J_h\phi\bigg(\frac{\varepsilon+J_{h}}{\sigma\sqrt{h}}\bigg)\right) \\
\label{eq:s2J}
&\quad - \sigma^2 h \wt\bE\left(\overline{\Phi}\bigg(\frac{\varepsilon -J_{h}}{\sigma\sqrt{h}}\bigg) + \overline{\Phi}\bigg(\frac{\varepsilon +J_{h}}{\sigma\sqrt{h}}\bigg)\right).
\end{align}
Recall that, under $\wt{\bP}$, $J_t=S_t+\tilde{\gamma}t$. Then, fixing $\tilde{\varepsilon}_{\pm}=\varepsilon\pm \tilde{\gamma}h$, we can apply \eqref{eq:Ephi} to deduce that:
\begin{align}%
\widetilde{\bE}\phi\bigg(\frac{\varepsilon\pm J_{h}}{\sigma\sqrt{h}}\bigg)&=\wt\bE\phi\bigg(\frac{\tilde{\varepsilon}_{\pm}\pm S_{h}}{\sigma\sqrt{h}}\bigg)= C_{\pm}\sigma h ^{\frac{3}{2}} \tilde{\varepsilon}_{\pm}^{-Y-1} + O\left(h^{\frac{5}{2}}\varepsilon^{-Y-3}\right)+O\bigg(e^{-\frac{\varepsilon^{2}}{2\sigma^{2}h}}\bigg).
\label{eq:Ephi0}
\end{align}
Similarly, by \eqref{eq:ESphi} and \eqref{eq:Ephibar},
\begin{align}%
\label{eq:EphiForJ}
\widetilde{\bE}\left(J_h\phi\bigg(\frac{\varepsilon\pm J_{h}}{\sigma\sqrt{h}}\bigg)\right)&={\wt\bE}\left(S_h\phi\bigg(\frac{\tilde{\varepsilon}_{\pm}\pm S_{h}}{\sigma\sqrt{h}}\bigg)\right)\, +\,\tilde{\gamma}h\,{\wt\bE}\left(\phi\bigg(\frac{\tilde{\varepsilon}_{\pm}\pm S_{h}}{\sigma\sqrt{h}}\bigg)\right)\\
&=\mp\varepsilon\,{\wt\bE}\left(\phi\bigg(\frac{\tilde{\varepsilon}_{\pm}\pm S_{h}}{\sigma\sqrt{h}}\bigg)\right)+O\left(h^{\frac{5}{2}} \varepsilon^{-Y-2}\right)+O\bigg(\varepsilon e^{-\frac{\varepsilon^{2}}{2\sigma^{2}h}}\bigg),
\nonumber
\\
\label{AsymBarPhi2}
    \widetilde{\bE}\overline{\Phi}\bigg(\frac{\varepsilon\pm J_{h}}{\sigma\sqrt{h}}\bigg) &= \widetilde{\bE}\overline{\Phi}\bigg(\frac{\tilde{\varepsilon}_{\pm}\pm S_{h}}{\sigma\sqrt{h}}\bigg) =\frac{C_\mp}{Y}h\tilde\varepsilon_{\pm}^{-Y} +  O\left(h^2\varepsilon^{-2-Y}\right)+{O\left(h\varepsilon^{2-2Y}\right)}\\
    \nonumber
    &\quad +O\left(\sqrt{h}\varepsilon^{-1}e^{-\frac{\varepsilon^2}{2\sigma^2 h}}\right) +{O\left(h^{-1/2} \varepsilon^{3-Y}  e^{-\frac{\varepsilon^2}{8\sigma^2 h}} \right).}
\end{align}
Together,  \eqref{eq:Ephi0}, \eqref{eq:EphiForJ}, and \eqref{AsymBarPhi2} imply that $I_1(h)$ has the following asymptotic behavior:
\begin{align}\label{AsympI1J}
I_1(h)  &= -\frac{ 2C_+ Y+C_-}{Y}\sigma^2 h ^{2} \tilde\varepsilon_+^{-Y}-\frac{2C_-Y+C_+}{Y}\sigma^2 h ^{2} \tilde\varepsilon_-^{-Y}+  O\left(h^3 \varepsilon^{-Y-2}\right)+{O\left(h^2\varepsilon^{2-2Y}\right)} \\
&\quad +O\bigg(\sqrt{h}\varepsilon e^{-\frac{\varepsilon^2}{2\sigma^2 h}}\bigg) %
+{O\left(h^{1/2} \varepsilon^{3-Y}  e^{-\frac{\varepsilon^2}{8\sigma^2 h}} \right)}.
\end{align}
Therefore, we obtain that 
\begin{align}\nonumber
e^{-\eta h} \, I_{1}(h) - I_{2}(h) 
&= I_{1}(h) - I_{2}(h) + I_1(h)O\left(h\right)
\\
\nonumber
&= -\sigma^{2}h^2\frac{2 C_+Y+C_-}{Y}\left( \tilde\varepsilon_+^{-Y}-\varepsilon^{-Y}\right)-\sigma^{2}h^2\frac{2C_-Y+C_+}{Y}\left( \tilde\varepsilon_-^{-Y}-\varepsilon^{-Y}\right)+O\left(h^3 \varepsilon^{-Y-2}\right)\\
\nonumber
&\quad+{O\left(h^2\varepsilon^{2-2Y}\right)}+O\bigg(\sqrt{h}\varepsilon e^{-\frac{\varepsilon^2}{2\sigma^2 h}}\bigg)
+{O\left(\sqrt{h} \varepsilon^{3-Y}  e^{-\frac{\varepsilon^2}{8\sigma^2 h}} \right)}\\
&= O\left(h^3 \varepsilon^{-Y-2}\right)+{O\left(h^2\varepsilon^{2-2Y}\right)}+O\bigg(\sqrt{h}\varepsilon e^{-\frac{\varepsilon^2}{2\sigma^2 h}}\bigg)
+{O\left(\sqrt{h} \varepsilon^{3-Y}  e^{-\frac{\varepsilon^2}{8\sigma^2 h}} \right)},
\label{eq:I12J}
\end{align}
where in the last equality we used that $h^{2}(\tilde\varepsilon_{\pm}^{-Y}-\varepsilon^{-Y})=O(h^3\varepsilon^{-Y-1})$.
Next, using the proof in the Step 1.2 and Step 3 of Lemma 3.1 of \cite{gong2021}, we know, as $h\rightarrow 0$,
\begin{align}
    I_3(h) &= \wt{\bE}\Big(\Big(e^{-\wt{U}_{h}}-1\Big)W_{1}^{2}\,{\bf 1}_{\{|\sigma\sqrt{h}W_{1}+J_{h}|>\varepsilon\}}\Big) = o\big(h^{1/Y}\big),\label{eq:I3J}\\
    I_4(h) &= \wt{\bE}\Big(\Big(e^{-\wt{U}_{h}}-1\Big)W_{1}J_{h}{\bf 1}_{\{|\sigma\sqrt{h}W_{1}+J_{h}|\leq\varepsilon\}}\Big) = O\big(h\varepsilon^{1-Y/2}\big).\label{eq:I4J}
\end{align}
Thus, combining \eqref{eq:Decompb1eps1J}, \eqref{eq:I12J}, \eqref{eq:I3J}, and \eqref{eq:I4J}, we have
\begin{align}
    \label{eq:first2diff} 
    \mathcal{R}_{n,1}&=
O\left(h^{3}\varepsilon^{-Y-2}\right)+{O\left(h^2\varepsilon^{2-2Y}\right)}+ O\big(h^{\frac{3}{2}}\varepsilon^{1-\frac{Y}{2}}\big) + o\big(h^{1+\frac{1}{Y}}\big)\\
&\quad 
+O\left(\sqrt{h}\varepsilon e^{-\frac{\varepsilon^2}{2\sigma^2h}}\right)+{O\left(\sqrt{h} \varepsilon^{3-Y}  e^{-\frac{\varepsilon^2}{8\sigma^2 h}} \right)}.
\end{align}

\smallskip
\noindent
\textbf{Step 2.} In this step, we will study the asymptotic behavior of the third term in \eqref{eq:Decompb1epsdiff}, as $h\rightarrow 0$. By \eqref{eq:DenTranTildePP}, \eqref{eq:DecompJZpm}, and \eqref{eq:DecompUpm}, we have
\begin{align}
\mathcal{R}_{n,2}&:=\bE\Big(J_{h}^{2}\,{\bf 1}_{\{|\sigma W_{h}+J_{h}|\leq\varepsilon\}}\Big) - \wt\bE\Big(S_{h}^{2}\,{\bf 1}_{\{|\sigma W_{h}+S_{h}|\leq\varepsilon\}}\Big)\\
&= \left(\wt{\bE}\Big(S_{h}^{2}\,{\bf 1}_{\{|\sigma W_{h}+S_{h}+\wt{\gamma}h|\leq\varepsilon\}}\Big) - \wt\bE\Big(S_{h}^{2}\,{\bf 1}_{\{|\sigma W_{h}+S_{h}|\leq\varepsilon\}}\Big)\right) \\
&\quad + e^{-\eta h}\, \wt{\bE}\Big(\Big(e^{-\wt{U}_{h}}-1\Big)S_{h}^{2}\,{\bf 1}_{\{|\sigma W_{h}+S_{h}+\wt{\gamma}h|\leq\varepsilon\}}\Big)+2\wt{\gamma}he^{-\eta h}\,\wt{\bE}\Big(e^{-\wt{U}_{h}}S_{h}{\bf 1}_{\{|W_{h}+S_{h}+\wt{\gamma}h|\leq\varepsilon\}}\Big) \\
&\quad +\wt{\gamma}^{2}h^{2}e^{-\eta h}\,\wt{\bE}\Big(e^{-\wt{U}_{h}}\,{\bf 1}_{\{|W_{h}+S_{h}+\wt{\gamma}h|\leq\varepsilon\}}\Big)+O(h)\cdot \wt{\bE}\Big(S_{h}^{2}\,{\bf 1}_{\{|\sigma W_{h}+S_{h}+\wt{\gamma}h|\leq\varepsilon\}}\Big)\\
 &=:I_{5,1}(h)  +e^{-\eta h}I_{5,2}(h)+2\wt{\gamma}he^{-\eta h}I_{6}(h)+O\big(h^{2}\big)+ O\left(\varepsilon^{1-Y}h^{5/2}e^{-\frac{\varepsilon^2}{2\sigma^2 h}}\right),%
\label{eq:Decompb1eps2J} 
\end{align}
where in the last equality we used  Lemma \ref{lemma:S2k_new}.
We begin with the analysis of $I_{5,1}(h)$. 
First note that, for large enough $n$,  
\[
	    \wt\bE\Big(S_{h}^{2}[{\bf 1}_{\{|\sigma W_{h}+S_{h}|\leq\varepsilon - |\tilde\gamma| h\}}-{\bf 1}_{\{|\sigma W_{h}+S_{h}|\leq\varepsilon\}}]\Big)\leq{}I_{5,1}(h)\leq\wt\bE\Big(S_{h}^{2}[{\bf 1}_{\{|\sigma W_{h}+S_{h}|\leq\varepsilon + |\tilde\gamma| h\}}-{\bf 1}_{\{|\sigma W_{h}+S_{h}|\leq\varepsilon\}}]\Big),
\]
and, since, by  Lemma \ref{lemma:S2k_new},
\begin{align}
&\wt\bE\Big(S_{h}^{2}\,{\bf 1}_{\{|\sigma W_{h}+S_{h}|\leq\varepsilon \pm |\tilde\gamma| h\}}\Big) - \wt\bE\Big(S_{h}^{2}\,{\bf 1}_{\{|\sigma W_{h}+S_{h}|\leq\varepsilon\}}\Big)\\
    &= \frac{\bar{C}}{2-Y} h \left(\left(\varepsilon\pm|\wt\gamma| h\right)^{2-Y} -\varepsilon^{2-Y}\right) + \frac{\bar{C}(1-Y)}{2} \sigma^2 h^{2} \left(\left(\varepsilon\pm|\wt\gamma| h\right)^{-Y} - \varepsilon^{-Y}\right) \\
    &\quad + O\left(h^3\varepsilon^{-2-Y}\right) + O\left(h^2\varepsilon^{2-2Y}\right) +  O\left(\varepsilon^{1-Y}h^{\frac{3}{2}}e^{-\frac{\varepsilon^2}{2\sigma^2 h}}\right)\\%
    &=   O\left(h^3\varepsilon^{-2-Y}\right)+O\left(h^2\varepsilon^{2-2Y}\right) +  O\left(h^{\frac{3}{2}} \varepsilon^{1-Y} e^{-\frac{\varepsilon^2}{2\sigma^2h}}\right),
\end{align}
we conclude that $I_{5,1}(h)=  O\left(h^3\varepsilon^{-2-Y}\right)+O\left(h^2\varepsilon^{2-2Y}\right) + O\left(h^{\frac{3}{2}} \varepsilon^{1-Y} e^{-\frac{\varepsilon^2}{2\sigma^2h}}\right)$. 
Next, in the proof of Theorem 3.1 of \cite{gong2021} (Steps 2.1 and 2.2 therein), it was shown that, as $h\rightarrow 0$,
\begin{align}\label{eq:Limitb1eps212J}
I_{5,2}(h)=O\big(h\,\varepsilon^{2-Y/2}\big),\quad
I_{6}(h)=O\big(h^{1/2}\varepsilon^{1-Y/2}\big).%
\end{align}
While the estimate for $I_{6}(h)$ will  suffice for our purpose, we need to develop a tighter bound for $I_{5,2}(h)$. In Appendix \ref{VeryTechProofs}, expression \eqref{e:I_{52}_improved}, we show the improved estimate:
\begin{align}\label{ImprovEstI652}
	I_{5,2}(h) =  O(h \varepsilon^{3-Y} \big(h \varepsilon^{-Y}\big)^{-\delta}),
\end{align}
valid for all small $\delta>0$.
Thus, 
we obtain that
\begin{align}\label{eq:thirddiff}
\mathcal{R}_{n,2}&=  O\left(h^3\varepsilon^{-2-Y}\right)+ O\left(h^2\varepsilon^{2-2Y}\right) +  O\left(h^{\frac{3}{2}} \varepsilon^{1-Y} e^{-\frac{\varepsilon^2}{2\sigma^2h}}\right) \\
    &\quad + O(h \varepsilon^{3-Y} \big(h \varepsilon^{-Y}\big)^{-\delta})
    + O\big(h^{3/2}\varepsilon^{1-Y/2}\big)\\
    &\quad+O\big(h^{2}\big)+ O\left(\varepsilon^{1-Y}h^{5/2}e^{-\frac{\varepsilon^2}{2\sigma^2 h}}\right)\\
&=  O\left(h^3\varepsilon^{-2-Y}\right)+O\left(h^2\varepsilon^{2-2Y}\right) + O(h \varepsilon^{3-Y} \big(h \varepsilon^{-Y}\big)^{-\delta}) + O\left(h^{\frac{3}{2}} \varepsilon^{1-Y} e^{-\frac{\varepsilon^2}{2\sigma^2h}}\right).
\end{align}
Finally, combining \eqref{eq:first2diff} and \eqref{eq:thirddiff}, we conclude the result.
\end{proof}
{
\begin{lemma}\label{prop:EX2}
Let $k$ be a positive integer. {As $h\to 0$,}
\begin{align*}
    \bE\left(\left(\Delta_i^n X\right)^{2k}\,{\bf 1}_{\{|\Delta_i^n X|\leq\varepsilon\}}\right) 
    &= (2k-1)!!\,\sigma^{2k} h^k + \frac{C_{+}+C_{-}}{2k-Y}\,h\varepsilon^{2k-Y}+  O\left(h^2 \varepsilon^{2k-Y-2}\right)
    +O\big(h\varepsilon^{2k-Y/2}\big)\\
    &\quad 
   + {O\left(h^{\frac{1}{2}}\varepsilon^{2k-1} e^{-\frac{\varepsilon^2}{2\sigma^2 h}}\right)}.
\end{align*}
\end{lemma}
\begin{proof}
By Lemmas \ref{lemma:W2k}, \ref{lemma:J2k}, and \ref{lemma:WJ}, we can compute%
\begin{align*}
    &\bE\left(\left(\Delta_i^n X\right)^{2k}\,{\bf 1}_{\{|\Delta_i^n X|\leq\varepsilon\}}\right) = \bE\Big(\big(\sigma W_{h}+J_{h}\big)^{2k}\,{\bf 1}_{\{|\sigma W_{h}+J_{h}|\leq\varepsilon\}}\Big)\\
    &\quad =\bE\left(\sigma^{2k}W_h^{2k}{\bf 1}_{\{|\sigma W_{h}+J_{h}|\leq\varepsilon\}}\right) + \bE\left(J_h^{2k}{\bf 1}_{\{|\sigma W_{h}+J_{h}|\leq\varepsilon\}}\right) \\
    & \quad + \bE\left(\sum_{j=1}^{k-1}\frac{(2k)!}{(2j)! \, (2k-2j)!} (\sigma W_h)^{2j} J_h^{2k-2j}{\bf 1}_{\{|\sigma W_{h}+J_{h}|\leq\varepsilon\}}\right)\\
    & \quad + \bE\left(\sum_{j=0}^{k-1}\frac{(2k)!}{(2j+1)! \, (2k-2j-1)!} (\sigma W_h)^{2j+1} J_h^{2k-2j-1} {\bf 1}_{\{|\sigma W_{h}+J_{h}|\leq\varepsilon\}}\right)\\
    &\quad= (2k-1)!!\,\sigma^{2k} h^k + 
    O\left(h^2 \varepsilon^{2k-Y-2}\right)+ o\left(h^{k+\frac{1}{Y}}\right)+ {O\left(h^{\frac{1}{2}}\varepsilon^{2k-1} e^{-\frac{\varepsilon^2}{2\sigma^2 h}}\right)}\\%
    &\qquad+ \frac{C_{+}\!+\!C_{-}}{2k-Y}h\varepsilon^{2k-Y} +  O\big(h^2\varepsilon^{2k-Y-2}\big) + O\big(h\varepsilon^{2k-Y/2}\big)%
   \\
    & \qquad + O\left(h^2 \varepsilon^{2k-Y-2}\right) + O\left(h^2 \varepsilon^{2k-Y-2}\right) + O\left(h^{3/2} \varepsilon^{2k-Y/2-1}\right).
\end{align*}
The result then follows. 
\end{proof}

\section{Technical lemmas}\label{VeryTechProofs}
Unless otherwise stated, throughout this section we continue to assume \eqref{e:basic_asymptotics}  in all asymptotic expressions.
The following lemma is used in the proof of Proposition \ref{lemma:Estable}.
\begin{lemma} \label{lemmaC1}
Fix  $\gamma_0\in \bR$, and let $J'_h =  S_h +\gamma_0 h$.  Then, %
as $h\to 0$,
\begin{align}%
{\wt\bE}\left(\phi\bigg(\frac{\varepsilon\pm J'_{h}}{\sigma\sqrt{h}}\bigg)\right) &= C_{\pm}\sigma h ^{\frac{3}{2}} (\varepsilon\pm \gamma_0h)^{-Y-1} + O\left(h^{\frac{5}{2}}\varepsilon^{-Y-3}\right)+O\bigg(e^{-\frac{\varepsilon^{2}}{2\sigma^{2}h}}\bigg).
\label{eq:Ephi_lemma}
\end{align}
\end{lemma}
\begin{proof}
Let $p_{J,h}^{\pm}$ be the density of $\pm J'_{h}$ under $\wt{\bP}$, and observe
\begin{align}
\big(\cF p_{J,h}^{\pm}\big)(u)%
&=\frac{1}{\sqrt{2\pi}}\exp\Big(c_{1}|u|^{Y}h\,{ \pm}\, ic_{2}|u|^{Y}h\,\text{sgn}(u)\mp iu\gamma_0h\Big),
\end{align} 
where
\begin{equation}\label{e:def_c1,c2}
c_{1}:=(C_{+}\!+C_{-})\,\cos\bigg(\frac{\pi Y}{2}\bigg)\Gamma(-Y)<0,\quad c_{2}:=(C_{-}-C_{+})\sin\bigg(\frac{\pi Y}{2}\bigg)\Gamma(-Y).
\end{equation}
 Also let
\begin{align*}
    \psi(x)&:=\bigg(\cF^{-1}\phi\bigg(\frac{\cdot}{\sigma\sqrt{h}}-\frac{\varepsilon}{\sigma\sqrt{h}}\bigg)\bigg)(x)
    =\frac{\sigma\sqrt{h}}{\sqrt{2\pi}}\exp\bigg(i\varepsilon x-\frac{1}{2}\sigma^{2}x^{2}h\bigg).
\end{align*}
Then, using Plancherel,
\begin{align}
    \wt{\bE}\phi\bigg(\frac{\varepsilon\pm J'_h}{\sigma\sqrt{h}}\bigg)
    &= \int_{\bR}(\cF\psi)(z)p_{J,h}^{\mp}(z)\,dz = \int_{\bR}\psi(u)\cF\big(p_{J,h}^{\mp}(z)\big)(u)\,du\\
    &= \frac{\sigma\sqrt{h}}{\pi} \int_{0}^{\infty}  e^{c_1u^Y h - \frac{1}{2}\sigma^2u^2 h} \, \cos \left( \mp c_2u^Y h + u(\varepsilon \pm {\gamma}_0 h)\right )du  \\
        &=\int_{0}^{\infty}g_1(\omega^{Y}h^{1-\frac{Y}{2}})e^{-\frac{\omega^{2}}{2}}\cos\bigg(\frac{\omega\varepsilon^{\pm}_0}{\sigma\sqrt{h}}\bigg)d\omega-
       \int_{0}^{\infty}g^\pm_2(\omega^{Y}h^{1-\frac{Y}{2}})e^{-\frac{\omega^{2}}{2}}\sin\bigg(\frac{\omega\varepsilon^{\pm}_0}{\sigma\sqrt{h}}\bigg)d\omega\\
       \label{TrmI2Pa}
       &=:I_1-I_2,
\end{align}
where
\[
	\varepsilon^{\pm}_0=\varepsilon\pm{\gamma_0}h,\quad g_1(u)=\frac{1}{\pi}e^{c_{1}u/\sigma^Y}\cos(c_{2}u/\sigma^Y),\quad
	g^\pm_2(u)= \mp \frac{1}{\pi}e^{c_{1}u/\sigma^Y}\sin(c_{2}u/\sigma^Y).
\]
Fix $\delta>0$ and set $D_{\delta,h}:=\{w\in(0,\infty): w^{Y}h^{1-Y/2}<(2\delta)^{Y/2}\}=(0,(2\delta)^{1/2}/h^{1/Y-1/2})$. Consider
\[	
	I_1=\left(\int_{D_{\delta,h}}+\int_{D_{\delta,h}^c}\right)g_1(\omega^{Y}h^{1-\frac{Y}{2}})e^{-\frac{\omega^{2}}{2}}\cos\bigg(\omega\frac{\varepsilon^{\pm}_0}{\sigma\sqrt{h}}\bigg)d\omega=:I_{11}+I_{12}.
\]
Clearly, since $c_1<0$, 
\[
	|I_{12}|\leq{}\frac{ 1}{\pi}\int_{(2\delta)^{1/2}/h^{1/Y-1/2}}^{\infty}e^{-w^{2}/2}dw=O\left(h^{\frac{2-Y}{2Y}}e^{-\delta h^{\frac{Y-2}{Y}}}\right).
\]
We decompose the integral on $D_{\delta,h}$ as follows:
\begin{align}\label{Term1I11}
	I_{11}&=\sum_{j=0}^{m-1}\frac{1}{j!}g_1^{(j)}(0)h^{j(1-\frac{Y}{2})}\int_{0}^{\infty}w^{jY}e^{-\frac{\omega^{2}}{2}}\cos\bigg(\omega\frac{\varepsilon_0^{\pm}}{\sigma\sqrt{h}}\bigg)d\omega\\
	\label{Term2I11}
	&\quad-\sum_{j=0}^{m-1} \frac{1}{j!}g_1^{(j)}(0)h^{j(1-\frac{Y}{2})}\int_{(2\delta)^{1/2}/h^{1/Y-1/2}}^{\infty}w^{jY}e^{-\frac{\omega^{2}}{2}}\cos\bigg(\omega\frac{\varepsilon_0^{\pm}}{\sigma\sqrt{h}}\bigg)d\omega\\
	\label{Term3I11}
	&\quad +\frac{h^{m(1-\frac{Y}{2})}}{m!}\int_{0}^{(2\delta)^{1/2}/h^{1/Y-1/2}}g_{1}^{(m)}(\theta_{w})w^{mY}e^{-\frac{\omega^{2}}{2}}\cos\bigg(\omega\frac{\varepsilon_0^{\pm}}{\sigma\sqrt{h}}\bigg)d\omega,
\end{align}
where $\theta_{\omega}\in(0,(2\delta)^{Y/2})$. The $j^{th}$ term in \eqref{Term2I11} is 
\[
	O\left(h^{j(1-\frac{Y}{2})}\int_{(2\delta)^{1/2}/h^{1/Y-1/2}}^{\infty}w^{jY}e^{-\frac{\omega^{2}}{2}}d\omega\right)
	=O\left(h^{\frac{2-Y}{2Y}}e^{-\delta h^{\frac{Y-2}{Y}}}\right),
\]
where above we expressed the integral in terms of the incomplete gamma function and then apply the standard asymptotics behavior for such a function (e.g., \cite[Section 8.35]{gradshteyn:ryzhik:2007}).  For \eqref{Term1I11}, the term corresponding to $j=0$ is clearly $O(e^{-\varepsilon^{2}/(2\sigma^{2}h)})$. For $j\geq{}1$, we can apply the same argument as in (A.11) of \cite{gong2021} (see also \cite[expression 13.7.2]{Olveretal}) to show that
\begin{align}\label{eq:KummerPowerCos0}
&h^{j(1-\frac{Y}{2})}\int_{0}^{\infty}\omega^{jY}\cos\bigg(\omega\frac{\varepsilon_0^{\pm}}{\sigma\sqrt{h}}\bigg)e^{-\omega^{2}/2}\,d\omega\\
&= h^{j(1-\frac{Y}{2})}2^{(jY-1)/2}\,\Gamma\bigg(\frac{jY+1}{2}\bigg) M\bigg(\frac{jY+1}{2},\frac{1}{2},-\frac{(\varepsilon_0^{\pm})^{2}}{2\sigma^{2}h}\bigg)\\
&=h^{j(1-\frac{Y}{2})}2^{\frac{jY-1}{2}}\,\Gamma\bigg(\frac{jY+1}{2}\bigg)\left(\frac{\Gamma(1/2)}{\Gamma\big(-jY/2\big)}\bigg(\frac{(\varepsilon_0^{\pm})^{2}}{2\sigma^{2}h}\bigg)^{-\frac{jY+1}{2}}\left(1+O\left(\frac{h}{\varepsilon^{2}}\right)\right)\right.\\
&\qquad\qquad\qquad\qquad\qquad\quad\qquad\,\,\left.+\frac{\Gamma(1/2)\,e^{-\varepsilon^{2}/(2\sigma^{2}h)}}{\Gamma\big((jY+1)/2\big)}\bigg(\frac{(\varepsilon^{\pm}_0)^{2}}{2\sigma^{2}h}\bigg)^{jY/2}\right)=O\bigg(\frac{h^{j+1/2}}{\varepsilon^{jY+1}}\bigg).
\end{align}
Therefore, the $j=1$ term in \eqref{Term1I11} is the leading order term,  and the second-order term therein is $O(h^{5/2}/\varepsilon^{2Y+1})$. Furthermore, by picking $m>5/(2-Y)$, we see the term in \eqref{Term3I11} -- which is   $O(h^{m(1-Y/2)})$ -- decays faster than the second-order term  in \eqref{Term1I11}. Putting together the above estimates and noting that $g_1'(0)=c_1/(\pi\sigma^Y)$, we conclude that 
\[
	I_{1}=\frac{c_1}{\pi}\sigma2^{Y}\frac{\Gamma(Y/2+1/2)\Gamma(1/2)}{\Gamma(-Y/2)}\frac{h^{3/2}}{(\varepsilon_0^{\pm})^{Y+1}}+O\left(e^{-\frac{\varepsilon^{2}}{2\sigma^{2}h}}\right)+O\bigg(\frac{h^{5/2}}{\varepsilon^{Y+3}}\bigg),%
\]
{where above we used that $h^{\frac{2-Y}{2Y}}e^{-\delta h^{\frac{Y-2}{Y}}}\ll h^{5/2}\varepsilon^{-(Y+3)}$.}
Applying the same analysis to the term $I_2$ in \eqref{TrmI2Pa}%
\[
	I_{2}=\pm \frac{c_2}{\pi}\sigma2^{Y+1}\frac{\Gamma(Y/2+1)\Gamma(3/2)}{\Gamma((1-Y)/2)}\frac{h^{3/2}}{(\varepsilon_0^{\pm})^{Y+1}}+O\left(e^{-\frac{\varepsilon^{2}}{2\sigma^{2}h}}\right)+{O\bigg(\frac{h^{5/2}}{\varepsilon^{Y+3}}\bigg)}.%
\]
We then get that
\begin{align*}
	 \wt{\bE}\phi\bigg(\frac{\varepsilon\pm J'_h}{\sigma\sqrt{h}}\bigg)
	& =\frac{\sigma 2^{Y}}{\pi}\left(c_1\frac{\Gamma(Y/2+1/2)\Gamma(1/2)}{\Gamma(-Y/2)}\, \mp\,c_2\frac{2\Gamma(Y/2+1)\Gamma(3/2)}{\Gamma((1-Y)/2)}\right)\frac{h^{3/2}}{(\varepsilon_0^{\pm})^{Y+1}}\\
	&\qquad+O\left(e^{-\frac{\varepsilon^{2}}{2\sigma^{2}h}}\right)+{O\bigg(\frac{h^{5/2}}{\varepsilon^{Y+3}}\bigg)}.%
\end{align*}
By applying the properties 
$\Gamma\left(x\right)\, \Gamma\left(x+1/2 \right) = \sqrt{\pi}\, 2^{1-2x}\,\Gamma\left(2x\right)$
and $\Gamma(1-x)\,\Gamma(x) = \frac{\pi}{\sin (\pi x)}$, we can simplify the coefficients as
\begin{align*}
	\frac{2^{Y}}{\pi}c_1\frac{\Gamma(Y/2+1/2)\Gamma(1/2)}{\Gamma(-Y/2)}=\frac{C_++C_-}{2},\quad  \frac{2^{Y+1}}{\pi}c_2\frac{\Gamma(Y/2+1)\Gamma(3/2)}{\Gamma((1-Y)/2)}=\frac{C_--C_+}{2}.
\end{align*}
We finally {obtain} \eqref{eq:Ephi_lemma}.
\end{proof}
The following lemma is used in the proofs of Lemmas \ref{lemma:W2k} and \ref{lemma:WJ}.
\begin{lemma} \label{lemma:4results}
Fix $\gamma_0\in \bR$, any integer $k\geq 0$, and let $J'_h =  S_h +\gamma_0 h$. Then, %
as $h\to 0$,
\begin{align}\label{e:J^kphi_asymptotics_1}
    \wt{\bE}\left(\left(\mp  J'_h \right)^k\,\phi\bigg(\frac{\varepsilon\pm J'_h}{\sigma\sqrt{h}}\bigg)\right) &= O \left(h^{\frac{3}{2}} \varepsilon^{k-Y-1}\right) +{O \left( \varepsilon^{k}e^{-\frac{\varepsilon^2}{2\sigma^2 h} } \right)},\\%
    \nonumber
    \wt{\bE}\left(\left(\frac{\varepsilon \pm  J'_h}{\sigma\sqrt{h}}\right)^k\,\phi\bigg(\frac{\varepsilon\pm J'_h}{\sigma\sqrt{h}}\bigg)\right) &= 
O \left(h^{\frac{3-k}{2}} \varepsilon^{k-Y-1}\right) +O \left(\varepsilon^{k}h^{\frac{-k}{2}}\,e^{-\frac{\varepsilon^2}{2\sigma^2 h}} \right).%
\end{align}
\end{lemma}
\begin{proof}
We use the same notation as in the proof of Lemma \ref{lemmaC1}.
We first establish the asymptotic bound  for any fixed $\ell=0,1,\dots$, 
\begin{align}\nonumber
\mathcal{I}(\ell)&:=\int_{\bR}  u^\ell \exp{\left (iu (\varepsilon \pm  \gamma_0 h) - \frac{1}{2}\sigma^2u^2 h +|u|^Y h \left(c_1 \,  \mp\, ic_2\sgn (u)\right) \right)} du\\
&
= O \left(h \varepsilon^{-Y-\ell-1}\right)+ O \left(h^{-\ell-\frac{1}{2}} \, \varepsilon^{\ell} e^{-\frac{\varepsilon^2}{2\sigma^2 h}} \right).%
\label{e:moments_of_FT_expression}
\end{align}
 We suppose $\ell\geq 0$ is even (the case for $\ell$ odd can be established similarly).  As in \eqref{TrmI2Pa}, we can write
\begin{align}
    \mathcal{I}(\ell)
    &= 2 \, (\sigma \sqrt{h})^{-\ell-1} \int_0^\infty w^\ell g_1(h^{1-Y/2} \omega^Y)e^{-\frac{\omega^2}{2}}\cos \left(\frac{\varepsilon_0^{\pm}}{\sigma \sqrt{h}}\omega\right) d\omega \\
    & \quad - 2 \, (\sigma \sqrt{h})^{-\ell-1} \int_0^\infty w^\ell g_2^{ \pm}(h^{1-Y/2} \omega^Y)e^{-\frac{\omega^2}{2}}\sin\left(\frac{\varepsilon_0^{\pm}}{\sigma \sqrt{h}}\omega\right) d\omega.
    \label{eq:inteven}
\end{align}
Applying the same arguments as in the proof of Lemma \ref{lemmaC1}, we can show that, for $\delta>0$ and positive integer $m$,
\begin{align}\label{T1NH1}
    \mathcal{I}(\ell)&=\sum_{j=0}^{m-1}O\left(h^{-\frac{\ell+1}{2}}h^{j(1-\frac{Y}{2})}\int_{(2\delta)^{1/2}/h^{1/Y-1/2}}^{\infty}w^{jY+\ell}e^{-\frac{\omega^{2}}{2}}d\omega\right)\\
    \label{T2NH1}
    &\quad+\sum_{j=1}^{m-1}O\left(h^{-\frac{\ell+1}{2}}h^{j(1-\frac{Y}{2})}\int_{0}^{\infty}\omega^{jY+\ell}\cos\bigg(\omega\frac{\varepsilon_0^{\pm}}{\sigma\sqrt{h}}\bigg)e^{-\omega^{2}/2}\,d\omega\right)\\
        \label{T2NH1b}
    &\quad+\sum_{j=1}^{m-1}O\left(h^{-\frac{\ell+1}{2}}h^{j(1-\frac{Y}{2})}\int_{0}^{\infty}\omega^{jY+\ell}\sin\bigg(\omega\frac{\varepsilon_0^{\pm}}{\sigma\sqrt{h}}\bigg)e^{-\omega^{2}/2}\,d\omega\right)\\
    \label{T3NH1}
    &\quad+O\left(h^{-\frac{\ell+1}{2}}h^{m(1-Y/2)}\right)+O \left(h^{-\ell-\frac{1}{2}} \, \varepsilon^{\ell} e^{-\frac{\varepsilon^2}{2\sigma^2 h}} \right).
 \end{align}
Each of the terms in \eqref{T1NH1} is $O(h^{( 1-Y-\ell)/Y}e^{-\delta h^{\frac{Y-2}{Y}}})$. For the terms in \eqref{T2NH1}-\eqref{T2NH1b}, we recall the estimates (see (A.11) and (A.12) in \cite{gong2021}):
\begin{align*} %
a_{j,r}^{\pm}(h)&:=h^{j(1-Y/2)}\int_{0}^{\infty}\omega^{jY+r}\cos\bigg(\omega\cdot\frac{\varepsilon}{\sigma\sqrt{h}}\bigg)e^{-\omega^{2}/2}\,d\omega = O\bigg(\frac{h^{j+(r+1)/2}}{\varepsilon^{jY+r+1}}\bigg),\\
d_{j,r}^{\pm}(h)&:=h^{j(1-Y/2)}\int_{0}^{\infty}\omega^{jY+r}\sin\bigg(\omega\cdot\frac{\varepsilon}{\sigma\sqrt{h}}\bigg)e^{-\omega^{2}/2}\,d\omega= O\bigg(\frac{h^{j+(r+1)/2}}{\varepsilon^{jY+r+1}}\bigg),
\end{align*}
valid when  $h\to{}0$ and $\varepsilon/\sqrt{h}\to{}\infty$. Then, we can conclude that the leading term in \eqref{T2NH1} and  \eqref{T2NH1b} corresponds to $j=1$ and is of order $O \left(h \varepsilon^{-Y-\ell-1}\right)$, which is asymptotically larger than every term in \eqref{T1NH1}. The first term in \eqref{T3NH1} can be made of higher order by choosing $m$ large enough. We then conclude \eqref{e:moments_of_FT_expression}.

Turning to \eqref{e:J^kphi_asymptotics_1}, 
we apply Plancherel to get:
\begin{align}
    &\wt{\bE}\left(\left(\mp  J'_h \right)^k\,\phi\bigg(\frac{\varepsilon\pm J'_h}{\sigma\sqrt{h}}\bigg)\right)\\
    &\quad= \int_{\bR}(\cF\psi)(z)z^k p_{J,h}^{\mp}(z)\,dz= (-i)^k \int_{\bR}\psi^{(k)}(u)\cF\big(p_{J,h}^{\mp}(z)\big)(u)\,du\\
    &\quad=  \varepsilon^k \, \frac{\sigma\sqrt{h}}{2\pi} \, \sum_{l=0}^k \sum_{m=0}^{\left \lfloor l/2 \right \rfloor} \frac{(-1)^{l+m}\,2^{l-2m}  \, k!}{(k-l)! \, m! \, (l-2m)!} \left(i\varepsilon\right)^{-l} \left(\frac{\sigma^2 h}{2}\right)^{l-m} \mathcal{I}({l-2m})\\
    &\quad=    O \left(h^{\frac{3}{2}} \varepsilon^{k-Y-1}\right)+O \left(\varepsilon^{k}\,e^{-\frac{\varepsilon^2}{2\sigma^2 h}} \right)+O\left(\varepsilon^{k}h^{\frac{2-Y}{2Y}}\,e^{-\delta h^{\frac{Y-2}{Y}}}\right),  \label{eq:EJk}
\end{align}
where on the last line we made use of the asymptotic expression \eqref{e:moments_of_FT_expression} for $\mathcal I(\ell)$.  This establishes \eqref{e:J^kphi_asymptotics_1}.  
For the second asymptotic bound, based on \eqref{eq:EJk}, we can compute
\begin{align*}
    & \wt{\bE}\left(\left(\frac{\varepsilon \pm  J'_h}{\sigma\sqrt{h}}\right)^k\,\phi\bigg(\frac{\varepsilon\pm J'_h}{\sigma\sqrt{h}}\bigg)\right) \\
    &\quad = (\sigma\sqrt{h})^{-k} \, \wt\bE\left(\sum_{m=0}^k\binom{k}{m}\varepsilon^{k-m}(\pm  J'_h)^m \phi\bigg(\frac{\varepsilon\pm J'_h}{\sigma\sqrt{h}}\bigg)\right)\\
    &\quad= 
    (\sigma\sqrt{h})^{-k} \sum_{m=0}^k \binom{k}{m}\varepsilon^{k-m} \left(O \left(h^{\frac{3}{2}} \varepsilon^{m-Y-1}\right)+O \left(\varepsilon^{m}\,e^{-\frac{\varepsilon^2}{2\sigma^2 h}} \right)+O\left(\varepsilon^{m}h^{\frac{2-Y}{2Y}}\,e^{-\delta h^{\frac{Y-2}{Y}}}\right)\right).
\end{align*}
This clearly implies the result.
\end{proof}

The following Lemma is used in the proof of Lemma \ref{prop:EX2}.
\begin{lemma} \label{lemma:W2k}
Fix a positive integer $k$. Then,  %
as $h\to 0$,
\begin{align*}
  \bE\left(W_h^{2k}\,{\bf 1}_{\{|\sigma W_{h}+J_{h}|\leq\varepsilon\}}\right)&=    {(2k-1)!!\,h^k + %
   O\left(h^2 \varepsilon^{2k-Y-2}\right)+ o\left(h^{k+1/Y}\right)+ O\left(h^{\frac{1}{2}}\varepsilon^{2k-1} e^{-\frac{\varepsilon^2}{2\sigma^2 h}}\right)}.%
\end{align*}
\end{lemma}
\begin{proof}
The case of $k=1$ has been proved in the Step 1 of Lemma 3.1 of \cite{gong2021}. Following a similar procedure, we generalize to the case of $k \geq 2$.  Note
\begin{align}
\bE\Big(W_{h}^{2k}\,{\bf 1}_{\{|\sigma W_{h}+J_{h}|\leq\varepsilon\}}\Big) &= (2k-1)!! \, h^k - \bE\Big(W_{h}^{2k}\,{\bf 1}_{\{|\sigma W_{h}+J_{h}|>\varepsilon\}}\Big)\\
&= (2k-1)!! \, h^k - \wt{\bE}\Big(e^{-\wt{U}_{h}-\eta h}\,W_{h}^{2k}\,{\bf 1}_{\{|\sigma W_{h}+J_{h}|>\varepsilon\}}\Big)\\
&= (2k-1)!! \, h^k - h^k e^{-\eta h}\,\wt{\bE}\Big(W_{1}^{2k}\,{\bf 1}_{\{|\sigma\sqrt{h}W_{1}+J_{h}|>\varepsilon\}}\Big)\\
&\quad -h^k e^{-\eta h}\,\wt{\bE}\Big(\Big(e^{-\wt{U}_{h}}-1\Big)W_{1}^{2k}\,{\bf 1}_{\{|\sigma\sqrt{h}W_{1}+J_{h}|>\varepsilon\}}\Big)\\
\label{eq:kDecompb1eps1} &=:(2k-1)!! \, h^k - h^k e^{-\eta h}I_{1}(h) - h^ke^{-\eta h}I_{2}(h).
\end{align}
We first analyze the asymptotic behavior of $I_1(h)$. Write
\begin{align}\label{eq:kDecompb1eps12}
I_{1}(h)=\wt{\bE}\Big(W_{1}^{2k}\,{\bf 1}_{\{\sigma\sqrt{h}W_{1}+J_{h}>\varepsilon\}}\Big)+\wt{\bE}\Big(W_{1}^{2k}\,{\bf 1}_{\{\sigma\sqrt{h}W_{1}-J_{h}>\varepsilon\}}\Big)=:I_{1}^{+}(h)+I_{1}^{-}(h).
\end{align}
Recalling $\phi$ and $\overline{\Phi}$ as in \eqref{e:def_phi,Phi-bar}, by repeating integration by parts, we have for all $x \in \bR$,
\begin{align*}
    \wt\bE\left(W_1^{2k}\,{\bf 1}_{\{W_1 >x\}}\right) = (2k-1)!!\, \overline{\Phi}(x) + \sum_{j=0}^{k-1}\frac{(2k-1)!!}{(2j+1)!!} \, x^{2j+1} \phi(x).
\end{align*}
Also, note that expression (A.21) of \cite{gong2021} shows
\begin{align}\label{eq:LimitExpPhi}
\wt{\bE}\left(\overline{\Phi}\bigg(\frac{\varepsilon}{\sigma\sqrt{h}}\pm\frac{J_{h}}{\sigma\sqrt{h}}\bigg)\right)=O\Big(e^{-\varepsilon^{2}/(2\sigma^{2}h)}\Big)+O\big(h\varepsilon^{-Y}\big),%
\quad\text{as }\,h\rightarrow 0.
\end{align}
Thus, by conditioning on $J_{h}$ and then applying Lemma \ref{lemma:4results},
\begin{align}\nonumber
I_{1}^{\pm}(h) &= \wt{\bE}\left((2k-1)!!\,  \overline{\Phi}\bigg(\frac{\varepsilon}{\sigma\sqrt{h}}\mp\frac{J_{h}}{\sigma\sqrt{h}}\bigg) + \sum_{j=0}^{k-1}\frac{(2k-1)!!}{(2j+1)!!} \, \bigg(\frac{\varepsilon}{\sigma\sqrt{h}}\mp\frac{J_{h}}{\sigma\sqrt{h}}\bigg)^{2j+1} \phi\bigg(\frac{\varepsilon}{\sigma\sqrt{h}}\mp\frac{J_{h}}{\sigma\sqrt{h}}\bigg) \right)\\
&= O\Big(e^{-\frac{\varepsilon^{2}}{2\sigma^{2}h}}\Big)+O\big(h\varepsilon^{-Y}\big) +  O \left(h^{2-k} \varepsilon^{2k-2-Y}\right)\\
&\quad + O\left(h^{\frac{1}{2}-k}\varepsilon^{2k-1} e^{-\frac{\varepsilon^2}{2\sigma^2 h}}\right)+O\left(h^{\frac{1}{Y}-k}\varepsilon^{2k-1} e^{-\delta h^{\frac{Y-2}{Y}}}\right) \\
&= O \left(h^{2-k} \varepsilon^{2k-Y-2}\right) + O\left(h^{\frac{1}{2}-k}\varepsilon^{2k-1} e^{-\frac{\varepsilon^2}{2\sigma^2 h}}\right)+O\left(h^{\frac{1}{Y}-k}\varepsilon^{2k-1} e^{-\delta h^{\frac{Y-2}{Y}}}\right). 
\label{eq:Decompb1eps12pm}
\end{align}
Using the same procedure as shown in the Step 1.2 of Theorem 3.1 of \cite{gong2021}, we have that  $I_2(h) = o(h^{1/Y})$. {Note that the last term \eqref{eq:Decompb1eps12pm} is not needed since this is of smaller order than the first term $O \left(h^{2-k} \varepsilon^{2k-Y-2}\right)$.} The result then follows.
\end{proof}
The following Lemma is used in the proofs of Propositions \ref{lemma:Estable}, \ref{lemma:2E}, and Lemma \ref{lemma:J2k}.
\begin{lemma} \label{lemma:S2k_new} 
Let $\gamma_0\in \bR$, and  consider $S_h$ as in \eqref{e:def_St}.  Assume that $\varepsilon \to 0$ with $\varepsilon \gg \sqrt{h}$.  Then, for any fixed $k\geq 1$,
\begin{align}
\wt\bE\left(S_h^{2k}\,{\bf 1}_{\{|\sigma W_{h}+S_{h} + h\gamma_0|\leq\varepsilon\}}\right) &= \frac{C_{+}\!+\!C_{-}}{2k-Y}h\varepsilon^{2k-Y} +  \frac{\sigma^2(C_++C_-)(2k-1-Y)}{2} h^{2}\,\varepsilon^{2k-Y-2}\\
& \quad +O\left(h^3\varepsilon^{2k-4-Y}\right) + O\big(h^{2}\,\varepsilon^{2k-2Y}\big) +  O\left(\varepsilon^{2k-1-Y}h^{\frac{3}{2}}e^{-\frac{\varepsilon^2}{2\sigma^2 h}}\right). \label{e:E-tilde_Sh_expansion}
\end{align}
\end{lemma}
\begin{proof}

First suppose $\gamma_0=0$. For simplicity, let $I(x)=({\frac{-\varepsilon  - x \sigma \sqrt{h}}{h^{1/Y}}},{\frac{\varepsilon -   x \sigma \sqrt{h}}{h^{1/Y}}})$ and also let $r=r(h) = \frac{\varepsilon}{\sigma \sqrt h}\to \infty$ as $h\to 0$.  Then, in terms of the density $p_S$ of $S_1$ under $\wt \bP$,
\begin{align}
\wt\bE\Big(S_{h}^{2k}\,{\bf 1}_{\{|{\sigma W_{h}}+{S_{h}}|\leq\varepsilon\}}\Big)&={h^{2k/Y}\wt\bE\Big(S_{1}^{2k}\,{\bf 1}_{\{|\sigma \sqrt h W_{1}+h^{1/Y}S_{1}|\leq\varepsilon\}}\Big)}\\
&=h^{2k/Y} \Big\{\int_{\{|x|<r\}} + \int_{\{|x|>r\}}\Big\}\int_{I(x)}u^{2k} p_S(u) du \phi(x) dx\\
&=:h^{2k/Y}(T_1 + T_2).\label{e:i1,i2}
\end{align}
We first focus on $T_1$. Write $\mathcal{D}(u):=p_S(u)-(C_-\mathbf 1_{\{u<0\}} + C_+\mathbf 1_{\{u>0\}}) |u|^{-1-Y}$, recall $\bar{C}=C_++C_-$, and observe
\begin{align}
T_1
 &= \frac{ \bar{C}}{2k-Y}(\varepsilon h^{-1/Y})^{2k-Y}\int_{-r}^r\Big( 1 - \frac{x}{r} \Big)^{2k-Y}  \phi(x)dx+ \int_{-r}^r\int_{I(x)} u^{2k} \mathcal{D}(u)du \phi(x) dx
  \label{e:C_term1}
 \end{align}
Letting $f(u) = (1-u)^{2k-Y}$, $q_n(u) = \sum_{\ell=0}^n \frac{f^{(\ell)}(0)}{\ell!} u^\ell$, and $R_n(u) = f(u)-q_n(u)$,
the integral in first term of  \eqref{e:C_term1} may then be written as
\begin{align}
\int_{-r}^r\Big( 1 - \frac{x}{r} \Big)^{2k-Y}\phi(x)dx  
& =\Big\{\int_{-r/2}^{r/2 } + \int_{(-r,r)\setminus(-r/2,r/2)}\Big\}f(x/r)\phi(x)dx \\
& =  \int_{-r/2}^{r/2} q_n(x/r) \phi(x) dx + \int_{-r/2}^{r/2} R_n(x/r) \phi(x)dx +O(r^{-1}e^{-r^2/ 8}).
\end{align}
For $|u|<\frac{1}{2}$,  there is a constant $K'=K'(n,Y)$ such that $|R_n(u)|\leq K'|u|^{n+1}$. Thus, 
\[ \int_{-r/2}^{r/2} |R_n(x/r)| \phi(x)dx \leq K'r^{-(n+1)} \int_{-r/2}^{r/2}x^{n+1}\phi(x)dx =O(r^{-(n+1)}).\]
Moreover, using that $\int_{r/2}^\infty x^n\phi(x) dx=O(r^{(n-1)} e^{-(r/2)^2/2})$,
$$
\int_{-r/2}^{r/2} q_n(x/r) \phi(x) dx  = \int_{-\infty}^\infty q_n(x/r) \phi(x) dx + O(r^{(n-1)} e^{-r^2/8}).
$$
Hence, taking $n=3$, 
\begin{align}\label{NIE0}
\int_{-r}^r\Big( 1 - \frac{x}{r} \Big)^{2k-Y}\phi(x)dx 
= 1 + \frac{(2k-Y)(2k-1-Y)}{2} \frac{1}{r^2} + O(r^{-4}),\quad r\to\infty,
\end{align}
where we used that $\int_\bR x^\ell \phi(x) dx = 0$ when $\ell$ is odd.
To study the second term on the right-hand side of   \eqref{e:C_term1}, we consider the cases $k=1$ and $k>1$ separately.  First suppose $k=1$.  By applying Lemma \ref{K0Is0} and using the symmetry of $\phi$,
\begin{align*}
 \int_{-r}^r\int_{I(x)} u^2\mathcal{D}(u)du \phi(x) dx=
- \int_{-r}^r \int_{\bR\setminus(-b(x),b(x))} u^2 \mathcal{D}(u) du\, \phi(x) dx,
 \end{align*}
 where $b(x) = \frac{\varepsilon + x \sigma \sqrt{h}}{h^{1/Y}}=\frac{\varepsilon(1+x/r)}{h^{1/Y}} \in (0,2\varepsilon h^{-1/Y})$.
Observe 
$0<b(x)<1$ if and only if $x\in \big(-\frac{\varepsilon}{\sigma \sqrt h},\frac{h^{1/Y}-\varepsilon}{\sigma \sqrt h}\big) =\big(-r, r(\frac{h^{1/Y} }{\varepsilon }-1)\big)$. So, we consider 
\begin{align}
    \int_{-r}^r\int_{b(x)}^\infty u^{2} |\mathcal{D}(u)|du\, \phi(x) dx = \left(\int_{-r}^{r(\frac{h^{1/Y} }{\varepsilon }-1)} + \int_{ r(\frac{h^{1/Y} }{\varepsilon }-1)}^{r}\right) \int_{b(x)}^\infty u^{2} |\mathcal{D}(u)|du\, \phi(x) dx.
\end{align}
Applying (\refeq{eq:1stOrderEstTailZ}), the first integral above can be estimated as follows:
\begin{align}
\nonumber
\int_{-r}^{(\frac{h^{1/Y} }{\varepsilon }-1)r} \int_{b(x)}^\infty u^2 |\mathcal{D}(u)|du \phi(x) dx
&\leq K\int_{-r}^{(\frac{h^{1/Y} }{\varepsilon }-1)r}\bigg( \int_{b(x)}^1 u^{1-Y} du + \int_1^\infty u^{1-2Y}du\bigg) \phi(x)dx\\
\nonumber
&\leq K\int_{-r}^{(\frac{h^{1/Y} }{\varepsilon }-1)r}\Big(\frac{1}{2-Y} + \frac{1}{|2-2Y|}\Big) \phi(x)dx\\
& \leq K\int_{-r}^{-r/2} \phi(x)dx \leq O(r^{-1} e^{-(r/2)^2/2}).
\label{e:i12,t0}
\end{align}
Now, for $x\in((\frac{h^{1/Y} }{\varepsilon }-1)r,r)$, $b(x)>1$ and, so, again applying (\refeq{eq:1stOrderEstTailZ}),
\begin{align}
\int_{(\frac{h^{1/Y} }{\varepsilon }-1)r}^r \int_{b(x)}^\infty u^{2}|\mathcal{D}(u)|du \phi(x) dx &\leq \int_{-r(1-\varepsilon^{-1}h^{1/Y})}^\infty \int_{b(x)}^\infty u^{2} |\mathcal{D}(u)|du \phi(x) dx\\
&\leq   
K (\varepsilon h^{-1/Y})^{2-2Y}\int_{(\frac{h^{1/Y} }{\varepsilon }-1)r}^\infty\Big(1 + \frac{x}{r}\Big)^{2-2Y} \phi(x) dx\\
&=  O((\varepsilon h^{-1/Y})^{2-2Y})\label{e:i12,t1}
\end{align}
Expressions \eqref{e:i12,t0}-\eqref{e:i12,t1} show that%
\[\int_{-r}^r\int_{b(x)}^\infty u^{2}|\mathcal{D}(u)|du\, \phi(x) dx = O\left(r^{-1} e^{-(r/2)^2/2}\right)+ O((\varepsilon h^{-1/Y})^{2-2Y}) =  O((\varepsilon h^{-1/Y})^{2-2Y}).\]
Similarly, one can show that 
$\int_{-r}^r \int_{-\infty}^{-b(x)} u^{2} |\mathcal{D}(u)|du\, \phi(x) dx =O((\varepsilon h^{-1/Y})^{2-2Y})$ and, thus,
\begin{equation}\label{e:i12,t2_keq1}
 \int_{-r}^r \int_{\bR\setminus(-b(x),b(x))} u^{2}\mathcal{D}(u) du\, \phi(x) dx=O((\varepsilon h^{-1/Y})^{2-2Y}).
\end{equation}
For the case $k>1$, using \eqref{eq:1stOrderEstTailZ},
\begin{align}
 \int_{-r}^r\int_{I(x)} u^{2k}| \mathcal{D}(u)|du \phi(x) dx &\leq  K\int_{-r}^r \int_{I(x)} u^{2k-2Y-1} du \phi(x)dx = O((\varepsilon h^{-1/Y})^{2k-2Y})\label{e:i12,t2_kgeq1}.
\end{align}
Plugging \eqref{NIE0}, \eqref{e:i12,t2_keq1}, and \eqref{e:i12,t2_kgeq1} into 
\eqref{e:C_term1}, we conclude that for $k\geq 1$,
\begin{equation}\label{e:I3_expansion}
T_1= \frac{\bar{C}}{2k-Y}\varepsilon^{2k-Y} h^{1-2k/Y} \Big( 1 + \frac{(2k-Y)(2k-1-Y)}{2} \frac{1}{r^2} +  O(r^{-4})\Big) + O((\varepsilon h^{-1/Y})^{2k-2Y}).
\end{equation}
Now we examine $T_2$ in \eqref{e:i1,i2}.
Again applying \eqref{eq:1stOrderEstTailZ},%
\begin{align}
\nonumber
 \int_{r}^\infty\int_{\frac{-\varepsilon - x \sigma \sqrt{h}}{h^{1/Y}}}^{\frac{\varepsilon - x \sigma \sqrt{h}}{h^{1/Y}}} |{u^{2k}} p_S(u) \phi(x) | du\, dx &\leq 
 K (\varepsilon h^{-1/Y})^{2k-Y}\int_r^\infty (1+x/r)^{2k-Y} \phi(x) dx\\
 \nonumber
  & \leq K  (\varepsilon h^{-1/Y})^{2k-Y}\int_r^\infty (2x/r)^{2k-Y} \phi(x) dx\\
  \nonumber
    & \leq K  (\varepsilon h^{-1/Y})^{2k-Y}(1/r)^{2k-Y} \times  r^{2k-Y-1} e^{-r^2/2}\\
  & = O\left((\varepsilon h^{-1/Y})^{2k-Y} r^{-1} e^{-r^2/2}\right).\label{e:I4_asymptotics}
  \end{align}
Analogously, 
\[\int_{-\infty}^{-r}\int_{\frac{-\varepsilon - x \sigma \sqrt{h}}{h^{1/Y}}}^{\frac{\varepsilon - x \sigma \sqrt{h}}{h^{1/Y}}} |{u^{2k}} p_S(u) \phi(x) | du\, dx= O\left((\varepsilon h^{-1/Y})^{2k-Y} r^{-1} e^{-r^2/2}\right).\]
 Thus, based on  \eqref{e:i1,i2}, \eqref{e:I3_expansion}, and \eqref{e:I4_asymptotics},
\begin{align}\label{eq:s2}
\bE\Big(S_{h}^{ 2k}\,{\bf 1}_{\{|\sigma W_{h}+S_{h}|\leq\varepsilon\}}\Big) 
&= \frac{\bar{C}}{2k-Y} h \varepsilon^{2k-Y} + \frac{\bar{C}(2k-1-Y)}{2} \sigma^2 {h^{2}\,\varepsilon^{2k-Y-2}}\\
&\quad + O\left(h^3\varepsilon^{2k-4-Y}\right) + O\big(h^{2}\,\varepsilon^{2k-2Y}\big) +  O\left(\varepsilon^{2k-1-Y}h^{3/2}e^{-\varepsilon^2/(2\sigma^2 h)}\right).
\end{align}
This completes the case $\gamma_0=0$.  For $\gamma_0 \neq 0$, take $h>0$ small enough so that $|\gamma_0|h<\varepsilon$, and observe %
\begin{align*}
\left|{\bf 1}_{\{|\sigma W_{h}+S_h + h\gamma_0|\leq\varepsilon\}}-{\bf 1}_{\{|\sigma W_{h}+S_h |\leq\varepsilon\}}\right| & \leq {\bf 1}_{\{\varepsilon- h|\gamma_0|<|\sigma W_{h}+S_h | \leq  \varepsilon\}}+{\bf 1}_{ \{\varepsilon<|\sigma W_{h}+S_h | \leq  \varepsilon + h|\gamma_0|\}}\\
& \leq 2 {\bf 1}_{\{\varepsilon- h|\gamma_0|<|\sigma W_{h}+S_h | \leq   \varepsilon+h|\gamma_0|\}}.
\end{align*}
Thus,
\begin{align}
 \wt\bE&\left(S_h^{2k}\,\left|{\bf 1}_{\{|\sigma W_{h}+S_h + h\gamma_0|\leq\varepsilon\}}-{\bf 1}_{\{|\sigma W_{h}+S_h |\leq\varepsilon\}}\right|\right) \\
 \ \ \ & \leq  2 h^{2k/Y} \wt\bE\left(S_1^{2k}\, {\bf 1}_{\{\varepsilon- h|\gamma_0|<|\sigma W_{h}+S_h | \leq   \varepsilon+h|\gamma_0|\}}\right) \\
 \ \ \  & =  K h^{2k/Y}\int_\bR\bigg\{  \int_{b(x)-\gamma_0h^{1-1/Y}}^{b(x)+\gamma_0h^{1-1/Y}} +  \int_{-b(x)-\gamma_0h^{1-1/Y}}^{-b(x)+\gamma_0h^{1-1/Y}}\bigg\}u^{2k} p_S(u) du \phi(x) dx \\
 \ \ \   & \leq  K h^{2k/Y} \int_\bR \int_{b(x)-\gamma_0h^{1-1/Y}}^{b(x)+\gamma_0h^{1-1/Y}}|u|^{2k-1-Y}  du \phi(x) dx \\
 \ \ \   & \leq  K h^{2k/Y+1-1/Y} \int_\bR \Big(\Big|\frac{\varepsilon + \sigma h^{1/2}x + h|\gamma_0|}{h^{1/Y}}\Big|^{2k-Y-1}\vee\Big|\frac{\varepsilon + \sigma h^{1/2}x - h|\gamma_0|}{h^{1/Y}}\Big|^{2k-Y-1}\Big) \phi(x) dx \\
 \ \ \     & \leq  K h^{2} \varepsilon^{2k-Y-1} \int_\bR \Big(1  + \frac{x}{r} +  o(1)\Big)^{2k-Y-1} \phi(x) dx\\
  \ \ \    &= O(h^{2} \varepsilon^{2k-Y-1}) = o(h^{2}\,\varepsilon^{2k-2Y}),
 \end{align}
 where we used the mean value theorem in the third inequality. Hence, 
\begin{align*}
 \wt\bE(S_h^{2k} {\bf 1}_{\{|\sigma W_{h}+S_h + h\gamma_0|\leq\varepsilon\}}) &= \wt\bE(S_h^{2k} {\bf 1}_{\{| \sigma W_{h}+S_h |\leq\varepsilon\}})+ \wt\bE\left(S_h^{2k}\,\left({\bf 1}_{\{|\sigma W_{h}+S_h + h\gamma_0|\leq\varepsilon\}}-{\bf 1}_{\{|\sigma W_{h}+S_h |\leq\varepsilon\}}\right)\right)\\
 & =  \wt\bE(S_h^{2k} {\bf 1}_{\{| \sigma W_{h}+S_h|\leq\varepsilon\}}) + o(h^{2}\,\varepsilon^{2k-2Y}).
 \end{align*}
In view of \eqref{eq:s2}, this implies the result for arbitrary $\gamma_0\in\bR$.
\end{proof}

The lemma below is similar to a result used in Step 2 of the proof of Theorem 3.1 of \cite{gong2021}; here we provide a sharper result and generalize to an arbitrary integer $k\geq 1$. The lemma is used in the proofs of Lemmas \ref{prop:EX2} and \ref{lemma:WJ}.
\begin{lemma} \label{lemma:J2k}
Assume that $\varepsilon \to 0$ with $\varepsilon \gg \sqrt{h}$. Let $k\geq 1$ be an integer.  Then, as $h\to 0$,
\begin{align*}
 \bE\left(J_h^{2k}\,{\bf 1}_{\{|\sigma W_{h}+J_{h}|\leq\varepsilon\}}\right)
  &= \frac{C_{+}\!+\!C_{-}}{2k-Y}h\varepsilon^{2k-Y} + O\big(h^2\varepsilon^{2k-Y-2}\big) + O\big(h\varepsilon^{2k-Y/2}\big).
\end{align*}
\end{lemma}
\begin{proof}

Recall $J_h = S_h + \wt\gamma h$, where $S_h$ is $Y$-stable under $\widetilde \bP$. %
Note that
\begin{align}
\bE\Big(J_{h}^{2k}\,{\bf 1}_{\{|\sigma W_{h}+J_{h}|\leq\varepsilon\}}\Big)&=\wt{\bE}\Big(e^{-\wt{U}_{h}-\eta h}\,J_{h}^{2k}\,{\bf 1}_{\{|\sigma W_{h}+J_{h}|\leq\varepsilon\}}\Big)\\
&=e^{-\eta h}\,\sum_{j=0}^{{2k}}\frac{(2k)!}{j!\,(2k-j)!}(\wt\gamma h)^{2k-j}\, \wt{\bE}\Big(e^{-\wt{U}_{h}}S_h^{j}{\bf 1}_{\{|W_{h}+S_h+\wt{\gamma}h|\leq\varepsilon\}}\Big).
\end{align}
Now, denoting $\wt{C}_{\ell}=\int_{\bR_{0}}\big(e^{-\ell\varphi(x)}-1+\ell\varphi(x)\big)\tilde{\nu}(dx)$, $\ell=1,2$, by {Assumption \ref{assump:Funtq}-(v)} we have
\begin{align} \label{eq:EwtU}
\wt{\bE}\left(\big(e^{-\wt{U}_{h}}-1\big)^{2}\right)=e^{\wt{C}_{2}h}-2e^{\wt{C}_{1}h}+1\sim\big(\wt{C}_{2}-2\wt{C}_{1}\big)h,\quad\text{as }\,h\rightarrow 0.
\end{align}
Therefore,  %
\begin{align}
&\wt{\bE}\Big(e^{-\wt{U}_{h}}S_{h}^{j}{\bf 1}_{\{|W_{h}+S_h+\wt{\gamma}h|\leq\varepsilon\}}\Big)\\
&=\wt{\bE}\Big(S_h^{j}\,{\bf 1}_{\{|\sigma W_{h}+S_h+\wt{\gamma}h|\leq\varepsilon\}}\Big)+\wt{\bE}\left(\big(e^{-\wt{U}_{h}}-1\big)S_h^{j}\,{\bf 1}_{\{|\sigma W_{h}+S_h+\wt{\gamma}h|\leq\varepsilon\}}\right)\\
&\leq \bigg(\wt{\bE}\Big(S_h^{2j}\,{\bf 1}_{\{|\sigma W_{h}+S_h+\wt{\gamma}h|\leq\varepsilon\}}\Big)\bigg)^{1/2}\left(1+\Big(\wt{\bE}\big(e^{-\wt{U}_{h}}-1\big)^{2}\Big)^{1/2}\right)\\
&=\bigg(\wt{\bE}\Big(S_h^{2j}\,{\bf 1}_{\{|\sigma W_{h}+S_h+\wt{\gamma}h|\leq\varepsilon\}}\Big)\bigg)^{1/2} \left( 1+ O\left(h^{1/2}\right)\right),\quad h\to 0. \label{e:E-tilde_exp(-U)S_h}%
\end{align}
In particular, applying Lemma \ref{lemma:S2k_new} with $\gamma_0=\wt\gamma$, based on expression \eqref{e:E-tilde_exp(-U)S_h}, for any $0\leq j \leq 2k-1$,  we see that
\begin{align*}
 h^{2k-j}\wt{\bE}\Big(e^{-\wt{U}_{h}}S_{h}^{j}{\bf 1}_{\{|W_{h}+S_h+\wt{\gamma}h|\leq\varepsilon\}}\Big) = h^{2k-j} O\left(\left(\wt{\bE}S_h^{2j}\,{\bf 1}_{\{|\sigma W_{h}+S_h+\wt{\gamma}h|\leq\varepsilon\}}\right)^{1/2}\right)
  =O(h^{3/2}\varepsilon^{2k-1-Y/2} ) .
\end{align*}

\noindent Thus, by Lemma \ref{lemma:S2k_new},%
\begin{align*}
    &\bE\Big(J_{h}^{2k}\,{\bf 1}_{\{|\sigma W_{h}+J_{h}|\leq\varepsilon\}}\Big)\\
    &=e^{-\eta h}\,\sum_{j=0}^{{2k}}\frac{(2k)!}{j!\,(2k-j)!}(\wt\gamma h)^{2k-j}\, \wt{\bE}\Big(e^{-\wt{U}_{h}}S_h^{j}{\bf 1}_{\{|W_{h}+S_h+\wt{\gamma}h|\leq\varepsilon\}}\Big)\\
    &= e^{-\eta h}\,\wt{\bE}\Big(e^{-\wt{U}_{h}}S_h^{2k}{\bf 1}_{\{|W_{h}+S_h+\wt{\gamma}h|\leq\varepsilon\}}\Big) + O(h^{3/2}\varepsilon^{2k-1-Y/2} )\\%
    &=e^{-\eta h}\,\wt{\bE}\Big(S_h^{2k}\,{\bf 1}_{\{|\sigma W_{h}+S_h+\wt{\gamma}h|\leq\varepsilon\}}\Big) + e^{-\eta h}\,\wt{\bE}\Big(\Big(e^{-\wt{U}_{h}}-1\Big)S_h^{2k}\,{\bf 1}_{\{|\sigma W_{h}+S_h+\wt{\gamma}h|\leq\varepsilon\}}\Big) + O\left(h^{3/2}\varepsilon^{2k-1-Y/2}\right)\\
        &= e^{-\eta h}\wt{\bE}\Big(S_h^{2k}\,{\bf 1}_{\{|\sigma W_{h}+S_h+\wt{\gamma}h|\leq\varepsilon\}}\Big) + O\left(h^{1/2}\varepsilon^{2k-Y/2}\right) \cdot O\left(h^{1/2}\right) + O\left(h^{3/2}\varepsilon^{2k-1-Y/2}\right)\\
    &= \big(1+O(h)\big)\wt{\bE}\Big(S_h^{2k}\,{\bf 1}_{\{|\sigma W_{h}+S_h+\wt{\gamma}h|\leq\varepsilon\}}\Big) + O\left(h\varepsilon^{2k-Y/2}\right),
\end{align*}
where in the last line we used that $h^{3/2}\varepsilon^{2k-1-Y/2}\ll h \varepsilon^{2k-Y/2}$ since $\sqrt h \ll \varepsilon$, and on the second-to-last line we applied Lemma \ref{lemma:S2k_new}.  The statement follows, since $h\wt{\bE}\Big(S_h^{2k}\,{\bf 1}_{\{|\sigma W_{h}+S_h+\wt{\gamma}h|\leq\varepsilon\}}\Big) = O(h^2 \varepsilon^{2k-Y}) = o\left(h\varepsilon^{2k-Y/2}\right)$ by Lemma \ref{lemma:S2k_new}.
\end{proof}

The following lemma is used in the proof of Lemma \ref{prop:EX2}.
\begin{lemma} \label{lemma:WJ}
Assume as $h\to 0$, $\varepsilon \to 0$ with $\varepsilon \gg  h^{1/2-\bar\delta}$ for some $\bar\delta>0$.
Let $a,b\geq 1$ be integers. Then,
\begin{equation}
     \bE\left(W_h^{a}\,J_h^{b}\,{\bf 1}_{\{|\sigma W_{h}+J_{h}|\leq\varepsilon\}}\right)\\
    =  \begin{cases}
    O\left(h^{2} \varepsilon^{a+b-Y-2}\right) + O\big(h^{1+a/2} \varepsilon^{b-Y/2}\big), & a \text{ odd},\\
    O\left(h^{2} \varepsilon^{a+b-Y-2}\right)+O(h^{1+a/2}\varepsilon^{b-Y}), & a,b \text{ even},\\
        O\left(h^{2} \varepsilon^{a+b-Y-2}\right)  +  O(h^{\frac{1+a}{2}}\varepsilon^{b-Y/2}),& a \text{ even}, b \text{ odd}.
    \end{cases}
    \end{equation}
\end{lemma}
\begin{proof}
First, we decompose $\bE\left(W_h^{a}\,J_h^{b}\,{\bf 1}_{\{|\sigma W_{h}+J_{h}|\leq\varepsilon\}}\right)$ as
\begin{align}\nonumber
\bE\left(W_h^{a}\,J_h^{b}\,{\bf 1}_{\{|\sigma W_{h}+J_{h}|\leq\varepsilon\}}\right) 
\nonumber
&= h^{a/2} e^{-\eta h}\Big[\wt{\bE}\Big(W_{1}^{a}\,J_h^{b}\,{\bf 1}_{\{|\sigma W_{h}+J_{h}|\leq\varepsilon\}}+\big(e^{-\wt{U}_{h}}-1\big) \, W_{1}^{a}\,J_h^{b}\,{\bf 1}_{\{|\sigma W_{h}+J_{h}|\leq\varepsilon\}}\Big)\Big]\\
&=: h^{a/2} e^{-\eta h}\big(T_{1}(h) + T_{2}(h)\big).\label{eq:DecompWJ} 
\end{align}
We first analyze the asymptotic behavior of $T_1(h)$.
Note that by repeating integration by parts, we have for all $x_1, x_2 \in \bR$, and $x_1<x_2$,
\begin{align}
    &\wt\bE\left(W_1^{a} \,{\bf 1}_{\{x_1<W_1<x_2\}}\right) = \begin{cases}
      \sum_{j=0}^{d} \frac{(a-1)!!}{(2j)!!} \left(x_1^{2j} \phi(x_1) - x_2^{2j} \phi(x_2)\right),&  a = 2d+1, \\
     \begin{array}{l}
     \sum_{j=0}^{d-1} \frac{(a-1)!!}{(2j+1)!!} \left(x_1^{2j+1} \phi(x_1) - x_2^{2j+1} \phi(x_2)\right)\\
     \quad+ (a-1)!!\,\left( \overline{\Phi}(x_1) - \overline{\Phi}(x_2)\right) ,
     \end{array}&  a = 2d,
      \end{cases} \label{e:W_moments_formula}
\end{align}
where $d\geq 0$ is an integer.
However, by applying Lemma \ref{lemma:4results} and noting that under the assumption $\varepsilon \gg  h^{1/2-\delta}$,  $\sqrt{h}\varepsilon^j e^{-\frac{\varepsilon ^2}{2\sigma^2 h}} \ll h^{\frac{3}{2}} \varepsilon^{j-Y-1} $ and $\sqrt{h}\varepsilon^j e^{-\delta h^{\frac{Y-2}{Y}}}\ll h^{\frac{3}{2}} \varepsilon^{j-Y-1} $, for any integer $j$, %
\begin{align}
\wt\bE\left((\mp J_h)^b \left(\frac{\varepsilon \mp J_h}{\sigma\sqrt{h}}\right)^{k} \phi \left(\frac{\varepsilon \pm J_h}{\sigma\sqrt{h}}\right) \right)& = (\sigma\sqrt{h})^{-k}\wt\bE\left( \sum_{m=0}^{k} \binom{k}{m} \varepsilon^{k-m}(\mp J_h)^{m+b} \, \phi \left(\frac{\varepsilon \pm J_h}{\sigma\sqrt{h}}\right) \right)\\
    &=  \left(\frac{\varepsilon}{\sigma\sqrt{h}}\right)^k  \sum_{m=0}^{k} \binom{k}{m} \varepsilon^{-m}\, O\left(h^{3/2} \varepsilon^{m+b-Y-1}\right)\\
    &= O\left(  h^{\frac{3-k}{2}} \varepsilon^{k+b-Y-1}\right). \label{e:moments_J_eq1}
\end{align}
Moreover, using Lemma \ref{lemma:J2k}, we see that
\begin{align}
&\wt\bE\left(J_h^b \, \overline{\Phi}\left(\frac{-\varepsilon - J_h}{\sigma\sqrt{h}}\right)\right) - \wt\bE\left(J_h^b\, \overline{\Phi}\left(\frac{\varepsilon - J_h}{\sigma\sqrt{h}}\right)\right)\\
    &\quad= \wt\bE\left(J_h^b \, {\bf 1}_{\{|\sigma W_{h}+J_{h}|\leq\varepsilon\}}\right) = \begin{cases}
    O(h\varepsilon^{b-Y}), & b \text{ even},\\
     O(h^{1/2}\varepsilon^{b-Y/2}),& b \text{ odd}.\label{e:moments_J_eq2}
    \end{cases}
\end{align}
 where we used Cauchy-Schwarz for the case when $b$ is odd. Thus, by conditioning on $J_h$, we may use the bounds \eqref{e:moments_J_eq1} and \eqref{e:moments_J_eq2} in expresssion \eqref{e:W_moments_formula} to obtain %
\begin{align}
    T_1(h) =  \begin{cases}
    O\left(h^{2-a/2} \varepsilon^{a+b-Y-2}\right),  & a \text{ odd},\\
    O\left(h^{2-a/2} \varepsilon^{a+b-Y-2}\right)+O(h\varepsilon^{b-Y}), & a,b \text{ even},\\
        O\left(h^{2-a/2} \varepsilon^{a+b-Y-2}\right)  +  O(h^{1/2}\varepsilon^{b-Y/2}),& a \text{ even}, b \text{ odd}.
    \end{cases} \label{e:eq_T1}
\end{align}
Now we turn to $T_2(h)$. Recall that by \eqref{eq:EwtU}, $\wt{\bE}(e^{-\wt{U}_{h}}-1\big)^{2}=O(h)$ as $h\rightarrow 0$. 
Then by Cauchy-Schwarz, %
\begin{align} 
    T_2(h) &= \wt{\bE}\Big(\big(e^{-\wt{U}_{h}}-1\big) \, W_{1}^{a}\,J_h^{b}\,{\bf 1}_{\{|\sigma\sqrt{h}W_{1}+J_{h}|\leq\varepsilon\}}\Big)\\
    &\leq \left( \wt{\bE}\left(W^{2a}_1\big(e^{-\wt{U}_{h}}-1\big)^2\,\right)  \right)^{1/2} \cdot \left( \wt\bE \left(J_h^{2b}\,{\bf 1}_{\{|\sigma\sqrt{h}W_{1}+J_{h}|\leq\varepsilon\}}\right) \right)^{1/2}\\
    &\leq K \left( \wt{\bE} \big(e^{-\wt{U}_{h}}-1\big)^2 \right)^{1/2} \left(O\big(h\varepsilon^{2b-Y}\big)\right)^{1/2}\\
    &=  O\big(h \varepsilon^{b-Y/2}\big), \label{e:eq_T2}
\end{align}
where we used the independence of $W_1$ and $U_t$ in the third line above.
Therefore, since $h \varepsilon^{b-Y/2} \ll h\varepsilon^{b-Y} $ and $h \varepsilon^{b-Y/2} \ll h^{1/2}\varepsilon^{b-Y/2}$, by \eqref{eq:DecompWJ}, \eqref{e:eq_T1} and \eqref{e:eq_T2}, the result holds.
\end{proof}

The following lemma is used in the proof of Proposition \ref{lemma:Estable}.%
\begin{lemma}\label{lemma:lambda}
As $h\to 0$,
\begin{align*}
    \wt \bE\left(\overline{\Phi}\bigg( \frac{\varepsilon}{\sigma\sqrt{h}}\pm\frac{S_{h}}{\sigma\sqrt{h}}\bigg)\right)  &= \frac{C_\mp}{Y}h\varepsilon^{-Y} + C_\mp(1+Y)\sigma^2 h^2\varepsilon^{-2-Y} + \frac{C_\mp (1+Y)(2+Y)(3+Y)}{8} \sigma^4 h^3\varepsilon^{-4-Y} \\
    &\quad   + O\left(h^{4} \varepsilon^{-Y-6}\right) + O\left(h^{3/2}\varepsilon^{-1-Y}e^{-\frac{\varepsilon^2}{2\sigma^2 h}}\right)\cdot\left(1+h\varepsilon^{-2}\right)\\
    &\quad +O(h\varepsilon^{2-2Y}) \cdot (1 + h^{1/2} \varepsilon^{-1} + h \varepsilon^{-2})\\%
    &\quad  + O\left(h \varepsilon^{3-2Y-Y/2}\right)+ O\left( h^{1/2}\varepsilon^{-1}e^{-\frac{\varepsilon^2}{2\sigma^2 h}}\right)+ O\left(h^{-1/2} \varepsilon^{3-Y}  e^{-\frac{\varepsilon^2}{8\sigma^2 h}} \right).
\end{align*}
\end{lemma}
\begin{proof}
Fix $0<\bar\delta<1/32$.  Let $ S^{\bar\delta}$ be a L\'evy process with no diffusion part, L\'evy measure $\wt \nu(dz){\bf 1}_{|z|\leq \bar\delta\varepsilon}$, and third component of the characteristic triplet  $\gamma_{\bar\delta} := {\tilde\gamma} - \int_{\bar\delta\varepsilon < |z| \leq 1} z\wt\nu(dz)$. %
Let $N^{\bar\delta}$ a standard Poisson process with intensity $\lambda_{\bar\delta}:= \wt\nu(\{z:|z|>{\bar\delta}\varepsilon\})$, $(\xi_i^{\bar\delta})_{i\geq 1}$ be a sequence of i.i.d. random variables with probability distribution $\wt\nu(dz){\bf 1}_{|z|> \bar\delta\varepsilon}/\lambda_{\bar\delta}$ (again under both $\wt\bP$ and $\bP$).
We also assume that $S^{\bar\delta}$, $N^{\bar\delta}$, and $(\xi_i^{\bar\delta})_{i\geq 1}$ are independent. Then, by construction, under $\wt\bP$, $S_h \eqDist S_h^{\bar\delta} + \sum_{i=1}^{N_h^{\bar\delta}} \xi_i^{\bar\delta}$.
Therefore, we can express $\wt \bE(\overline{\Phi}((\varepsilon\pm S_{h})/{\sigma\sqrt{h}})) = \wt \bP\left(\sigma W_h \mp S_h \geq \varepsilon\right)=\bP(\sigma W_h \mp S_h^{\bar\delta} \mp \sum_{i=1}^{N_h^{\bar\delta}} \xi_i^{\bar\delta} \geq \varepsilon)$ and conditioning on $N_{h}^{\bar\delta}$,
\begin{align}
\nonumber
     \wt \bE\left(\overline{\Phi}\bigg(\frac{\varepsilon}{\sigma\sqrt{h}}\pm\frac{S_{h}}{\sigma\sqrt{h}}\bigg)\right)
    &= e^{-\lambda_{\bar\delta} h} \, \bP\left(\sigma W_h \mp S_h^{\bar\delta} \geq \varepsilon\right)\\
    \label{eq:decomp}
    & \quad +\lambda_{\bar\delta} h\, e^{-\lambda_{\bar\delta} h} \, \bP\left(\sigma W_h \mp S_h^{\bar\delta} \mp \xi_1^{\bar\delta}\geq \varepsilon\right)\\
    & \quad + e^{-\lambda_{\bar\delta} h}\sum_{k\geq 2}\frac{(\lambda_{\bar\delta} h\,)^k}{k!} \, \bP\left(\sigma W_h \mp S_h^{\bar\delta} \mp \sum_{i=1}^k\xi_i^{\bar\delta}\geq \varepsilon\right).
\end{align}
To estimate the first term in (\refeq{eq:decomp}), by a Taylor approximation,
\begin{align}
    \bP\left(\sigma W_h \mp S_h^{\bar\delta} \geq \varepsilon\right) &= \bE\left(\overline{\Phi}\left( \frac{\varepsilon}{\sigma\sqrt{h}}\pm\frac{S_{h}^{\bar\delta}}{\sigma\sqrt{h}}\right)\right)\\
    \nonumber
    &= \overline{\Phi}\left(\frac{\varepsilon}{\sigma\sqrt{h}}\right) \pm \overline{\Phi}'\left(\frac{\varepsilon}{\sigma\sqrt{h}}\right)\,\bE \left(\frac{S_h^{\bar\delta}}{\sigma\sqrt{h}}\right) + \bE\left(\frac{(S_h^{\bar\delta})^2}{\sigma^2 h}\int_0^1 (1-\theta)  \overline{\Phi} '' \left(\frac{\varepsilon\pm\theta S_h^{\bar\delta}}{\sigma\sqrt{h}}\right) d\theta \right) \\
    \label{e:firstterm_decomp}
    &= \bP\left(\sigma W_h \geq \varepsilon\right) \, \mp\,\phi\left(\frac{\varepsilon}{\sigma\sqrt{h}}\right)\frac{\gamma_{\bar\delta} h}{\sigma \sqrt{h}} \\
    \nonumber
    & \quad + \sigma^{-3} h^{-3/2} (2\pi)^{-1/2} \, \bE\left((S_h^{\bar\delta})^2 \int_0^1 (1-\theta)\, e^{-\frac{(\varepsilon \pm \theta S_h^{\bar\delta})^2}{2\sigma^2 h}}  (\varepsilon \pm\theta S_h^{\bar\delta}) d\theta\right).
\end{align}
To estimate the last term in \eqref{e:firstterm_decomp}, first note that for any integer $k\geq 2$,
\begin{align*}
    \bE\left((S_h^{\bar\delta})^k\right)\sim h\int_{|z|\leq {\bar\delta} \varepsilon} z^k\wt\nu(dz) = O(h\varepsilon^{k-Y}).
\end{align*}

Since $\bar \delta<1/32$, by similar arguments to those shown in expression (46) of \cite{mijatovic2016new}, it holds that
    $\bP\left(|S_h^{\bar\delta}| > \frac{\varepsilon}{2}\right) \leq C_0(h\varepsilon^{-Y})^4$,
    for some constant $C_0$.
Thus, we can compute
\begin{align}
    & \bE\left((S_h^{\bar\delta})^2 \int_0^1 (1-\theta)\, e^{-\frac{(\varepsilon\pm\theta S_h^{\bar\delta})^2}{2\sigma^2 h}} d\theta\right) \\
    &= \bE\left((S_h^{\bar\delta})^2\, {\bf 1}_{\{|S_h^{\bar\delta}| \leq \frac{\varepsilon}{2}\}} \int_0^1 (1-\theta)\, e^{-\frac{(\varepsilon\pm\theta S_h^{\bar\delta})^2}{2\sigma^2 h}} d\theta\right) + \bE\left((S_h^{\bar\delta})^2\,{\bf 1}_{\{
    |S_h^{\bar\delta}| > \frac{\varepsilon}{2}\}} \int_0^1 (1-\theta)\, e^{-\frac{(\varepsilon\pm\theta S_h^{\bar\delta})^2}{2\sigma^2 h}} d\theta\right)\\
    & \leq  \bE\left((S_h^{\bar\delta})^2\right) \, e^{-\frac{\varepsilon^2}{8\sigma^2 h}} + \bE\left((S_h^{\bar\delta})^2\,{\bf 1}_{\{|S_h^{\bar\delta}| > \frac{\varepsilon}{2}\}} \right)\\
    &\leq  O\left(h \varepsilon^{2-Y}  e^{-\frac{\varepsilon^2}{8\sigma^2 h}} \right) + \bE\left((S_h^{\bar\delta})^4\right)^{1/2}  \bP\left(|S_h^{\bar\delta}| > \frac{\varepsilon}{2}\right)^{1/2}\\
    &=  O\left(h \varepsilon^{2-Y}  e^{-\frac{\varepsilon^2}{8\sigma^2 h}} \right) + O\left(h^{5/2} \varepsilon^{2-2Y-Y/2}\right). \label{e:firstterm_remainder_term1}
\end{align}
Similarly, 
\begin{align}
     \bE\left((S_h^{\bar\delta})^3 \int_0^1 (1-\theta)\theta e^{-\frac{(\varepsilon\pm\theta S_h^{\bar\delta})^2}{2\sigma^2 h}} d\theta\right)
    & \leq  \bE\left((S_h^{\bar\delta})^3\right) \, e^{-\frac{\varepsilon^2}{8\sigma^2 h}} + \bE\left((S_h^{\bar\delta})^3\,{\bf 1}_{\{|S_h^{\bar\delta}| > \frac{\varepsilon}{2}\}} \right)\\
    &=  O\left(h \varepsilon^{3-Y}  e^{-\frac{\varepsilon^2}{8\sigma^2 h}} \right) + O\left(h^{5/2} \varepsilon^{3-2Y-Y/2}\right). \label{e:firstterm_remainder_term2}
\end{align}
Then, based on \eqref{e:firstterm_decomp}, using $\bP\left(\sigma W_h \geq \varepsilon\right) \pm \phi\left(\frac{\varepsilon}{\sigma\sqrt{h}}\right)\frac{\gamma_{\bar\delta} h}{\sigma \sqrt{h}} = O(h^{1/2}\varepsilon^{-1}e^{-\frac{\varepsilon^2}{2\sigma^2 h}})$ together with  \eqref{e:firstterm_remainder_term1} and  \eqref{e:firstterm_remainder_term2}, we obtain
\begin{align}\label{eq:lambda1}
    \bP\left(\sigma W_h \mp S_h^{\bar\delta} \geq \varepsilon\right) = O\left( h^{1/2}\varepsilon^{-1}e^{-\frac{\varepsilon^2}{2\sigma^2 h}}\right) + O\left(h^{-1/2} \varepsilon^{3-Y}  e^{-\frac{\varepsilon^2}{8\sigma^2 h}} \right) + O\left(h \varepsilon^{3-2Y-Y/2}\right).
\end{align}
This establishes the asymptotic behavior of the first term in \eqref{eq:decomp}. To estimate the second term in (\refeq{eq:decomp}), again by a Taylor approximation,
\begin{align}
\label{e:secondterm_decomp1a}
    \bP\left(\sigma W_h \mp S_h^{\bar\delta} \mp\xi_1^{\bar\delta} \geq \varepsilon\right) 
    &= \bP\left(\sigma W_h  \mp \xi_1^{\bar\delta} \geq \varepsilon\right) \mp \bE\left(\phi\left(\frac{\varepsilon \pm \xi_1^{\bar\delta}}{\sigma\sqrt{h}}\right)\right) \frac{\gamma_{\bar\delta} h}{\sigma \sqrt{h}} \\
    \nonumber
    & \quad +  \sigma^{-3} h^{-3/2} (2\pi)^{-1/2} \, \bE\left((S_h^{\bar\delta})^2 \int_0^1 (1-\theta)\, e^{-\frac{(\varepsilon \pm \xi_1^{\bar\delta} \pm\theta S_h^{\bar\delta})^2}{2\sigma^2 h}}  (\varepsilon \pm \xi_1^{\bar\delta} \pm\theta S_h^{\bar\delta}) d\theta\right)\\
    &= \bP\left(\sigma W_h  \mp \xi_1^{\bar\delta} \geq \varepsilon\right) + O\left(h^{1/2} \varepsilon^{1-Y}\right) + O\left( \varepsilon^{2-Y} \right), \label{e:secondterm_decomp1}
\end{align}
where the last equality is because
\begin{align*}
    \bE\left((S_h^{\bar\delta})^2 \int_0^1 (1-\theta)\, e^{-\frac{(\varepsilon \mp \xi_1^{\bar\delta} \mp\theta S_h^{\bar\delta})^2}{2\sigma^2 h}}  {|\varepsilon \mp \xi_1^{\bar\delta} \mp\theta S_h^{\bar\delta}|} d\theta\right)& \leq\sigma \sqrt 2\sup_{x} |x| e^{ -x^2}\cdot \sqrt h \bE \left((S_h^{\bar\delta})^2\right)
    = O(h^{3/2}\varepsilon^{2-Y}).%
\end{align*}
For the first term in \eqref{e:secondterm_decomp1}, note
\begin{align}
    \lambda_{\bar\delta} \bP\left(\sigma W_h  + \xi_1^{\bar\delta} \geq \varepsilon\right) 
    &= \int_{-\infty}^{-{\bar\delta} \varepsilon} \bP(\sigma W_h \geq \varepsilon -z) \wt \nu (dz) + \int_{{\bar\delta}\varepsilon}^\infty \bP(\sigma W_h \geq \varepsilon -z) \wt \nu (dz)\\
    &\quad - \int_\varepsilon^\infty {\bf 1}_{\{|z|> {\bar\delta} \varepsilon\}}\wt \nu (dz) + \lambda_{\bar\delta} \bP\left( \xi_1^{\bar\delta} \geq \varepsilon\right).\label{eq:probdecomp}
\end{align}
Now, for the first term in (\refeq{eq:probdecomp}), we employ the tail bound $\bP(|W_1|>x)\leq \frac{1}{x\sqrt{2\pi}}\exp\{-x^2/2\}$:
\begin{align}
    \int_{-\infty}^{-{\bar\delta} \varepsilon} \bP(\sigma W_h \geq \varepsilon -z) \wt \nu (dz)   
    &\leq \frac{\sigma \sqrt{h}}{\varepsilon\sqrt{2 \pi}} e^{-\frac{\varepsilon^2}{2\sigma^2 h}} \cdot \frac{C_-}{Y}({\bar\delta} \varepsilon)^{-Y}= O\left(h^{1/2}\varepsilon^{-1-Y}e^{-\frac{\varepsilon^2}{2\sigma^2 h}}\right). \label{eq:probdecomp_0}
\end{align}
Then we estimate the second and the third term in (\refeq{eq:probdecomp}),
\begin{align}
    &\int_{{\bar\delta} \varepsilon}^\infty \bP(\sigma W_h \geq \varepsilon -z) \wt\nu(dz)- \int_\varepsilon^\infty \wt\nu(dz)\\
    &= \int_{{\bar\delta} \varepsilon}^\varepsilon \bP(\sigma W_h \geq \varepsilon -z) \wt\nu(dz) - \int_\varepsilon^\infty \bP(\sigma W_h \leq \varepsilon -z) \wt\nu(dz) \\
    &= C_+\int_{0}^{\varepsilon (1-{\bar\delta})} \bP(\sigma W_h \geq u) \, (\varepsilon-u)^{-1-Y} du - C_+\int_0^\infty \bP(\sigma W_h \leq -u) \, (\varepsilon+u)^{-1-Y} du\\
    &= C_+\int_{0}^{\varepsilon (1-{\bar\delta})} \bP(\sigma W_h \geq u) \, \big((\varepsilon-u)^{-1-Y} - (\varepsilon+u)^{-1-Y}\big)du  \\
    & \quad - C_+\int_{\varepsilon (1-{\bar\delta})}^\infty \bP(\sigma W_h \leq -u) \, (\varepsilon+u)^{-1-Y} du=: T_1-T_2.\label{e:secondterm_decomp}
\end{align}
For the first term $T_1$ in \eqref{e:secondterm_decomp}, for simplicity let $r= r(h)= \frac{\varepsilon}{\sigma \sqrt h}\to \infty$ as $h\to 0$, and also let $f(u) =(1- u)^{-1-Y} - (1+u)^{-1-Y}$.  Also, recall $\bE\left(|W_1|^{k+1}\right) =2k \int_0^\infty  x^k \bP\left(W_1>x\right)\,dx$.  By a 4th order Taylor approximation of $f(u)$ at $u=0$,
\begin{align}
    T_1&= C_+\sigma h^{1/2} \varepsilon^{-1-Y}\int_0^{r(1-{\bar\delta})} \bP( W_1 \geq x) f(x/r) dx\\
    & = C_+\sigma h^{1/2} \varepsilon^{-1-Y}\int_0^{r(1-{\bar\delta})} \bP( W_1 \geq x) \Big(f'(0)(x/r)   + \frac{f^{(3)}(0) (x/r)^3}{3!} + \frac{f^{(5)}(\vartheta(x)) (x/r)^5}{5!} \Big) dx\\
                \nonumber
        & = C_+\sigma h^{1/2} \varepsilon^{-1-Y} \Bigg(\frac{f'(0) \bE W_1^2 }{r}   + \frac{f^{(3)}(0)  \bE W_1^4 }{r^3\cdot3!} + O(r^{-5}) - O\Big(\int_{r(1-{\bar\delta})}^\infty \bP( W_1 \geq x) f'(0)(x/r)dx \Big) \Bigg)\\
                    \nonumber
            & = C_+\sigma h^{1/2} \varepsilon^{-1-Y} \bigg(\frac{f'(0) \bE W_1^2 }{r}   + \frac{f^{(3)}(0)  \bE W_1^4 }{r^3\cdot3!} \bigg) + O\left(h^{3} \varepsilon^{-Y-6}\right)-  O\left(h^{3/2} \varepsilon^{-3-Y} e^{-\frac{\varepsilon^2(1-\delta)^2}{2\sigma^2 h}}\right) \\   
            \nonumber  
    &= C_+(1+Y)\sigma^2 h \varepsilon^{-2-Y} + \frac{C_+(1+Y)(2+Y)(3+Y)}{8}  \sigma^4 h^2\varepsilon^{-4-Y}\\
    &\qquad + O\left(h^{3/2} \varepsilon^{-3-Y} e^{-\frac{\varepsilon^2(1-\delta)^2}{2\sigma^2 h}}\right) + O\left(h^{3} \varepsilon^{-Y-6}\right). \label{e:secondterm_decomp_term1}
\end{align}
The second term in \eqref{e:secondterm_decomp} can be estimated as
\begin{align}
|T_2|
    & \leq C_+ \, \bP(\sigma W_h \leq \varepsilon ({\bar\delta} -1)) \int_{\varepsilon (1-{\bar\delta})}^\infty  \, (\varepsilon+u)^{-1-Y} du
    = O\left(h^{1/2} \varepsilon^{-1-Y} e^{-\frac{\varepsilon^2(1-\delta)^2}{2\sigma^2 h}}\right). \label{e:secondterm_decomp_term2}
\end{align}
Finally, for the fourth term in (\refeq{eq:probdecomp}),
\begin{align}
    \lambda_{\bar\delta} \bP\left( \xi_1^{\bar\delta} \geq \varepsilon\right) = \int_{ \varepsilon}^\infty \wt\nu(dz) = \frac{C_+}{Y}\varepsilon^{-Y} \label{e:fourthterm}
\end{align}
Then, combining \eqref{eq:probdecomp_0}, \eqref{e:secondterm_decomp_term1},  \eqref{e:secondterm_decomp_term2}, and \eqref{e:fourthterm}, and by repeating analogous arguments for $ \lambda_{\bar\delta} \bP\left(\sigma W_h  - \xi_1^{\bar\delta} \geq \varepsilon\right) $, based on \eqref{eq:probdecomp} we obtain%
\begin{align*}
    \lambda_{\bar\delta} \bP\left(\sigma W_h  \mp \xi_1^{\bar\delta} \geq \varepsilon\right) &= \frac{C_\mp}{Y}\varepsilon^{-Y} + C_\mp(1+Y)\sigma^2 h \varepsilon^{-2-Y} + \frac{C_\mp(1+Y)(2+Y)(3+Y)}{8} \sigma^4 h^2\varepsilon^{-4-Y}\\
    &\quad + O\left(h^{3} \varepsilon^{-Y-6}\right) +  O\left(h^{1/2}\varepsilon^{-1-Y}e^{-\frac{\varepsilon^2}{2\sigma^2 h}}\right)\cdot\left(1+h\varepsilon^{-2}\right).
\end{align*}
Therefore, by \eqref{e:secondterm_decomp1}, and using that $\lambda_{\bar\delta} =O(\varepsilon^{-Y})$,
\begin{align}\label{eq:lambda2}
    & \lambda_{\bar\delta} h\, e^{-\lambda_{\bar\delta} h} \, \bP\left(\sigma W_h \mp S_h^{\bar\delta} \mp \xi_1^{\bar\delta}\geq \varepsilon\right)\\ %
    &= \frac{C_\mp}{Y}h\varepsilon^{-Y} + C_\mp(1+Y)\sigma^2 h^2\varepsilon^{-2-Y} + \frac{C_\mp (1+Y)(2+Y)(3+Y)}{8} \sigma^4 h^3\varepsilon^{-4-Y}\\
    &\quad + O\left(h^{4} \varepsilon^{-Y-6}\right) + O\left(h^{3/2}\varepsilon^{-1-Y}e^{-\frac{\varepsilon^2}{2\sigma^2 h}}\right)\cdot\left(1+h\varepsilon^{-2}\right) + {O(h\varepsilon^{2-2Y}) \cdot (1 + h^{1/2} \varepsilon^{-1} )}.
\end{align}
This establishes the asymptotic behavior of the second term in \eqref{eq:decomp}.  For the third term in \eqref{eq:decomp}, simply note
\begin{equation}\label{eq:lambda3}
e^{-\lambda_{\bar\delta} h}\sum_{k\geq 2}\frac{(\lambda_{\bar\delta} h\,)^k}{k!} \, \bP\left(\sigma W_h \mp S_h^{\bar\delta} \mp \sum_{i=1}^k\xi_i^{\bar\delta}\geq \varepsilon\right)\leq \sum_{k\geq 2}\frac{(\lambda_{\bar\delta} h\,)^k}{k!} = O(h^2\varepsilon^{-2Y}).
\end{equation}
Finally, combining (\refeq{eq:lambda1}), (\refeq{eq:lambda2}) and (\refeq{eq:lambda3}), based on \eqref{eq:decomp} we obtain the result.
\end{proof}

The following lemma is used in the proof of Lemma \ref{lemma:S2k_new}.
\begin{lemma} \label{K0Is0}
	Let $\mathcal{D}(u):=p_S(u)-(C_-\mathbf 1_{\{u<0\}} + C_+\mathbf 1_{\{u>0\}}) |u|^{-1-Y}$. Then, 
\begin{align}\label{K0Is0Prf}
	K_S:=\int_{-\infty}^\infty u^2 \mathcal{D}(u) du =0.
\end{align}
\end{lemma}	
\begin{proof}
{Let $ \varphi_S(\omega)$ denote the characteristic function of $S_1$ under $\widetilde \bP$.  Then, with $\varsigma$ and $\rho$ denoting the scale and skewness parameters of $S_1$, respectively, we have}
\begin{align}
\varphi_S(\omega) &  = \exp\Big(- |\varsigma \omega|^Y\big[1-i\rho\hspace{0.5mm}\sgn(\omega)\tan(\pi Y/2)\big]\Big)\\
 &=\exp\!\left(\!-(C_{+}\!+C_{-})\,\bigg|\!\cos\bigg(\!\frac{\pi Y}{2}\!\bigg)\!\bigg|\Gamma(-Y)|\omega|^{Y} \bigg(\!1 - i\frac{C_{+}\!-C_{-}}{C_{+}\!+C_{-}}\tan\!\bigg(\!\frac{\pi Y}{2}\!\bigg)\text{sgn}(\omega)\!\bigg)\!\right).%
\end{align}
Without loss of generality, suppose $\varsigma=1.$ For simplicity, let $\mathcal E(u):= u^2\mathcal D(u)$, and observe $\mathcal E \in L^1(\bR)$ since $ Y\in(1,2)$, and hence $\hat {\mathcal E}(\omega) := \int_\bR e^{i\omega u}\mathcal E(u)du$ exists.
Observe that 
$ \int_\bR\mathcal E(u) du =  \hat {\mathcal E}(0).$
 Now, {recall that, for any $\alpha\in(0,1)$,}
\[ \int_0^\infty e^{ix} x^{\alpha-1} dx = e^{i\pi \alpha /2} \Gamma(\alpha), \]
{(see \cite{gradshteyn:ryzhik:2007},  eq.~3.381.7)}, where the integral can be understood as the limit $\lim_{R\to\infty}\int_0^Re^{ix} x^{\alpha-1} dx.$  Thus,
\begin{equation}\label{e:ft_x^(1-Y)}
\int_0^\infty e^{iu\omega}  u^{1-Y}du = - |\omega|^{Y-2} e^{-\text{sgn}(\omega)i\pi Y/2}\Gamma(2-Y), \quad Y\in(1,2).
\end{equation}
This implies
\begin{align*}
 C_+ \int_{0}^\infty e^{ix\omega}|x| ^{1-Y}dx +& C_-\int_{-\infty}^0 e^{ix\omega }|x| ^{1-Y}dx \\
    &= -|\omega|^{Y-2}\, \Gamma(2-Y) \left(C_+ e^{-i \,\text{sgn}(\omega)\pi Y/2}+ C_- e^{i\,\text{sgn}(\omega)\pi Y/2}\right)\\
    &= -|\omega|^{Y-2}\, \Gamma(2-Y) \left(\left(C_+ + C_-\right) \cos(\pi Y/2) - \left(C_+ - C_-\right) i\sin(\pi Y/2)\, \text{sgn}(\omega) \right)\\
    &= |\omega|^{Y-2} (C_{+}\!+C_{-})\,\Big|\cos\bigg(\!\frac{\pi Y}{2}\!\bigg) \Big| \Gamma(2-Y)\left(\!1 - i\frac{C_{+}\!-C_{-}}{C_{+}\!+C_{-}}\tan\!\bigg(\!\frac{\pi Y}{2}\!\bigg)\text{sgn}(\omega)\!\right)\\
    & = |\omega|^{Y-2} Y(Y-1) \Big(1-i\rho\hspace{0.5mm}\sgn(\omega)\tan(\pi Y/2)\Big).%
\end{align*}
However, it can be shown that%
$$
 \int_\bR e^{iu\omega} u^2p_S(u)du = - \varphi_S''(\omega), \quad  \omega \neq 0,
$$
where the integral is interpreted as the limit {$\lim_{R\to\infty}\int_{-R}^R e^{iu\omega}u^2p_S(u)du.$}  Thus, by differentiating {$-\varphi_S(\omega)$} twice, we obtain
\begin{align*}
 \int_\bR e^{iu\omega} u^2p_S(u)du &= \Big(Y(Y-1)|\omega|^{Y-2} - Y^2 |\omega|^{2Y-2}\Big)\Big(1-i\rho\hspace{0.5mm}\sgn(\omega)\tan(\pi Y/2)\Big)\varphi_S(\omega).%
 \end{align*}
Therefore,
\begin{align}
    \hat {\mathcal E}(\omega) &=  -\hat p''(\omega) + |\omega|^{Y-2} Y(Y-1) \left(1-i\rho\hspace{0.5mm}\sgn(\omega)\tan(\pi Y/2)\right)= O\left(|\omega|^{2Y-2}\right),\quad \omega \to 0.
\end{align}
In particular, $K_S=\int_\bR\mathcal E(x) dx =\hat{\mathcal{E}}(0)=0,\quad Y\in (1,2).$
\end{proof}

\begin{proof}[Proof Of (\refeq{ImprovEstI652})]
Let us start with the decomposition
\begin{align}
	I_{5,2}=\wt{\bE}\Big(\Big(e^{-\wt{U}_{h}}-1+\wt{U}_{h}\Big)S_{h}^{2}\,{\bf 1}_{\{|\sigma W_{h}+S_{h}+\wt{\gamma}h|\leq\varepsilon\}}\Big)-\wt{\bE}\Big(\wt{U}_{h}S_{h}^{2}\,{\bf 1}_{\{|\sigma W_{h}+S_{h}+\wt{\gamma}h|\leq\varepsilon\}}\Big)
	=:T_1-T_2.
\end{align}
Since $|\sigma W_{h}+S_{h}+\wt{\gamma}h|\leq\varepsilon$ implies that $|S_h|<\varepsilon+\sigma|W_h|+|\wt{\gamma}|h$ and $e^{-x}-1+x\geq{}0$ for all $x$, we can estimate the first term as follows:
\begin{align}\label{DBNH0}
0\leq{}T_1&\leq{}\wt{\bE}\Big(\left(e^{-\wt{U}_{h}}-1+\wt{U}_{h}\right)\left(\varepsilon+\sigma|W_h|+|\widetilde{\gamma}| h\right)^{2}\Big)\\
	&=O(\varepsilon^{2})\wt{\bE}\Big(e^{-\wt{U}_{h}}-1+\wt{U}_{h}\Big)=O(\varepsilon^{2}h).
\end{align}
For $T_2$, we consider the following decomposition
\begin{align*}
\wt{U}_{h}=\int_{0}^{h}\int_{\bR_{0}}\big(\varphi(x)+\alpha_{\text{sgn}(x)}x\big)\wt{N}(ds,dx)-\int_{0}^{h}\int_{\bR_{0}}\alpha_{\text{sgn}(x)}x\wt{N}(ds,dx)=:\wt{U}^{\text{BV}}_{h}-\alpha_{+}{S}_{h}^{+}-\alpha_{-}{S}^{-}_{h},
\end{align*}
where the first integral is well-defined in light of Assumption \ref{assump:Funtq}-(i) \& (ii). Then, we have 
\begin{align}\nonumber
	T_2&=\wt{\bE}\Big(\wt{U}^{BV}_{h}S_{h}^{2}\,{\bf 1}_{\{|\sigma W_{h}+S_{h}+\wt{\gamma}h|\leq\varepsilon\}}\Big)\\
	&\quad-\alpha_+\wt{\bE}\Big(S_h^+S_{h}^{2}\,{\bf 1}_{\{|\sigma W_{h}+S_{h}+\wt{\gamma}h|\leq\varepsilon\}}\Big)
	-\alpha_-\wt{\bE}\Big(S_h^-S_{h}^{2}\,{\bf 1}_{\{|\sigma W_{h}+S_{h}+\wt{\gamma}h|\leq\varepsilon\}}\Big).
	\label{IANTH}
\end{align}
For the first term we proceed as in \eqref{DBNH0}:
\begin{equation}
	\wt{\bE}\Big(|\wt{U}^{BV}_{h}|S_{h}^{2}\,{\bf 1}_{\{|\sigma W_{h}+S_{h}+\wt{\gamma}h|\leq\varepsilon\}}\Big)\leq{}
	\wt{\bE}\Big(|\wt{U}^{BV}_{h}| \left(\varepsilon+\sigma|W_h|+|\widetilde{\gamma}| h\right)^{2}\Big)
	=O(\varepsilon^2)\wt{\bE}\Big(|\wt{U}^{BV}_{h}| \Big)
	=O(\varepsilon^2 h) \label{e:T2_term1}
\end{equation}
since $\wt{\bE}\Big(\big|\wt{U}^{\text{BV}}_{h}\big|\Big)\leq 2h\int_{\bR_{0}}\big|\varphi(x)+\alpha_{\text{sgn}(x)}x\big|\,\wt{\nu}(dx)<\infty$ due to Assumption \ref{assump:Funtq}.
For the expectations in \eqref{IANTH}, we apply H\"older's inequality to obtain
\begin{align}\label{HInq}
	\wt{\bE}\Big(|S_h^{\pm}|S_{h}^{2}\,{\bf 1}_{\{|\sigma W_{h}+S_{h}+\wt{\gamma}h|\leq\varepsilon\}}\Big) \leq \bigg(\wt{\bE}\left|S_h^{\pm}\right|^p\bigg)^{1/p} \bigg({\wt \bE } \Big(S_{h}^{2}\,{\bf 1}_{\{|\sigma W_{h}+S_{h}+\wt{\gamma}h|\leq\varepsilon\}}\Big)^q\bigg)^{1/q},
\end{align}
where $p\in(1,Y)$ and $q>1$ is such that $p^{-1} + q^{-1}=1$. %
However, by Lemma \ref{lemma:S2k_new},
\begin{align*}
\bigg({\wt \bE }\Big(S_{h}^{2}\,{\bf 1}_{\{|\sigma W_{h}+S_{h}+\wt{\gamma}h|\leq\varepsilon\}}\Big)^q\bigg)^{1/q} &=  \Big(O \big (h\varepsilon^{2q-Y} \big)\Big)^{1/q}= O\big(h^{1/q} \varepsilon^{2-Y/q} \big). %
\end{align*}
Therefore, plugging the above estimate into expression \eqref{HInq}, we get
\begin{align*}
\wt{\bE}\Big(|S_h^{\pm}|S_{h}^{2}\,{\bf 1}_{\{|\sigma W_{h}+S_{h}+\wt{\gamma}h|\leq\varepsilon\}}\Big) &\leq
\big(O(h^{p/Y})\Big)^{1/p} \times O\big(h^{1/q} \varepsilon^{2-Y/q} )\\
\quad  &= O(h^{1/Y +1/q} \varepsilon^{2-Y/q}).
\end{align*}
Note $p\in (1,Y)$ implies $\frac{1}{q}\in (0,1- \frac{1}{Y})$. Therefore,  for all small $\delta>0$,
\begin{align} \label{e:E(SpmS^2)}
\wt{\bE}\Big(|S_h^{\pm}|S_{h}^{2}\,{\bf 1}_{\{|\sigma W_{h}+S_{h}+\wt{\gamma}h|\leq\varepsilon\}}\Big)=O(h^{1/Y}\varepsilon^2 \big(h \varepsilon^{-Y}\big)^{(1-\frac{1}{Y})-\delta}) = O(h \varepsilon^{3-Y} \big(h \varepsilon^{-Y}\big)^{-\delta}).
\end{align}
{Finally, combining \eqref{DBNH0}, \eqref{e:T2_term1}, and \eqref{e:E(SpmS^2)},} we get that 
\begin{align} \label{e:I_{52}_improved}
	I_{5,2}=O(\varepsilon^{2}h)+O(h \varepsilon^{3-Y} \big(h \varepsilon^{-Y}\big)^{-\delta})=O(h \varepsilon^{3-Y} \big(h \varepsilon^{-Y}\big)^{-\delta}).
\end{align}
\end{proof}
}

\bibliographystyle{plain}

\end{document}